\documentclass[11pt]{article} 
\usepackage{amsfonts,amsmath,amssymb,amsthm}
\usepackage{booktabs}
\usepackage{mathrsfs}
\usepackage[round,authoryear]{natbib}
\usepackage[margin=1.0in]{geometry}
\usepackage[sc,center,noindentafter,medium]{titlesec}
\usepackage[onehalfspacing,nodisplayskipstretch]{setspace}
\usepackage{graphicx}
\usepackage{paralist}
\usepackage{tabularx} 
\usepackage[T1]{fontenc}
\usepackage{lmodern}
\usepackage[hypertexnames=false, colorlinks=true, linkcolor=blue, citecolor=blue, bookmarksnumbered]{hyperref}
\usepackage[font=small,labelfont=sc]{caption}
\newtheorem{theorem}{\normalfont\scshape Theorem}[section]

\newtheorem{corollary}{\normalfont\scshape Corollary}[section]
\newtheorem{lemma}{\normalfont\scshape Lemma}[section]
\newtheorem{assumption}{\normalfont\scshape Assumption}[section]
\newtheorem{assnLetter}{\normalfont\scshape Assumption}
\expandafter
\def \expandafter \normalsize \expandafter{\normalsize \setlength \abovedisplayskip{10pt plus 2pt minus 7pt}}
\expandafter
\def \expandafter \normalsize \expandafter{\normalsize \setlength \abovedisplayshortskip{0pt plus 2pt}}
\expandafter
\def \expandafter \normalsize \expandafter{\normalsize \setlength \belowdisplayskip{10pt plus 2pt minus 7pt}}
\expandafter
\def \expandafter \normalsize \expandafter{\normalsize \setlength \belowdisplayshortskip{5pt plus 2pt minus 3pt}}

\numberwithin{equation}{section}
\pltopsep=\medskipamount
\DeclareMathOperator*{\argmin}{\arg\!\min}
\DeclareMathOperator{\var}{\mathbb{V}}

\DeclareMathOperator{\E}{\mathbb{E}}
\DeclareMathOperator{\PP}{\mathbb{P}}
\DeclareMathOperator{\RR}{\mathbb{R}}
\DeclareMathOperator{\diag}{\mathrm{diag}}

\DeclareMathOperator*{\plowto}{\to_{\mathrm{p}}}

\DeclareMathOperator*{\dto}{\overset{d}{\to}}
\DeclareMathOperator*{\dlowto}{\to_{\mathrm{d}}}
\DeclareMathOperator*{\dstarto}{\overset{d*}{\to_{\mathrm{p}}}}

\DeclareMathOperator*{\plim}{\mathrm{plim}}
\DeclareMathOperator*{\op}{o_{\mathrm{p}}}
\DeclareMathOperator*{\Op}{O_{\mathrm{p}}}
\DeclareMathOperator*{\opstar}{o_{\mathrm{p}^\ast}}

\newcommand{\CI}{\mathrm{CI}_n}
\newcommand{\CIrbc}{\mathrm{CI}_{\mathsf{RBC},n}}
\newcommand{\CIboot}{\mathrm{CI}_{\mathsf{Boot},n}}
\newcommand{\CIp}{\mathrm{CI}_{\mathsf{P}\hspace{-0.5pt},n}}
\newcommand{\CIprbc}{\mathrm{CI}_{\mathsf{P} \textsf{-} \mathsf{RBC},n}}

\newcommand{\CIgp}{\mathrm{CI}_{\mathsf{GP}\hspace{-0.5pt},n}}
\newcommand{\CIpgp}{\mathrm{CI}_{\mathsf{PGP}\hspace{-0.5pt},n}}

\newcommand{\CIlp}{\mathrm{CI}_{\mathsf{LP}\hspace{-0.5pt},n}}
\newcommand{\CIplp}{\mathrm{CI}_{\mathsf{PLP}\hspace{-0.5pt},n}}
\newcommand{\CIplprbc}{\mathrm{CI}_{\mathsf{PLP} \textsf{-} \mathsf{RBC},n}}

\newcommand{\CImplp}{\mathrm{CI}_{\mathsf{mPLP}\hspace{-0.5pt},n}}
\newcommand{\CImplprbc}{\mathrm{CI}_{\mathsf{mPLP} \textsf{-} \mathsf{RBC},n}}

\newcommand{\CIrd}{\mathrm{CI}_{\mathsf{rd}\hspace{-0.5pt},n}}
\newcommand{\CIrdrbc}{\mathrm{CI}_{\mathsf{rd} \textsf{-} \mathsf{RBC},n}}

\makeatletter
\renewcommand*{\eqref}[1]{\hyperref[{#1}]{\textup{\tagform@{\ref*{#1}}}}}
\makeatother

\begin{document}

\date{March 3, 2026}

\title{\textsc{improved inference for nonparametric regression\\\vspace*{-0.25cm}and regression-discontinuity designs}\thanks{We are grateful for comments from Matias Cattaneo, Bruce Hansen, Phillip Heiler, Guido Imbens, Michael Jansson, Sid Kankanala, Dennis Kristensen, Xinwei Ma, Mikkel S{\o}lvsten, Guo Yan, seminar participants at Aarhus~U, Bank of Portugal, Harvard-MIT, MacQuarie~U, Queen's~U, Rutgers~U, Tilburg~U, TU~Dortmund, U~Bologna, UC3M, U~Connecticut, U~Copenhagen, U~Exeter, UNSW, U~Sydney, and participants at several conferences and workshops. 
 Cavaliere and Zanelli acknowledge support from the Italian Ministry of University and Research (PRIN 2020 Grant 2020B2AKFW and PRIN 2022 Grant 2020B2AKFW). Nielsen and Zanelli are grateful for support from the Danish National Research Foundation (DNRF Chair grant DNRF154) and the Aarhus Center for Econometrics (ACE) funded by the Danish National Research Foundation grant DNRF186. Gon\c{c}alves acknowledges support from the Natural Science and Engineering Research Council of Canada (NSERC) grant RGPIN-2021-02663.}}
\author{Giuseppe Cavaliere\\University of Bologna, Italy\\\& Exeter Business School, UK
\and S\'{\i}lvia Gon\c{c}alves\\McGill University, Canada\\\& CIREQ, Canada
\and Morten {\O}rregaard Nielsen\\Aarhus Center for Econometrics\\Aarhus University, Denmark
\and Edoardo Zanelli\\Aarhus Center for Econometrics\\Aarhus University, Denmark}

\maketitle

\begin{abstract}
Nonparametric regression and regression-discontinuity designs suffer from smoothing bias that distorts conventional confidence intervals.  Solutions based on robust bias correction (\textsf{RBC}) are now central to the economist's toolbox. In this paper, we establish a novel connection between \textsf{RBC} methods and bootstrap prepivoting. Revisiting RBC through the lens of bootstrapping allows us to develop a novel bias correction procedure which delivers improved nonparametric inference. The resulting confidence intervals are 17\% shorter than the usual intervals employed in curve estimation and regression discontinuity designs, without compromising asymptotic coverage. This holds regardless of evaluation point location, bandwidth choice, or regressor and error distribution.

\bigskip\noindent\textsc{Keywords}: Asymptotic bias; bootstrap; local polynomial estimation; nonparametric regression; regression-discontinuity; robust bias correction.

\bigskip\noindent\textsc{JEL codes}: C14; C21.
\end{abstract}

\newpage

\section{introduction}

Economists routinely rely on nonparametric regression and regression-discontinuity designs (RDD) to estimate causal effects. Inference in this setting is challenging due to the presence of a bias component affecting the nonparametric estimators and statistics usually employed by practitioners, even asymptotically. Several solutions to this inference problem have been proposed in the econometrics and statistics literatures. These include the so-called undersmoothing and bias correction approaches; see, inter alia, Hall (1992), Imbens and Lemieux (2008), Calonico, Cattaneo, and Titiunik (2014), Calonico, Cattaneo and Farrell (2018), and the references therein. A recent leading solution is the robust bias correction (\textsf{RBC}) approach of Calonico et al.\ (2014, 2018), which explicitly resolves this problem by debiasing the reference nonparametric estimator, while adjusting its standard error to account for the additional uncertainty introduced by the bias correction. Bootstrap and resampling methods, though widely used by economists as a bias correction tool, are rarely employed in the present setting due to the fact that they generally fail to correct for smoothing bias (e.g.,\ Hall,~1992). 

The main objective of this paper is to propose a new robust procedure for inference in the context of nonparametric regression and~RDD. Our approach is based on a non-standard application of the bootstrap that implicitly corrects for the bias, thus challenging the above fact. Standard bootstrap methods fail in the present context because the bootstrap cannot correctly mimic the asymptotic bias of the nonparametric estimator, leading to confidence intervals with incorrect coverage and bootstrap p-values which are not uniformly distributed (see also Cavaliere and Georgiev,~2020). We show that prepivoting, originally proposed by Beran (1987) as a means to obtain refinements for asymptotically unbiased estimators, can provide a solution to this failure, as motivated by recent work on asymptotically biased estimators in Cavaliere et al.\ (2024). The idea is that if the asymptotic distribution of the boostrap p-value can be consistently estimated, then we can use it to restore validity by transforming (or prepivoting) a non-uniformly distributed bootstrap p-value into one that is uniformly distributed. In our context, we can use its quantiles to adjust the original bootstrap confidence intervals. 

In this framework, our first contribution is to show that prepivoting performs an implicit bias correction and at the same time adjusts the standard error of the nonparametric regression estimator to account for the additional uncertainty introduced by debiasing. That is, the prepivoted bootstrap interval is asymptotically equivalent to an \textsf{RBC}-style interval. We also show that, for certain bootstrap schemes, bootstrap inference based on prepivoting can be computationally straightforward, as it does not require any resampling algorithm. This desirable property occurs because both the bootstrap bias correction and the standard error adjustment are known functions of the regressors (and the kernel). When the external random variable used to construct the bootstrap confidence interval is drawn from a normal distribution, the prepivoted bootstrap interval is in fact identical to an \textsf{RBC}-style interval.

As our second contribution, we show that the \textsf{RBC} interval of Calonico et al.\ (2014, 2018) is asymptotically equivalent to prepivoting a confidence interval based on the bootstrap scheme considered by Bartalotti et al.\ (2017) and He and Bartalotti (2020) for RDD. For inference based on a local linear estimator at a given evaluation point (e.g., the cutoff for RDD), this bootstrap method generates bootstrap observations using a local quadratic estimator estimated at that point. For this reason, we call this the `global polynomial' (GP) bootstrap. The GP bootstrap is asymptotically invalid when used to construct conventional bootstrap intervals. Bartalotti et al.\ (2017) and He and Bartalotti (2020) estimate the bias by a first level bootstrap, and then use a second level bootstrap to obtain standard errors. However, we show that prepivoting solves the invalidity problem directly by modifying the bootstrap quantiles to reflect the presence of bias. In particular, we show that the prepivoted GP bootstrap (hereafter, \textsf{PGP}) interval is asymptotically equivalent to the \textsf{RBC} interval of Calonico et al.\ (2014, 2018) based on local quadratic estimates of the second-order derivative that enters the bias component. More generally, by changing the estimator of the conditional mean function used in the bootstrap DGP, we obtain different bias corrections and different studentizations.

By exploiting the asymptotic equivalence between prepivoting and \textsf{RBC}-style intervals, our third and main contribution is to revisit the classical bootstrap in nonparametric statistics (e.g., Härdle and Bowman, 1988, Härdle and Marron, 1991, and Hall and Horowitz, 2013). This bootstrap smooths the conditional mean function at each regressor value using a local polynomial estimator, and we label it the `local polynomial' (LP) bootstrap. The inconsistency of naive, i.e., non-prepivoted, implementations of the LP bootstrap when used with `large' (including MSE-optimal) bandwidths is well documented in the statistics literature. Different solutions have been proposed, and they typically require choosing additional bandwidth or tuning parameters. We propose and analyze new prepivoted LP bootstrap schemes (hereafter, \textsf{PLP}), and we show that they (i) deliver novel confidence intervals of the \textsf{RBC} style, not explored in the existing literature, and, more importantly, (ii) lead to efficiency gains (i.e., shorter confidence intervals) with respect to the classic \textsf{RBC} interval. Specifically, we show that the bias correction mechanism implicitly generated by prepivoting these bootstrap methods is more efficient than the bias estimator employed in the \textsf{RBC} procedure, resulting in confidence intervals of shorter length and correct coverage, asymptotically. Importantly, \textsf{PLP} allows using the same bandwidth for point estimation and in the bootstrap data generating process~(DGP). As far as we are aware, ours is the first approach in the nonparametrics literature that delivers a valid and efficient implementation of the LP bootstrap without the need for additional tuning parameters. 

As is usually the case for nonparametric estimators, the asymptotic distribution of \textsf{PLP} statistics depends on whether the evaluation point is interior or boundary. For interior points, prepivoting can be applied directly as discussed above. However, for boundary points, the implicit bias correction mimics the leading bias term of the original statistic only up to a multiplicative factor. The consequence is that the standard prepivoting approach studied by Cavaliere et al.\ (2024) requires modification in this context. Since the multiplicative term depends on the regressors (and the kernel) only, and is therefore known, our solution involves reweighting the bootstrap statistic before applying the prepivoting approach. We show that this `modified' \textsf{PLP} approach (hereafter, \textsf{mPLP}) adapts to the nature of the evaluation point and yields asymptotically valid inference across the entire support of the regressor, including boundary points.

Importantly, the new \textsf{mPLP} method not only has correct asymptotic coverage, but it also delivers shorter interval length compared to the \textsf{RBC} interval. Intuitively, \textsf{mPLP} is based on a convolution of the original observations, and such convolution introduces an additional layer of smoothing. We show that this additional smoothing induces a more efficient bias-correction in the sense that the overall variance of the debiased statistic is smaller. Crucially, this efficiency result is not tied to a specific DGP, as the asymptotic relative lengths of the confidence intervals based on \textsf{mPLP} and \textsf{RBC} depend only on the choice of kernel in the nonparametric regression and on whether the evaluation point is interior or boundary. Table~\ref{tab:lengths} gives the asymptotic relative lengths of the \textsf{mPLP} bootstrap and \textsf{RBC} intervals for five common kernels for the special case of the local linear estimator. For the popular Epanechnikov kernel, \textsf{mPLP} bootstrap intervals are 17\% shorter than \textsf{RBC} intervals for both interior and boundary evaluation points. These efficiency results are supported by our Monte Carlo simulation experiments.

\begin{table}[t]
\caption{Comparison of relative (asymptotic) interval lengths}
\label{tab:lengths}
\vskip -6pt
\begin{tabular*}{\textwidth}{@{\extracolsep{\fill}}lccccc}
\toprule
 & Triangular & Uniform & Epanechnikov  & Biweight & Triweight \\
\midrule
Interior point & 0.84 & 0.86 & 0.83  & 0.85 & 0.86 \\
Boundary point & 0.84 & 0.86 & 0.83 & 0.84 & 0.85 \\
\bottomrule
\end{tabular*}
\vskip 4pt
{\footnotesize \emph{Notes}: Asymptotic relative length of confidence intervals based on \textsf{RBC} and \textsf{mPLP} for five different kernels for the local linear estimator. Entries less than one imply that \textsf{mPLP} is shorter.}
\end{table}

The remainder of this paper is organized as follows. Section~\ref{section RBC and prepivoting} presents some general results on prepivoting in nonparametric regression and its relation to \textsf{RBC}-style intervals, as well as to the GP bootstrap scheme. In Section~\ref{Section main}, we discuss how improved inference can be based on the LP bootstrap. We show asymptotic validity of the corresponding prepivoted confidence interval and analyze its efficiency, in terms of length, compared with the \textsf{RBC} interval. In Section~\ref{Section boundary} we extend our approach to boundary problems and, in particular, to (sharp)~RDD. In Section~\ref{Section Monte Carlo} we assess the performance of our methods in finite samples via Monte Carlo simulation. Section~\ref{Section guidance} provides guidance on practical implementation for applied researchers. Finally, Section~\ref{Section conclusions} concludes. All notation, technical derivations, and proofs are included in the appendix and in the accompanying supplemental material. \textsf{R} packages that implement the procedures in this paper are available at \url{https://pppackages.github.io}.

\section{robust bias correction and prepivoting}
\label{section RBC and prepivoting}

\subsection{preliminaries}
\label{section preliminaries}

Given a random sample $\mathscr{D}_n := \{(y_i, x_i): i=1,\dots,n\}$, consider the problem of inference on the conditional expectation $g(x):=\E[y_i|x_i=x]$ at a fixed point $x=\mathsf{x}$, using a local ($p$-th order, $p$ odd) polynomial estimator $\hat{g}_n(x): = \hat{g}_n(x;h,K)$, where $h=h_n>0$ is the bandwidth and $K$ the kernel function; that is, with $r_p(u)=(1,u,\ldots ,u^{p})^{\prime }$,
\begin{equation}
\label{eq LP estimator}
\hat{g}_n(x):=\iota _0^{\prime }\hat{\beta}_{p,n}(x),\quad \hat{\beta}_{p,n}(x):=\argmin_{b\in \RR ^{p+1}}\sum_{i=1}^{n}
\big(y_i-r_p(x_i-x)^{\prime }b\big) ^2 K\big( \tfrac{x_i-x}{h}\big) .
\end{equation}
The estimator admits the linear closed-form expression $\hat{g}_n(x) = \sum_{i=1}^n w_i (x) y_i /(nh)$ with weights $w_i(x)$ depending on the data, the polynomial order $p$, and the kernel function~$K$; see Appendix~\ref{app:definition}. Inference based on $\hat{g}_n(\mathsf{x})$ is challenging due to the presence of an asymptotic bias. In particular, we typically have
\begin{equation} 
\label{biasedCLT}
T_n:= \sqrt{nh} (\hat{g}_n(\mathsf{x}) - g(\mathsf{x})) \dto N(B,v^2_1),
\end{equation}
where, letting $\mathscr{X}_n:=\{x_i : i=1,\dots,n\}$, the asymptotic bias $B:=\plim_{n \to \infty} B_n$, $B_n:=\E[T_n|\mathscr{X}_n]$, is non-zero when bandwidths with MSE-optimal rate are employed. Sufficient conditions for \eqref{biasedCLT} are given by Assumptions~\ref{Ass_g}--\ref{Ass_K,h} below; e.g., Ruppert and Wand (1994) and Fan and Gijbels (1996). 

\begin{assumption}\label{Ass_g}
$(y_i,x_i)$ are i.i.d.\ and $x_i$ has bounded support $\mathbb{S}_x$ and continuous density~$f$. Let $\mathsf{N}$ be an open neighborhood of~$\mathsf{x}$. Then, on $\mathsf{N} \cap \mathbb{S}_x$: (i)~$f(x)>0$; (ii)~$\sigma^2 (x):=\var[y_i|x_i=x]>0$ is continuous and $\sup_x \E[y_i^4|x_i=x]<\infty$; (iii)~$g^{(p+1)}$ is H\"{o}lder continuous with exponent~$\eta>0$.  
\end{assumption}
\begin{assumption}\label{Ass_K,h}
(i)~The kernel $K$ has support $(-1,1)$ on which it is symmetric, positive, and Lipschitz continuous; (ii)~the bandwidth $h=h_n$ is such that $h + (nh)^{-1} \log n \to 0$ and $nh^{2p+3}\to\kappa \in [0,+\infty)$.
\end{assumption}
Assumption~\ref{Ass_K,h} covers the most widely used kernel functions, e.g., the uniform, Epanechnikov, and triangular kernels, as well as `large' ($\kappa >0$) and undersmoothing ($\kappa = 0$) bandwidths.  Under Assumptions~\ref{Ass_g}--\ref{Ass_K,h}, the bias term $B_n$ satisfies 
\begin{equation} \label{eq:Bn}
B_n  = \sqrt{nh^{2p+3}}g^{(p+1)}(\mathsf{x})C_n /(p+1)!+\op (1),
\end{equation}
where $C_n =C_n (\mathsf{x})$ is a random variable satisfying $C_n \plowto C$ for a non-zero constant $C$ defined in Appendix~\ref{app:definition}. Note that for undersmoothing bandwidths, $B_n=\op (1)$ and $B=0$.

The presence of bias invalidates confidence intervals that ignore it. For instance, with $\alpha \in (0,1)$, a $100(1-\alpha)$\% nominal level conventional interval that ignores the bias is
\begin{equation*} 
    \CI := \big[ \hat{g}_n(\mathsf{x}) \pm z_{1-\alpha/2}(nh)^{-1/2}\hat{v}_{1n}\big],
\end{equation*}
where $\hat{v}^2_{1n}$ is a consistent estimator of~$v^2_1$. However, $\PP ( g(\mathsf{x}) \in  \CI )$ does not converge to $1-\alpha$ as $n\to \infty$. To restore asymptotic validity of the confidence interval, the \textsf{RBC} approach of Calonico et al.\ (2014, 2018) recenters $\CI$ by an estimate of the leading bias term and adjusts the standard error to account for the variability of the bias estimator. With $\hat{B}_{\textsf{RBC},n}$ and $\hat{v}_{\textsf{RBC},n}$ denoting the bias estimate and the adjusted standard error, respectively, the $100(1-\alpha)$\% nominal level \textsf{RBC} interval is 
\begin{equation} 
\label{eq RBC int}
     \CIrbc  := \big[ (\hat{g}_n(\mathsf{x}) - (nh)^{-1/2}\hat{B}_{\textsf{RBC},n}) \pm z_{1-\alpha/2}(nh)^{-1/2} \hat{v}_{\textsf{RBC},n} \big].
\end{equation}
For local polynomial estimators, $\hat{B}_{\textsf{RBC},n}$ is based on nonparametric estimators of higher-order derivatives, which may involve the choice of additional bandwidths. It has the form $\hat{B}_{\textsf{RBC},n} = \sqrt{nh^{2p+3}}\hat{g}_n^{(p+1)}(\mathsf{x})C_n/(p+1)!$, where $C_n$ is as introduced previously, and $\hat{g}_n^{(p+1)}(\mathsf{x})$ is an estimator of~$g^{(p+1)}(\mathsf{x})$. The estimator $\hat{v}^2_{\textsf{RBC},n}$ estimates the variance of the debiased statistic $\sqrt{nh}\hat{g}_n(\mathsf{x})-\hat{B}_{\textsf{RBC},n}$. As such, it is equal to $\hat{v}^2_{1n}$ (the variance estimator of $\sqrt{nh}\hat{g}_n(\mathsf{x})$) plus an additional term that estimates the variance of $\hat{B}_{\textsf{RBC},n}$ and the covariance between $\hat{B}_{\textsf{RBC},n}$ and $\sqrt{nh}\hat{g}_n(\mathsf{x})$; see Calonico et al.\ (2018, Section~3.1).

\subsection{an equivalence result} 
\label{section equivalence}

As we show next, \textsf{RBC} intervals can be derived by prepivoting specific (invalid) bootstrap schemes. In Section~\ref{Section main} we exploit the link between the two approaches to derive new \textsf{RBC}-type intervals with improved length properties.

Let $\hat{g}^*_n(\mathsf{x}): = \hat{g}^*_n(\mathsf{x};h,K)$ denote a bootstrap analogue of $\hat{g}_n(\mathsf{x})$ based on the same kernel function $K$ and the same bandwidth~$h$. Conventional percentile bootstrap intervals are typically based on the quantiles of the bootstrap distribution of $T^*_n$, a given bootstrap analogue of $T_n$ (for instance, $T^*_n = \sqrt{nh} ( \hat{g}^*_n(\mathsf{x}) - \hat{g}_n(\mathsf{x}))$, see Section~\ref{Section main}). Letting $\hat{L}_n(u):= \PP^* ( T_n^* \leq u )$, a standard $100(1-\alpha)$\% nominal level equal-tailed bootstrap interval for $g(\mathsf{x})$ is
\begin{equation} \label{eq general invalid BS}
\CIboot := \big[ \hat{g}_n(\mathsf{x}) - (nh)^{-1/2} \hat{L}^{-1}_n(1-\alpha/2),  \hat{g}_n (\mathsf{x}) - (nh)^{-1/2} \hat{L}^{-1}_n(\alpha/2) \big] ,
\end{equation}
which, however, does not contain $g(\mathsf{x})$ with probability converging to $1-\alpha$. Just as the original nonparametric estimator $\hat{g}_n(\mathsf{x})$ is biased, its bootstrap analogue $\hat{g}^*_n(\mathsf{x})$ is also biased, implying the invalidity of~$\CIboot$. As discussed by Cavaliere et al.\ (2024, Remark~3.2), the root of the problem lies in the fact that the standard bootstrap p-value $\hat{p}_n:=\PP^*(T^*_n\leq T_n)$ is not asymptotically distributed as $U_{[0,1]}$ (standard uniform) in the presence of asymptotic bias.

Prepivoting solves this problem by changing the level of the bootstrap quantiles. Let $\hat{H}_n$ be a (uniformly) consistent estimator of $H$, the asymptotic cumulative distribution function (cdf) of~$\hat{p}_n$. A $100(1-\alpha)$\% nominal level equal-tailed bootstrap interval based on prepivoting is
\begin{equation} \label{BTprepCI}
\CIp := \big[ \hat{g}_n (\mathsf{x}) - (nh)^{-1/2} \hat{L}_n^{-1} (\hat{H}_n^{-1}(1-\alpha/2)),  \hat{g}_n (\mathsf{x}) - (nh)^{-1/2} \hat{L}_n^{-1} (\hat{H}_n^{-1}(\alpha/2 ))\big].
\end{equation}
When $\hat{p}_n\dlowto U_{[0,1]}$ then $H(u)=u$ and $\hat{H}_n^{-1}(\alpha )\plowto\alpha$, in which case the prepivoted interval $\CIp$ is asymptotically equivalent to $\CIboot$. But when asymptotic uniformity of $\hat{p}_n$ fails, $\CIp$ remains asymptotically valid while $\CIboot$ becomes invalid. The main reason why prepivoting restores validity is that if $\hat{H}_n$ is a uniformly consistent estimator of $H$ (regardless of whether the latter is the uniform cdf or not), the prepivoted p-value $\tilde{p}_n:=\hat{H}_n(\hat{p}_n)$ is asymptotically distributed as $U_{[0,1]}$ by the probability integral transform. This implies asymptotic validity of $\CIp$, as shown by Cavaliere et al.\ (2024, Remark~3.4); that is, for any $\alpha \in (0,1)$,
\begin{align}
    \PP ( g(\mathsf{x}) \in \CIp ) &= \PP ( \hat{L}_n^{-1} ( \hat{H}_n^{-1}(\alpha/2 ))\leq T_n \leq  \hat{L}_n^{-1} ( \hat{H}_n^{-1}(1 - \alpha/2 )) ) \nonumber \\
    &=  \PP ( \hat{H}_n^{-1}(\alpha/2 )\leq \hat{p}_n \leq  \hat{H}_n^{-1}(1 - \alpha/2 ) ) 
\to 1- \alpha. 
\label{CIp coverage}
\end{align}

Prepivoting avoids an explicit bias correction whenever $\hat{H}_n$ does not require estimating~$B$. Denote by $\hat{B}_n :=\E^*[T_n^*]$ and $\hat{v}^2_n:=\var^*[T_n^*]$ the bias and variance, respectively, of the bootstrap statistic $T^*_n$ under the bootstrap probability measure. As we show next, we may regard $\hat{B}_n$ as a bias estimator implicitly induced by the bootstrap. Suppose the following high level conditions hold.

\setcounter{assnLetter}{7}
\begin{assnLetter}
\label{assn 1}
(i) $T^*_n-\hat{B}_n \dstarto N(0,v^2)$ with $v^2=\plim \hat v^2_n>0$; (ii) $(T_n - B_n, \hat{B}_n - B_n)^\prime \dto N(0,V)$, where $V$ is positive definite.
\end{assnLetter}
Assumption~\ref{assn 1}(i) requires the bootstrap statistic $T^*_n$ to be conditionally asymptotically Gaussian after we remove the bootstrap bias~$\hat{B}_n$. Note that the bootstrap does not need to replicate the asymptotic variance of~$T_n$, i.e., $v^2$ may be different from~$v^2_1$. Under Assumption~\ref{assn 1}(i), as in Theorem~3.4 of Cavaliere et al.\ (2024), we can show that $\hat{p}_n=\Phi(\hat{v}_n^{-1}(T_n-\hat{B}_n))+\op(1)$. If in addition Assumption~\ref{assn 1}(ii) holds, then $T_n-\hat{B}_n\dto N(0,v^2_\mathsf{P})$ with $v_{\mathsf{P}}^2>0$, and the asymptotic cdf of $\hat p_n$ is $H (u):= \Phi (m^{-1} \Phi^{-1} (u))$, $m:= {v}_{\mathsf{P}}/{v}$. A consistent (plug-in) estimator of $H$ can be obtained as
\begin{equation}\label{Hhat_def}
  \hat{H}_n(u):=\Phi(\hat{m}_n^{-1}\Phi^{-1}(u)), \quad \hat{m}_n:=\hat{v}_{\mathsf{P},n}/\hat{v}_n,
\end{equation}
where $\hat{v}^2_{\mathsf{P},n}$ is a consistent estimator of~$v^2_{\mathsf{P}}$. Different bootstrap methods induce different $\hat{B}_n$ and consequently require different estimators~$\hat{v}^2_{\mathsf{P},n}$.

Although $\hat{H}_n$ of \eqref{Hhat_def}, and hence $\CIp$, does not involve $\hat{B}_n$ explicitly, we can show that $\CIp$ is asymptotically equivalent to an interval that bears a close resemblance to the \textsf{RBC} interval $\CIrbc$ in that it too contains a bias correction and adjusts the standard error for the additional uncertainty appropriately. We describe this result next. 

For simplicity, suppose that $T^*_n$ is conditionally Gaussian; that is, conditionally on $\mathscr{D}_n$, $T^*_n \sim N(\hat{B}_n,\hat{v}^2_n)$ and hence $\hat{L}_n(u) = \Phi ((u-\hat{B}_n )/\hat{v}_n )$. In this case, $\CIp$ in \eqref{BTprepCI} is \emph{exactly} equal to 
\begin{equation} \label{CIprbc}
    \CIprbc =\big[ (\hat{g}_n(\mathsf{x}) - (nh)^{-1/2}\hat{B}_n) \pm z_{1-\alpha/2}(nh)^{-1/2} \hat{v}_{\mathsf{P},n}\big].
\end{equation}
In general, $T^*_n$ is Gaussian only asymptotically, and we obtain the following result.

\begin{theorem} \label{th 1}
Let $\CIp$, $\hat{H}_n$, and $\CIprbc$ be defined in \eqref{BTprepCI}, \eqref{Hhat_def}, and \eqref{CIprbc}, respectively.
\begin{itemize}
\item[(i)] If Assumption~\ref{assn 1}(i) holds then $\CIp=\CIprbc+\op((nh)^{-1/2})$.
\item[(ii)] If Assumption~\ref{assn 1}(i) holds and $T_n^*$ is Gaussian conditional on~$\mathscr{D}_n$, then $\CIp=\CIprbc$~a.s.
\item[(iii)] If Assumption~\ref{assn 1} holds and $\hat{v}^2_{\mathsf{P},n}\plowto v^2_{\mathsf{P}}$, then $ \PP ( g(\mathsf{x}) \in \CIp ) \to 1- \alpha$.
\end{itemize}
\end{theorem}

An interesting implication of Theorem~\ref{th 1} is that we can obtain $\CIprbc$ without any resampling as it depends only on $\hat{B}_n$ and $\hat{v}^2_{\mathsf{P},n}$ (which we derive analytically for different residual-based bootstrap methods in the next sections), and standard normal critical values~$z_\alpha$. More importantly, this asymptotic equivalence result implies that we can view \textsf{RBC} through the lens of prepivoting (or vice versa), as we show in the next section.

\subsection{standard rbc inference as bootstrap prepivoting}
\label{section GP}

Assume that the bootstrap data $\mathscr{D}_n^*:=\{(y_i^*,x_i^*): i=1,\dots, n\}$ are generated by setting $x_i^* = x_i$ for $i=1,\dots, n$ (`fixed regressor' bootstrap) and
\begin{equation}
y_i^{\ast}=\tilde{g}_{\mathsf{x},n}(x_i)+\varepsilon _i^{\ast}\text{,\quad }%
i=1,\ldots ,n , \label{btsDGPLQ}
\end{equation}
where $\varepsilon _i^{\ast}:=\tilde\varepsilon_i e_i^{\ast}$ with $\tilde\varepsilon_i$ being the bias-corrected HC\emph{k} residuals considered in Calonico et al.\ (2018) and $e_i^{\ast}$ being (conditionally on the data)\ i.i.d.\ with $\E^{\ast}[e_i^{\ast}]=0$ and $\E^{\ast}[e_i^{\ast 2}]=1$. The bootstrap conditional expectation $\E^{\ast}[y_i^{\ast}|x^*_i=x_i]$ in \eqref{btsDGPLQ} is 
\begin{equation*}
\tilde{g}_{\mathsf{x},n} (x_i):=  r_{p+1}(x_i - \mathsf{x})'\hat{\beta}_{p+1,n}(\mathsf{x})=\sum_{j=0}^{p+1}\iota _{j}^{\prime }\hat{\beta}_{p+1,n}(\mathsf{x})(x_i-\mathsf{x})^{j},
\end{equation*}
where $\hat{\beta}_{p+1,n}(\mathsf{x})$ is obtained as in \eqref{eq LP estimator} using local polynomial regression of order $p+1$ at the \emph{fixed} point~$\mathsf{x}$; see Bartalotti et al.\ (2017) and He and Bartalotti (2020) for the case $p=1$. Since $\iota _{j}^{\prime }\hat{\beta}_{p+1,n}(\mathsf{x})$ is an estimator of $g^{(j)}(\mathsf{x})/j!$, the function $\tilde{g}_{\mathsf{x},n} (x)$ is an estimator of a ($p+1$)-th order Taylor expansion of the conditional expectation $g(x)=\E[y_i|x_i=x]$ around the fixed point~$\mathsf{x}$. That is, to generate the bootstrap data we use \emph{the same} function $\tilde{g}_{\mathsf{x},n} (x)$ estimated locally at $\mathsf{x}$ but evaluated globally at each $x=x_i$ for $i =1,\dots, n$. Hence, we use `global polynomial' (GP) bootstrap to label the DGP~\eqref{btsDGPLQ}. Importantly, this method generates bootstrap data using a polynomial order ($p+1$) that is larger than the order $p$ used in estimation. This mismatch is key to inducing an asymptotic bias under the bootstrap probability measure that satisfies Assumption~\ref{assn 1}.

Let $\hat{g}_n^{\ast}(x)$ be the local ($p$-th order) polynomial estimator applied to the bootstrap sample generated as \eqref{btsDGPLQ}. The GP bootstrap analogue of $T_n$ is $T_{\mathsf{GP},n}^{\ast}=\sqrt{nh}(\hat{g}_n^{\ast}(\mathsf{x})-\tilde{g}_{\mathsf{x},n} (\mathsf{x}))$, and its bootstrap bias and variance are denoted $\hat{B}_{\mathsf{GP},n}:=\E^*[T_{\mathsf{GP},n}^\ast]$ and $\hat{v}_{\mathsf{GP},n}^2:=\var^*[T_{\mathsf{GP},n}^\ast]$, respectively. Because the GP bootstrap is a fixed-design scheme (i.e., the regressors are fixed across bootstrap samples), these moments can be computed in closed form. Specifically, we show that
\begin{equation}
\hat{B}_{\mathsf{GP},n} =\sqrt{nh^{2p+3}}\hat{g}_n^{(p+1)}(\mathsf{x})\tfrac{C_n}{(p+1)!},
\label{BhatGP}
\end{equation}
where $\hat{g}_n^{(p+1)}(\mathsf{x}):=(p+1)!\iota_{p+1}' \hat{\beta}_{p+1,n}(\mathsf{x})$ is an estimator of~$g^{(p+1)}(\mathsf{x})$, clearly implying that $\hat{B}_{\mathsf{GP},n}$ is an estimator of the dominant term of $B_n$ in~\eqref{eq:Bn}. Interestingly, $\hat{B}_{\mathsf{GP},n}$ is identical to $\hat{B}_{\mathsf{RBC},n}$ as given in Calonico et al.\ (2018, Section~3); see the discussion after~\eqref{eq RBC int}.


As explained in the previous section, a prepivoted bootstrap confidence interval \eqref{BTprepCI} based on $(T_n,T_{\mathsf{GP},n}^{\ast})$, say $\CIpgp$, can be constructed by choosing $\hat{v}^2_{\mathsf{PGP},n}$ as a consistent estimator of $v^2_{\mathsf{PGP}}$, which is the asymptotic variance of $T_n-\hat{B}_{\mathsf{GP},n}$. Because $\hat{B}_{\mathsf{GP},n}=\hat{B}_{\mathsf{RBC},n}$, the natural plug-in estimator for the GP case coincides with the variance estimator in Calonico et al.\ (2018). In Theorem~\ref{Th RBC PGP equivalence} below we show (i)~validity of such confidence interval, (ii)~its asymptotic equivalence to the \textsf{RBC} interval $\CIrbc$ in~\eqref{eq RBC int}, and (iii)~its almost sure equality to the \textsf{RBC} interval for Gaussian bootstraps.
The proof requires verifying that Assumption~\ref{assn 1} holds for $(T_n,T_{\mathsf{GP},n}^{\ast})$ under Assumptions~\ref{Ass_g} and~\ref{Ass_K,h}. The theorem then follows by Theorem~\ref{th 1}.
\begin{theorem} \label{Th RBC PGP equivalence}
Under Assumptions~\ref{Ass_g} and~\ref{Ass_K,h}, for any $\mathsf{x}\in \mathbb{S}_{x}$, 
$\PP(g(\mathsf{x})\in \CIpgp )\to 1-\alpha$ and $\CIpgp = \CIrbc + \op ((nh)^{-1/2})$. If, in addition, $T_{\mathsf{GP},n}^{\ast}$ is Gaussian conditional on~$\mathscr{D}_n$, then $\CIpgp=\CIrbc$~a.s.
\end{theorem}

\section{improved nonparametric inference}
\label{Section main}

\begin{figure}[t]
\includegraphics[width=\textwidth]{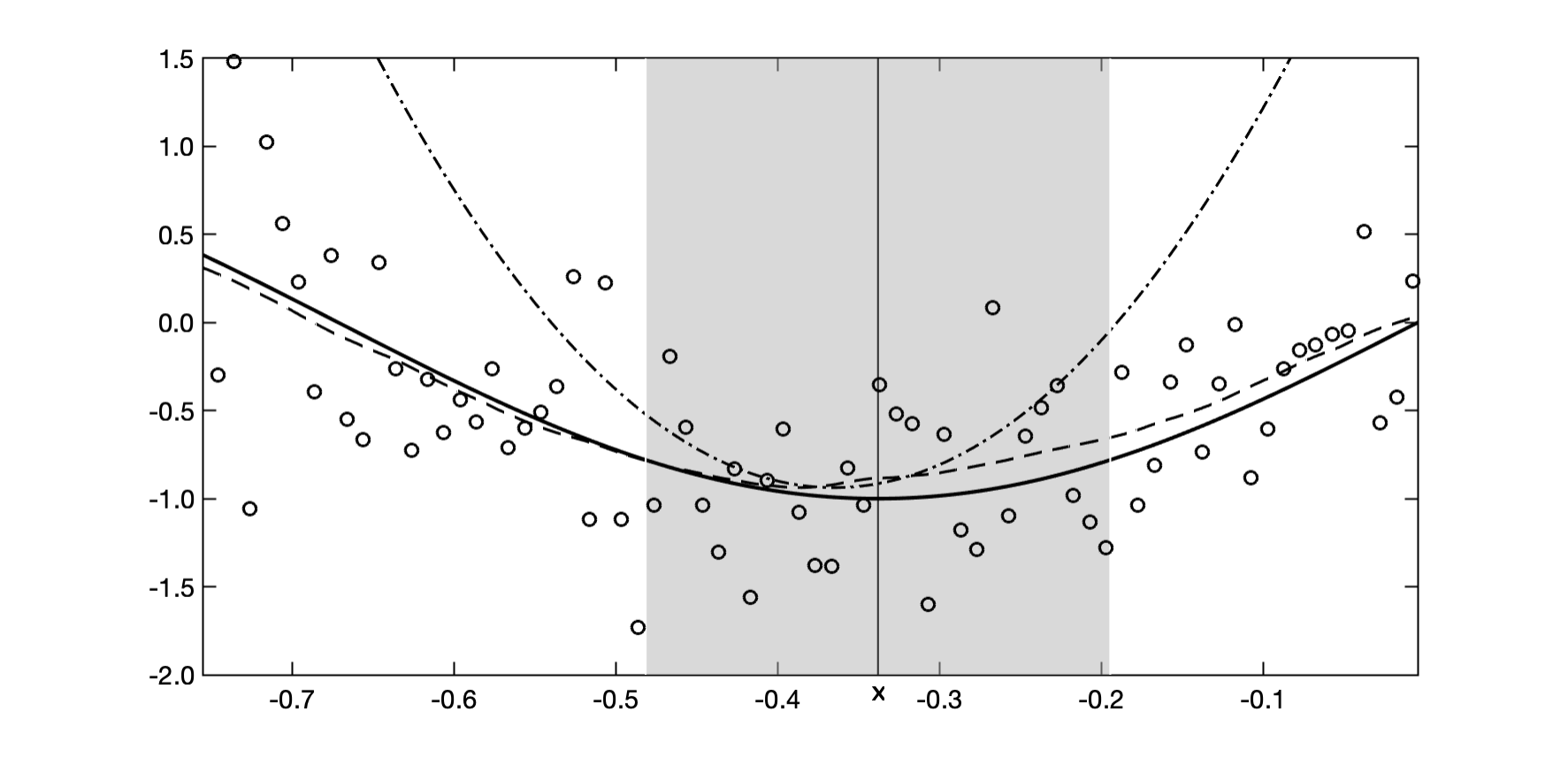}
\caption{Nonparametric curves:\ original (\rule[0.5ex]{1em}{0.8pt}), LP bootstrap (\rule[0.5ex]{0.3em}{0.8pt}\hspace{0.3em}\rule[0.5ex]{0.3em}{0.8pt}\hspace{0.3em}\rule[0.5ex]{0.3em}{0.8pt}), and GP bootstrap (\rule[0.5ex]{0.3em}{0.8pt}\hspace{0.3em}\rule[0.5ex]{0.1em}{0.8pt}\hspace{0.3em}\rule[0.5ex]{0.3em}{0.8pt}\hspace{0.3em}\rule[0.5ex]{0.1em}{0.8pt})}
\vskip 2pt
{\footnotesize \emph{Notes}: The vertical bar corresponds to the evaluation point $\mathsf{x}=-1/3$; the shaded area corresponds to the selected bandwidth; $n=1,000$ observations are summarized into $n/5$ bins (circles); polynomial order $p=1$.}
\label{figure curves}
\end{figure}

The main feature of the GP bootstrap discussed above is that the conditional mean function under the bootstrap probability measure is based on a \emph{single} local approximation of the curve~$g$, using a polynomial expansion of $g$ around~$\mathsf{x}$, which is then applied globally to all~$x_i$'s. In contrast, in this section we consider the `local polynomial' (LP) bootstrap scheme traditionally employed in the statistics literature, where $g$ is approximated locally for each of the $x_i$'s, implying that the bootstrap conditional expectation function resembles $g$ more closely over the entire support (Härdle and Marron, 1988, Härdle and Bowman, 1991, Hall and Horowitz, 2013). This point is illustrated in Figure~\ref{figure curves}, where we contrast the true regression curve with the GP and LP bootstrap curves under the setup considered in Section~\ref{Section Monte Carlo}.

In this section, we show how the prepivoted LP bootstrap leads to improved inference. We focus here on the case where $\mathsf{x}$ is an interior point. Boundary points will be considered in Section~\ref{Section boundary}.

\subsection{the local polynomial bootstrap}
\label{Section main-LP}

The LP bootstrap sets the bootstrap conditional expectation to $\E^{\ast}[y_i^{\ast}|x^*_i=x] = \hat{g}_n(x)$, where $\hat{g}_n(x)$ is as in \eqref{eq LP estimator}. The bootstrap data $\mathscr{D}_n^*:=\{(y_i^*,x_i^*): i=1,\dots, n\}$ satisfy $x_i^* = x_i$, $i=1,\dots, n$, and 
\begin{equation}
y_i^{\ast}=\hat{g}_n(x_i)+\varepsilon _i^{\ast}\text{,\quad }i=1,\ldots ,n,  \label{btsDGPLP}
\end{equation}
where $\varepsilon_i^* := \tilde\varepsilon_i e_i^*$ with $\tilde\varepsilon_i$ and $e_i^*$ as defined in Section~\ref{section GP} to facilitate comparison with the existing \textsf{RBC} approach. Other choices are possible, e.g., $\tilde\varepsilon_i = y_i - \hat{g}_n(x_i)$, without changing our results. Bootstrap DGPs of this or similar forms have been widely applied to nonparametric regression without prepivoting; see, e.g., Härdle and Marron (1988), Härdle and Bowman (1991), and Hall and Horowitz (2013).

The LP bootstrap uses a $p$-th order local polynomial regression estimate of $g(x)$ calculated at each sample point $x=x_i$ to generate $y^*_i$, rather than extrapolating values for $g(x_i)$ from a single $(p+1)$-th order local polynomial regression estimated at $x=\mathsf{x}$, as for the GP bootstrap. In this section, we exploit this key difference and show the following. First, we demonstrate that prepivoting can also be used to restore asymptotic validity of the LP bootstrap without requiring undersmoothing or the choice of additional bandwidths, a result that stands in contrast to the existing bootstrap literature. Second, we show that the resulting prepivoted LP bootstrap (\textsf{PLP}) generates a novel RBC-type confidence interval which differs from the existing \textsf{RBC} intervals in that it does not rely on a higher-order local polynomial estimate of the derivatives entering the bias term. Third, and most importantly, we show that in large samples, the new \textsf{PLP} interval is shorter than the classical \textsf{RBC} CIs in Calonico et al.\ (2014, 2018), while maintaining correct coverage. Put differently, with respect to the usual bias estimators used in \textsf{RBC} procedures, the bias estimator implicitly generated by the \textsf{PLP} method leads to improved inference on the regression curve.

\subsection{theory}
\label{Section main-theory}

Consider the prepivoted bootstrap confidence interval \eqref{BTprepCI} based on $(T_n,T_{\mathsf{LP},n}^{\ast})$, say $\CIplp$, where $T_{\mathsf{LP},n}^{\ast}=\sqrt{nh}(\hat{g}_n^{\ast}(\mathsf{x})-\hat{g}_n(\mathsf{x}))$ and $\hat{g}_n^{\ast}(\mathsf{x})$ is the local ($p$-th order) polynomial estimator applied to the bootstrap sample generated as~\eqref{btsDGPLP}. The bootstrap bias and variance of $T_{\mathsf{LP},n}^{\ast}$ are $\hat{B}_{\mathsf{LP},n}:=\E^*[T_{\mathsf{LP},n}^\ast]$ and $\hat{v}_{\mathsf{LP},n}^2:=\var^*[T_{\mathsf{LP},n}^\ast]$, respectively. 

Recall from Section~\ref{section equivalence} that in order to construct $\CIplp$, a consistent estimator $\hat{v}_{\mathsf{PLP},n}^2$ of the asymptotic variance $v^2_{\mathsf{PLP}}$ of $T_n - \hat{B}_{\mathsf{LP},n}$ is required. 
To find $\hat{v}_{\mathsf{PLP},n}^2$, we first derive the bootstrap bias~$\hat{B}_{\mathsf{LP},n}$. It follows from the closed-form expression for $\hat{g}^{\ast}_n(\mathsf{x})$ that
\begin{align} \label{eq:BhatLP0}
\hat{B}_{\mathsf{LP},n} 
= \sqrt{nh} \Big( \frac{1}{nh} \sum_{i=1}^n w_i  ( \mathsf{x})\hat{g}_n (x_i) - \hat{g}_n  ( \mathsf{x}) \Big).
\end{align}
Since $B_n = \sqrt{nh}\big((nh)^{-1}\sum_{i=1}^n w_i(\mathsf{x}) g(x_i) - g(\mathsf{x})\big)$, \eqref{eq:BhatLP0} makes clear that the LP bootstrap bias $\hat{B}_{\mathsf{LP},n}$ differs from $\hat{B}_{\mathsf{RBC},n}$ by targeting the bias directly, rather than estimating the $(p+1)$-th derivative that appears in its asymptotic expansion. We can rewrite $\hat{B}_{\mathsf{LP},n}$ as
\begin{equation}\label{eq Bhat LP}
\hat{B}_{\mathsf{LP},n} = \frac{1}{\sqrt{nh}} \sum_{i=1}^n  w_{\mathsf{LP} \text{-}\mathsf{bc} ,i}  (\mathsf{x})y_i
\end{equation}
with weights defined by the convolution $w_{\mathsf{LP}\text{-}\mathsf{bc},i} (\mathsf{x}):= (nh)^{-1} \sum_{j=1}^n w_{j}  ( \mathsf{x}) w_i  ( x_j)- w_i( \mathsf{x})$. Since $T_n = \sqrt{nh}\big( (nh)^{-1} \sum_{i=1}^n w_i(\mathsf{x})y_i - g(\mathsf{x})\big)$, we obtain
\begin{equation*}
    T_n - \hat{B}_{\mathsf{LP},n} =  \sqrt{nh} \Big( \frac{1}{{nh}} \sum_{i=1}^n w_{\mathsf{PLP},i}(\mathsf{x}) y_i  - g(\mathsf{x})\Big), \hspace{0.5em} w_{\mathsf{PLP},i} (\mathsf{x}) :=  w_i(\mathsf{x}) - w_{\mathsf{LP}\text{-}\mathsf{bc},i} (\mathsf{x}) .
\end{equation*}
Let $v^2_{\mathsf{PLP}}$ denote the asymptotic variance of~$T_n - \hat{B}_{\mathsf{LP},n}$. Because the weights are measurable with respect to $\mathscr{X}_n$ and $\var [y_i | \mathscr{X}_n] = \sigma^2 (x_i)$, by conditioning on the regressors we have $\var [T_n - \hat{B}_{\mathsf{LP},n}| \mathscr{X}_n] = (nh)^{-1} \sum_{i=1}^n  w_{\mathsf{PLP},i}(\mathsf{x})^2 \sigma^2 (x_i)$. This suggests the estimator
\begin{equation} \label{vPest}
\hat{v}_{\mathsf{PLP},n}^2:=  \frac{1}{nh} \sum_{i=1}^n  w_{\mathsf{PLP},i}(\mathsf{x})^2 \tilde\varepsilon_i^2 .
\end{equation}

A formal proof of consistency of $\hat{v}_{\mathsf{PLP},n}^2$ under Assumptions~\ref{Ass_g} and~\ref{Ass_K,h} is provided next.

\begin{lemma} \label{Lemma v2plp}
Under Assumptions~\ref{Ass_g} and~\ref{Ass_K,h}, $\hat{v}_{\mathsf{PLP},n}^2 \plowto v_{\mathsf{PLP}}^2 >0$.
\end{lemma}

In view of Lemma~\ref{Lemma v2plp}, an application of Theorem~\ref{th 1} implies the asymptotic validity of the \textsf{PLP} interval~$\CIplp$. In addition, $\CIplp$ is asymptotically equivalent to
\begin{equation*}
\CIplprbc = [(\hat{g}_n(\mathsf{x})-(nh)^{-1/2} \hat{B}_{\mathsf{LP},n})\pm z_{1-\alpha /2}(nh)^{-1/2}\hat{v}_{\mathsf{PLP},n}],
\end{equation*}
which is a new $\mathsf{RBC}$-type interval that does not require resampling. The following theorem states these results.

\begin{theorem} \label{Th RBC PLP equivalence}
Let $\mathsf{x}$ be an interior point. Under Assumptions~\ref{Ass_g} and~\ref{Ass_K,h}, $\PP(g(\mathsf{x})\in \CIplp)\to 1-\alpha$ and $\CIplp=\CIplprbc + \op ((nh)^{-1/2})$. If, in addition, $T_{\mathsf{LP},n}^{\ast}$ is Gaussian conditional on~$\mathscr{D}_n$, then $\CIplp=\CIplprbc$~a.s.
\end{theorem}

\subsection{improved inference}
\label{Section main-improved}

In this section, we compare the asymptotic efficiency of the new prepivoted \textsf{PLP} intervals and the existing \textsf{RBC}-type intervals proposed by Calonico et al.\ (2014, 2018). We show that the former are asymptotically shorter than the latter. This efficiency result is based on the asymptotic equivalence between prepivoting and the \textsf{RBC} approach in Section~\ref{section equivalence}. In particular, Theorems~\ref{Th RBC PGP equivalence} and~\ref{Th RBC PLP equivalence} imply that the asymptotic relative length of $\CIplp$ compared to $\CIrbc$ is equal to the ratio between the asymptotic standard deviations of their respective debiased statistics, i.e., $v_{\mathsf{PLP}}/v_{\mathsf{RBC}}$.

To shed some light on this ratio, let $\mathsf{w}(u), u\in \RR$, be the so-called `equivalent kernel' associated with $K$ and~$p$.  Specifically, for $\mathcal{X}=\mathbb{R}$,  $\mathsf{w}(u) := \iota_0^{\prime}(\int_{\mathcal{X}} r_p(s) r_p^{\prime}(s)K(s) ds)^{-1} r_p(u) K(u)$, the asymptotic analogue of the sample weights~$w_i(x)$; see Appendix~\ref{app:definition}. Similarly, define the asymptotic analogue of $w_{\mathsf{PLP},i}(x)$ as $\mathsf{w}_{\mathsf{PLP}}(u) := 2 \mathsf{w}(u) - \int_{\mathcal{X}}  \mathsf{w}(u-r) \mathsf{w}(r) dr$. This contrasts with that of the \textsf{RBC} method given by $\mathsf{w}_{\mathsf{RBC}}(u) := \mathsf{w}(u) - C\iota_{p+1}^{\prime}(\int_{\mathcal{X}} r_{p+1}(s) r_{p+1}^{\prime}(s)K(s) ds)^{-1} r_{p+1}(u) K(u)$; see Lemma~\ref{lemmaKernelConstants} and Calonico et al.\ (2022, Section~4.2).

\begin{corollary} \label{corollary v^2_P comparisons}
Under Assumptions~\ref{Ass_g} and~\ref{Ass_K,h}, 
\begin{equation*}
(v^2_{\mathsf{PLP}}, v^2_{\mathsf{RBC}})= \frac{\sigma^2 (\mathsf{x})}{f(\mathsf{x})} (\mathcal{K}_{\mathsf{PLP}}, \mathcal{K}_{\mathsf{RBC}}) ,
\end{equation*}
where 
\begin{equation}
\label{eq K PLP}
\mathcal{K}_{\mathsf{PLP}} := \int_{\mathcal{X}} \mathsf{w}_{\mathsf{PLP}}(u)^2 du >0
\quad\text{and}\quad
\mathcal{K}_{\mathsf{RBC}} := \int_{\mathcal{X}} \mathsf{w}_{\mathsf{RBC}}(u)^2 du >0 
\end{equation}
are functions only of the kernel $K$ and the polynomial order~$p$.
\end{corollary}

As shown in Section~\ref{Section main-theory}, $v^2_{\mathsf{PLP}}$ is the asymptotic variance of a (scaled) weighted average of~$y_i$ with weights~$w_{\mathsf{PLP},i}(\mathsf{x})$. Using fairly standard arguments in the nonparametric literature (e.g., Fan and Gijbels, 1996, p.~66), Corollary~\ref{corollary v^2_P comparisons} shows that this variance is proportional to $\sigma^2 (\mathsf{x})/f(\mathsf{x})$ where the constant of proportionality is~$\mathcal{K}_{\mathsf{PLP}}$, which is a known function of the kernel and the polynomial order,~$p$. A similar proportionality holds for $v^2_{\mathsf{RBC}}$ with the difference that the constant of proportionality is~$\mathcal{K}_{\mathsf{RBC}}$, which is a different function of the kernel and~$p$. Hence, the relative asymptotic length of the $\mathsf{PLP}$ and $\mathsf{RBC}$ intervals is the square root of $\mathcal{K}_{\mathsf{PLP}}/\mathcal{K}_{\mathsf{RBC}}$.

\begin{table}[t] 
\caption{Relative (asymptotic) interval length comparison}
\label{table relative variances}
\vskip -6pt
\begin{tabular*}{\textwidth}{@{\extracolsep{\fill}}lccc}
\toprule
Kernel   & $\mathcal{K}_{\mathsf{PLP}}$       & $\mathcal{K}_{\mathsf{RBC}}$   & $(\mathcal{K}_{\mathsf{PLP}}/\mathcal{K}_{\mathsf{RBC}})^{1/2}$     \\
\midrule
Triangular   & 0.95 & 1.33 & 0.84\\
Uniform      & 0.83 & 1.13 & 0.86\\
Epanechnikov & 0.85 & 1.25 & 0.83\\
Biweight     & 1.01 & 1.41 & 0.85\\
Triweight    & 1.15 & 1.55 & 0.86 \\
\bottomrule
\end{tabular*}
\vskip 4pt
{\footnotesize \emph{Notes}: Polynomial order $p=1$.}
\end{table}

\begin{figure}[t] 
\vskip -6pt
\includegraphics[trim={2.5cm 0 2cm 0},clip,width=1\textwidth]{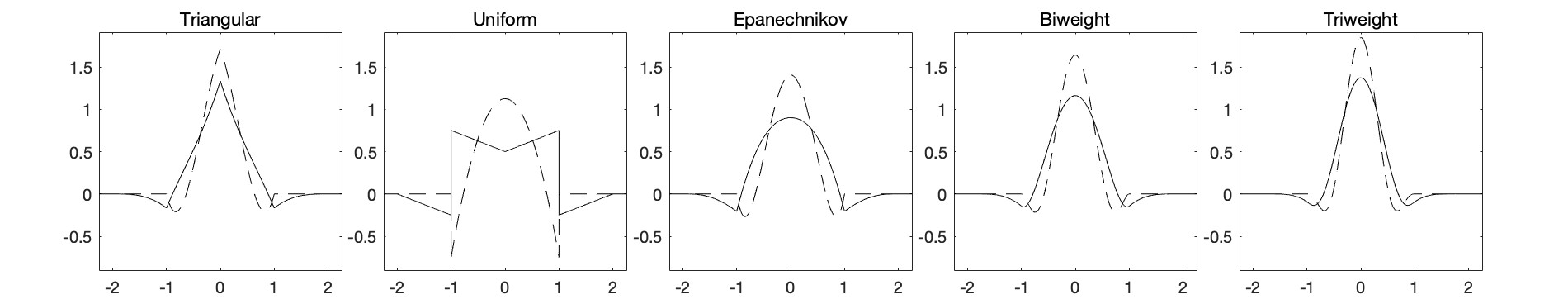}
\vskip -4pt
\caption{The functions $\mathsf{w}_{\mathsf{PLP}}$ (\rule[0.5ex]{1em}{0.8pt}) and $\mathsf{w}_{\mathsf{RBC}}$  (\rule[0.5ex]{0.3em}{0.8pt}\hspace{0.3em}\rule[0.5ex]{0.3em}{0.8pt}\hspace{0.3em}\rule[0.5ex]{0.3em}{0.8pt}) for five kernels} 
{\footnotesize \emph{Notes}: Polynomial order $p=1$.}
\label{figure equiv_ker_int}
\end{figure}

Table~\ref{table relative variances} reports the values of $\mathcal{K}_{\mathsf{PLP}}$, $\mathcal{K}_{\mathsf{RBC}}$, and the square root of their ratio for five common kernels and $p=1$, which is by far the most commonly used value in practice. For all the kernels considered, the decrease in interval length is substantial, ranging from $14\%$ to~$17\%$. Figure~\ref{figure equiv_ker_int} provides a comparison of the functions $\mathsf{w}_{\mathsf{PLP}}$ and $\mathsf{w}_{\mathsf{RBC}}$ for the five kernels previously considered and $p=1$. While neither function is always closer to zero than the other, $\mathsf{w}_{\mathsf{PLP}}(u)$ clearly shows less variation around zero, resulting in $\mathcal{K}_{\mathsf{PLP}}<\mathcal{K}_{\mathsf{RBC}}$.


\section{inference at the boundary and rdd}
\label{Section boundary}

In this section, we extend the results in Section~\ref{Section main} to address inference on $g(\mathsf{x})$ when $\mathsf{x}$ is a boundary point. This includes standard applications of nonparametric regression to RDD, which is prominent in empirical work. As it turns out, the key challenge when $\mathsf{x}$ is on the boundary is that the LP bootstrap bias is no longer centered around the original bias~$B_n$, not even asymptotically. Therefore, Assumption~\ref{assn 1}(ii) breaks down and Theorem~\ref{Th RBC PLP equivalence} no longer applies; see Section~\ref{sec failure}. We show that a simple modification of the prepivoting method ($\mathsf{mPLP}$) restores the validity results established in Section~\ref{Section main} for interior points. Importantly, confidence intervals based on the proposed $\mathsf{mPLP}$ procedure automatically account for cases where $\mathsf{x}$ is at the boundary, while remaining asymptotically equivalent to the $\mathsf{PLP}$ procedure from Section~\ref{Section main} when $\mathsf{x}$ is an interior point. The proposed modification is presented in Section~\ref{sec adaptive}, with its application to regression discontinuity designs discussed in Section~\ref{sec RDD}.

\subsection{Challenges of standard prepivoting at the boundary} 
\label{sec failure}

A crucial assumption underlying the results of Section~\ref{Section main-theory} is Assumption~\ref{assn 1}(ii), which requires that $\hat{B}_{\mathsf{LP},n} - B_n$ converges in distribution to a mean-zero Gaussian random variable. This assumption fails when $\mathsf{x}$ is a boundary point since $\hat{B}_{\mathsf{LP},n} - B_n$ is no longer (asymptotically) centered at zero. The reason is that, in contrast with the interior case, $C_n$ and its LP bootstrap analogue $C_{\mathsf{LP},n}$ do not converge to the same constant $C$ when $\mathsf{x}$ is on the boundary. Instead, Lemma~\ref{lemmaLPbias} shows that
\begin{equation}
\label{eq Bhat An boundary}
\hat{B}_{\mathsf{LP},n} - B_n - A_n = \xi_{2n,\mathsf{LP}} +\op (1), \quad A_n \plowto A := \sqrt{\kappa} g^{(p+1)}(\mathsf{x}) (C_{\mathsf{LP}} - C) /(p+1)!,
\end{equation}
where $\xi_{2n,\mathsf{LP}}:= (nh)^{-1/2} \sum_{i=1}^n w_{\mathsf{LP}\text{-}\mathsf{bc},i} (\mathsf{x}) \varepsilon_i$ has zero mean, $A_n:=\sqrt{nh^{2p+3}}{g^{(p+1)}(\mathsf{x})}({C}_{\mathsf{LP},n}-C_n)/(p+1)!$, $C_{\mathsf{LP},n}:=(nh)^{-1}\sum_{i=1}^{n}w_i(\mathsf{x})C_n(x_i)$ is a smoothed version of~$C_n$, and $g^{(p+1)}(\mathsf{x})$ denotes the (directional) ($p+1$)-order derivative at~$\mathsf{x}$. Importantly, $C \neq C_{\mathsf{LP}}$ when $\mathsf{x}$ is on the boundary, but they are uniquely determined by the kernel~$K$ and the polynomial order~$p$; see Lemma~\ref{lemmaKernelConstants}. 

Our modified method restores validity of the prepivoted bootstrap confidence interval by simply rescaling the bootstrap statistic $T^*_{\mathsf{LP},n}$ with a known function of the data, say~$Q_n$. The scaling factor $Q_n$ is chosen such that the term $A_n$ is eliminated from the asymptotic distribution of the bootstrap bias. As we shall see, it does not involve any additional tuning parameters or unknown quantities.

\subsection{a boundary-adaptive prepivoting method}
\label{sec adaptive}

Let $T^*_{\mathsf{mLP},n}:=Q_nT^*_{\mathsf{LP},n}$, where $Q_n:=C_n/C_{\mathsf{LP},n}$ depends only on $K$, $h$, and~$\mathscr{X}_n$. By definition, $\hat{B}_{\mathsf{mLP},n} :=\E^* [T^*_{\mathsf{mLP},n}] = Q_n \hat{B}_{\mathsf{LP},n}$ with $\hat{B}_{\mathsf{LP},n}$ as in Section~\ref{Section main-theory}. In contrast with $T^*_{\mathsf{LP},n}$, Assumption~\ref{assn 1}(ii) holds for $T^*_{\mathsf{mLP},n}$ for any $\mathsf{x}$, including boundary points, because we have defined $Q_n$ precisely such that $Q_n C_{\mathsf{LP},n} = C_n$. Specifically,
\begin{equation}
\label{eq:BmLP-B}
\hat{B}_{\mathsf{mLP},n}-B_n=\sqrt{nh^{2p+3}}\tfrac{g^{(p+1)}(\mathsf{x})}{(p+1)!}(Q_nC_{\mathsf{LP},n}-C_n) + Q_n\xi_{2n,\mathsf{LP}}+\op (1)=Q_n\xi_{2n,\mathsf{LP}}+\op (1).
\end{equation}
Consequently, the modified LP bootstrap bias $\hat{B}_{\mathsf{mLP},n}$ is asymptotically centered at~$B_n$, as required in Assumption~\ref{assn 1}(ii). Crucially, this result holds for any $\mathsf{x} \in \mathbb{S}_x$, including boundary points.

Taken together, these results imply that an asymptotically valid modified confidence interval, denoted $\CImplp$, can be constructed provided a consistent estimator $\hat{v}_{\mathsf{mPLP},n}^2$ is available for the asymptotic variance $v^2_{\mathsf{mPLP}}$ of $T_n-\hat{B}_{\mathsf{mLP},n}$. By relying on arguments similar to those used for $\hat{v}^2_{\mathsf{PLP},n}$ in \eqref{vPest}, we can show that a consistent estimator of $v_{\mathsf{mPLP}}^2$ is given by
\begin{align} \label{eq:hatvmPLP}
\hat{v}_{\mathsf{mPLP},n}^2:=  \frac{1}{{nh}} \sum_{i=1}^n  w_{\mathsf{mPLP},i}^2  (\mathsf{x}) \tilde\varepsilon_i^2 ,\quad w_{\mathsf{mPLP},i} (\mathsf{x}) :=  w_i(\mathsf{x}) - w_{\mathsf{mLP}\text{-}\mathsf{bc},i} (\mathsf{x}),
\end{align}
where $w_{\mathsf{mLP}\text{-}\mathsf{bc},i} (\mathsf{x}):=Q_n w_{\mathsf{LP}\text{-}\mathsf{bc},i} (\mathsf{x})$ with $w_{\mathsf{LP}\text{-}\mathsf{bc},i} (\mathsf{x})$ and $\tilde\varepsilon_i$ as defined in Section~\ref{Section main-LP}. The following lemma formalizes this result.
\begin{lemma} \label{Lemma v2mplp}
Under Assumptions~\ref{Ass_g} and~\ref{Ass_K,h}, $\hat{v}_{\mathsf{mPLP},n}^2 \plowto v_{\mathsf{mPLP}}^2 >0$.
\end{lemma}

Because $T^*_{\mathsf{mLP},n}$ satisfies Assumption~\ref{assn 1}, we can apply Theorem~\ref{th 1} to conclude that the \textsf{mPLP} interval~$\CImplp$ is asymptotically valid and equivalent to
\begin{equation*}
\CImplprbc = [(\hat{g}_n(\mathsf{x})-(nh)^{-1/2} \hat{B}_{\mathsf{mLP},n})\pm z_{1-\alpha /2}(nh)^{-1/2}\hat{v}_{\mathsf{mPLP},n}],
\end{equation*}
which is an RBC-type interval involving no resampling. This interval is based on a new bootstrap bias correction, $\hat{B}_{\mathsf{mLP},n}$, and a new studentization, $\hat{v}_{\mathsf{mPLP},n}$, and it is valid for both interior and boundary points as stated in the next theorem.

\begin{theorem} \label{Th RBC mPLP equivalence}
Under Assumptions~\ref{Ass_g} and~\ref{Ass_K,h}, for any $\mathsf{x} \in \mathbb{S}_{x}$, $\PP(g(\mathsf{x})\in \CImplp)\to 1-\alpha$ and $\CImplp=\CImplprbc + \op ((nh)^{-1/2})$. If, in addition, $T_{\mathsf{LP},n}^{\ast}$ is Gaussian conditional on~$\mathscr{D}_n$, then $\CImplp=\CImplprbc$~a.s.
\end{theorem} 

An important feature of the \textsf{mPLP} approach is that it automatically adapts to the location of the point~$\mathsf{x}$. That is, it coincides with the \textsf{PLP} approach asymptotically when $\mathsf{x}$ lies in the interior of $\mathbb{S}_x$ (where $Q_n\plowto Q=1$), and it provides a valid generalization when $\mathsf{x}$ is on the boundary. 

Finally, as in Section~\ref{Section main-improved}, we can show that confidence intervals for $g(\mathsf{x})$ are asymptotically shorter when based on \textsf{mPLP} compared to the existing \textsf{RBC} intervals. This result, which holds for both interior and boundary points, follows by comparing the asymptotic variances $v^2_{\mathsf{mPLP}}$ and~$v^2_{\mathsf{RBC}}$. 

\begin{corollary} \label{corollary v^2_mP comparisons}
Let Assumptions~\ref{Ass_g} and~\ref{Ass_K,h} hold.
If $\mathsf{x}$ is an interior point then $v^2_{\mathsf{mPLP}} = v^2_{\mathsf{PLP}}$, and the conclusions of Corollary~\ref{corollary v^2_P comparisons} hold.
If $\mathsf{x}$ is a boundary point, then
\begin{equation*}
(v^2_{\mathsf{mPLP}}, v^2_{\mathsf{RBC}})= \frac{\sigma^2 (\mathsf{x})}{f(\mathsf{x})} (\mathcal{K}_{\mathsf{mPLP}}, \mathcal{K}_{\mathsf{RBC}}) ,
\end{equation*}
where $\mathcal{K}_{\mathsf{mPLP}}$ and $\mathcal{K}_{\mathsf{RBC}}$ are functions only of the kernel $K$ and the polynomial order~$p$.
\end{corollary}

In contrast to Corollary~\ref{corollary v^2_P comparisons}, Corollary~\ref{corollary v^2_mP comparisons} applies to any evaluation point~$\mathsf{x}$, including both interior and boundary points. The values of the constants $\mathcal{K}$ depend on the kernel function~$K$, the polynomial order~$p$, as well as on whether $\mathsf{x}$ is in the interior or at the boundary of~$\mathbb{S}_x$. When $\mathsf{x}$ is in the interior of~$\mathbb{S}_x$, these values coincide with those of $\mathcal{K}_{\mathsf{PLP}}$ and $\mathcal{K}_{\mathsf{RBC}}$ in Corollary~\ref{corollary v^2_P comparisons}, implying that Corollary~\ref{corollary v^2_P comparisons} is a special case of Corollary~\ref{corollary v^2_mP comparisons}. 

\begin{table}[t]
\caption{Relative (asymptotic) interval length comparison -- boundary case}
\label{table relative variances boundary}
\vskip -6pt
\begin{tabular*}{\textwidth}{@{\extracolsep{\fill}}lccc}
\toprule
Kernel   & $\mathcal{K}_{\mathsf{mPLP}}$       & $\mathcal{K}_{\mathsf{RBC}}$   & $(\mathcal{K}_{\mathsf{mPLP}}/\mathcal{K}_{\mathsf{RBC}})^{1/2}$     \\
\midrule
Triangular   & 7.17 & 10.29 & 0.84\\
Uniform & 6.62 & 9.00 & 0.86\\
Epanechnikov & 6.78  & 9.82 & 0.83\\
Biweight     & 7.67 & 10.87 & 0.84\\
Triweight    & 8.54 & 11.87 & 0.85 \\
\bottomrule
\end{tabular*}
\vskip 4pt
{\footnotesize \emph{Notes}: Polynomial order $p=1$.}
\end{table}

When $\mathsf{x}$ is a boundary point, $\mathcal{K}_{\mathsf{RBC}}$ takes different values compared with the interior point case, and $\mathcal{K}_{\mathsf{mPLP}}$ is different from~$\mathcal{K}_{\mathsf{PLP}}$. Table~\ref{table relative variances boundary} is the boundary analogue of Table~\ref{table relative variances} and confirms that, for the five kernels previously considered, the modified prepivoted interval~$\CImplp$ is asymptotically shorter than $\CIrbc$. As in Figure~\ref{figure equiv_ker_int}, we also plot in Figure~\ref{figure equiv_ker_bnd} the equivalent kernels of \textsf{mPLP} and \textsf{RBC} (see Appendix~\ref{app:kernels}). The comparison reveals that the conclusions from Section~\ref{Section main-improved} also hold for \textsf{mPLP}, i.e., $\mathcal{K}_{\mathsf{mPLP}}<\mathcal{K}_{\mathsf{RBC}}$.

\begin{figure}[t]
\vskip -6pt
\includegraphics[trim={2.5cm 0 2cm 0},clip,width=1\textwidth]{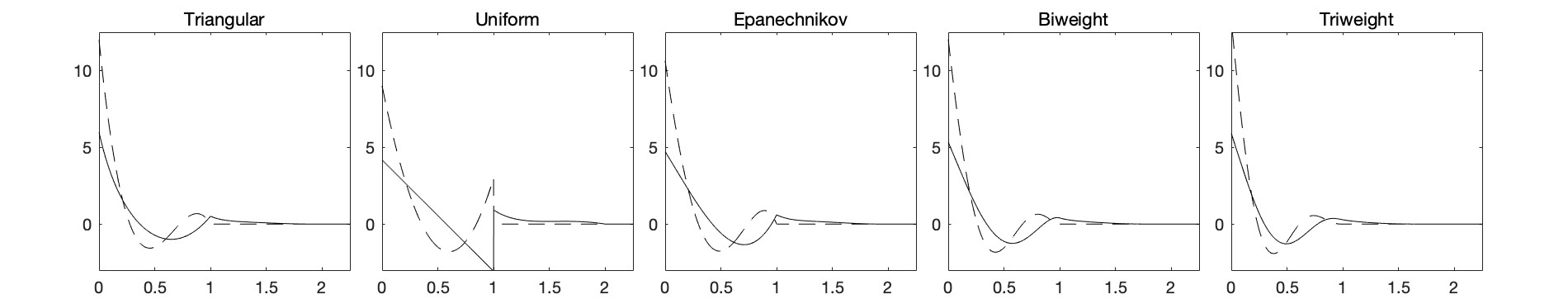}
\vskip -4pt
\caption{The functions $\mathsf{w}_{\mathsf{mPLP}}$ (\rule[0.5ex]{1em}{0.8pt}) and $\mathsf{w}_{\mathsf{RBC}}$  (\rule[0.5ex]{0.3em}{0.8pt}\hspace{0.3em}\rule[0.5ex]{0.3em}{0.8pt}\hspace{0.3em}\rule[0.5ex]{0.3em}{0.8pt}) for five kernels -- boundary case} 
{\footnotesize \emph{Notes}: Polynomial order $p=1$.}
\label{figure equiv_ker_bnd}
\end{figure}

\subsection{application to regression-discontinuity designs} 
\label{sec RDD}

In this section, we apply the modified prepivoting approach from Section~\ref{sec adaptive} to the problem of inference in~RDD. Adopting the standard potential outcome framework, let $y_i(1)$ and $y_i(0)$ denote the potential outcomes for individual $i$ ($i=1, \ldots,n$) with and without the treatment, respectively. For each individual, we observe the associated treatment indicator $d_i$, which equals 1 if unit $i$ is treated (0 otherwise), and the observed outcome, $y_i=y_i(0)+(y_i(1)-y_i(0))d_i$. We also observe the `forcing' variable~$x_i$, a scalar covariate which is not affected by the treatment and determines whether~$d_i=1$. Hence, the observed data are $\mathscr{D}_n := \{(y_i, x_i, d_i): i=1,\dots,n\}$.

We consider (sharp) RDD, where $d_i$ is determined by $x_i$ being above a given known cutoff $\mathsf{x}$; that is, $d_i:=\mathbb{I}_{\{x_i\geq \mathsf{x}\}}$. Interest is in estimating the average treatment effect at the cutoff, namely $\mathsf{ATE}(\mathsf{x}) := \E[y_i(1)-y_i(0)|x_i= \mathsf{x}] = g_{+}(\mathsf{x})-g_{-}(\mathsf{x}) $, where $g_{+}(x) := \E[y_i(1)|x_i = x]$ and $g_{-}(x) := \E[y_i(0)|x_i = x]$ are the regression functions of the potential outcomes. We consider the following modification of Assumption~\ref{Ass_g}, where $\sigma^2_{+}(x) := \var[y_i(1)|x_i = x]$ and $\sigma^2_{-}(x) := \var[y_i(0)|x_i = x]$. 

\begin{assumption}\label{Ass_g_RDD}
$(y_i,x_i,d_i)$ are i.i.d.\ and $x_i$ has bounded support $\mathbb{S}_x$ and continuous density~$f$. For all $x$ in an open neighborhood of $\mathsf{x}$, it holds that (i)~$f(x)>0$; (ii)~$\sigma_{+}^2$ and $\sigma_{-}^2$ are continuous and $\sup_{x\in\mathbb{S}_x}\E[y_i^4|x_i=x]<\infty$; (iii)~$g_{+}^{(p+1)}$ and $g_{-}^{(p+1)}$ are H\"{o}lder continuous with exponent~\mbox{$\eta>0$}.  
\end{assumption}

This assumption, which will be used for the asymptotic analysis, also identifies $\mathsf{ATE}(\mathsf{x})$ as the jump in the regression function $g$ at the cutoff~$\mathsf{x}$, i.e., as $\tau(\mathsf{x}):= \lim_{x\to \mathsf{x}^{+}}g(x)-\lim_{x\to  \mathsf{x}^{-}}g(x)=g_{+}(\mathsf{x})-g_{-}(\mathsf{x}) $; see Hahn et al.\ (2001) and Imbens and Kalyanaraman (2012). Hence, $\tau(\mathsf{x})$ is the parameter of interest, and the average treatment effect can be estimated as the difference of two local polynomial regressions at the cutoff~$\mathsf{x}$. Formally, we can write this estimator~as
\begin{equation} \label{LPrd}
\hat{\tau}_n (\mathsf{x}) := \hat{g}_{+,n} (\mathsf{x}) - \hat{g}_{-,n}(\mathsf{x}),
\end{equation}
where $\hat{g}_{+,n} (x)$ and $\hat{g}_{-,n} (x)$ are defined as in \eqref{eq LP estimator} with $K((x_i-x)/h)$ replaced by $K_{+}((x_i-x)/h):=K((x_i-x)/h)\mathbb{I}_{\{x_i \geq \mathsf{x}\}}$ and $K_{-}((x_i-x)/h):=K((x_i-x)/h)\mathbb{I}_{\{x_i < \mathsf{x}\}}$, respectively.

To construct valid confidence intervals using the modified prepivoting method of Section~\ref{sec adaptive}, let $T_n:=(nh)^{1/2}(\hat{\tau}_n(\mathsf{x})-\tau (\mathsf{x}))$ and its LP bootstrap analogue $T_n^{\ast}:=(nh)^{1/2}(\hat{\tau}_n^{\ast}(\mathsf{x})-\hat{\tau}_n(\mathsf{x}))$, where $\hat{\tau}_n^{\ast}(\mathsf{x})$ is calculated from bootstrap data, $\mathscr{D}_n^*:=\{(y_i^*,x_i^*,d_i^*): i=1,\dots, n\}$. Specifically, $(x_i^*,d_i^*) = (x_i, d_i)$, $i=1,\dots, n$, and the $y^*_i$'s are as in \eqref{btsDGPLP} with $\hat{g}_n(x_i)$ replaced by $\hat{g}_{+,n}(x_i)\mathbb{I}_{\{x_i \geq \mathsf{x}\}}+\hat{g}_{-,n}(x_i)\mathbb{I}_{\{x_i < \mathsf{x}\}}$.
 
Next, we define the modified bootstrap statistic,~$T_{\mathsf{rd},n}^{\ast}$. Since estimation of $\tau$ requires fitting two distinct LP regressions, it is useful to decompose $T_n=T_{+,n}-T_{-,n}$, where $T_{+,n}:=(nh)^{1/2}(\hat{g}_{+,n}(\mathsf{x})-g_{+}(\mathsf{x}))$ and $T_{-,n}:=(nh)^{1/2}(\hat{g}_{-,n}(\mathsf{x})-g_{-}(\mathsf{x}))$. In the same way, we can decompose $T_n^{\ast}=T_{+,n}^{\ast}-T_{-,n}^{\ast}$ and define the scaling factors $Q_{+,n}$ and $Q_{-,n}$ as in Section~\ref{sec adaptive} with $K$ replaced by $K_{+}$ and $K_{-}$, respectively. Thus, the modified bootstrap statistic is $T_{\mathsf{rd},n}^{\ast}:=Q_{+,n}T_{+,n}^{\ast}-Q_{-,n}T_{-,n}^{\ast}$.

The bootstrap bias is 
\begin{equation}
\hat{B}_{\mathsf{rd},n}:=\E^{\ast}[T_{\mathsf{rd},n}^{\ast}]=Q_{+,n}\hat{B}_{+,n}-Q_{-,n}\hat{B}_{-,n},
\end{equation}
where $\hat{B}_{+,n}:=\E^{\ast}[T_{+,n}^{\ast}]$ and $\hat{B}_{-,n}:=\E^{\ast}[T_{-,n}^{\ast}]$ are computed as in \eqref{eq Bhat LP} with $K$ replaced by $K_{+}$ and $K_{-}$, respectively, in the definition of the weights~$w_i(x)$. Finally, we need a consistent estimator $\hat{v}_{\mathsf{rd},n}^2$ of the asymptotic variance $v_{\mathsf{rd}}^2$ of $T_n-\hat{B}_{\mathsf{rd},n}$. To this end, we note that $\var [T_n-\hat{B}_{\mathsf{rd},n}|\mathscr{X}_n] = \var [T_{+,n}-Q_{+,n}\hat{B}_{+,n}|\mathscr{X}_n]+\var [T_{-,n}-Q_{-,n}\hat{B}_{-,n}|\mathscr{X}_n]$. Consistent estimators of these two variance components can be obtained as in \eqref{eq:hatvmPLP} with $K$ replaced by $K_{+}$ and $K_{-}$, respectively, thus yielding~$\hat{v}_{\mathsf{rd},n}^2$.

We show in the next theorem that $T^{\ast}_{\mathsf{rd},n}$ satisfies Assumption~\ref{assn 1}, so we can apply Theorem~\ref{th 1} to conclude that a modified confidence interval for $\tau (\mathsf{x})$, denoted $\CIrd$ and constructed using $\hat{v}_{\mathsf{rd},n}^2$ as in Sections~\ref{section equivalence} and~\ref{Section main-theory}, is asymptotically valid and equivalent to
\begin{equation*}
\CIrdrbc = [(\hat{\tau}_n(\mathsf{x})-(nh)^{-1/2} \hat{B}_{\mathsf{rd},n})\pm z_{1-\alpha /2}(nh)^{-1/2}\hat{v}_{\mathsf{rd},n}],
\end{equation*}
which is an RBC-type interval involving no resampling.

\begin{theorem} \label{Th RBC mPLP equivalence rdd}
Under Assumptions~\ref{Ass_K,h} and~\ref{Ass_g_RDD} it holds that (i)~$\hat{v}_{\mathsf{rd},n}^2\plowto v_{\mathsf{rd}}^2$, (ii)~$\PP( \tau (\mathsf{x})\in \CIrd)\to 1-\alpha$, (iii)~$\CIrd=\CIrdrbc + \op ((nh)^{-1/2})$, and (iv)~if, in addition, $T_{\mathsf{rd},n}^{\ast}$ is Gaussian conditional on~$\mathscr{D}_n$, then $\CIrd=\CIrdrbc$~a.s.
\end{theorem} 

Finally, we note that all the results in this section can be generalized to allow for different kernel and bandwidth choices on each side of the cutoff. 
This changes the efficiency results in Sections~\ref{Section main-improved} and~\ref{sec adaptive} only slightly. Specifically, let $\mathcal{K}_{+,\mathsf{rd}}$ and $\mathcal{K}_{-,\mathsf{rd}}$ denote $\mathcal{K}_{\mathsf{mPLP}}$ for the kernel used to the right and to the left of the cutoff, respectively. We similarly define $\mathcal{K}_{+,\mathsf{RBC}}$ and $\mathcal{K}_{-,\mathsf{RBC}}$. By the i.i.d.\ assumption, the following result then follows from Theorem~\ref{Th RBC mPLP equivalence} and Corollary~\ref{corollary v^2_mP comparisons}.

\begin{corollary} \label{cor RDD}
Under Assumptions~\ref{Ass_K,h} (applied to both sides of the cutoff) and~\ref{Ass_g_RDD},
\begin{equation*}
\frac{v^2_{\mathsf{rd}}}{v^2_{\mathsf{RBC}\textsf{-}\mathsf{rd}}} 
= \frac{\sigma_{+}^2(\mathsf{x})\mathcal{K}_{+,\mathsf{rd}}+\sigma_{-}^2(\mathsf{x})\mathcal{K}_{-,\mathsf{rd}}}{\sigma_{+}^2(\mathsf{x})\mathcal{K}_{+,\mathsf{RBC}}+\sigma_{-}^2(\mathsf{x})\mathcal{K}_{-,\mathsf{RBC}}},
\end{equation*}
where $v^2_{\mathsf{RBC}\textsf{-}\mathsf{rd}}$ is the asymptotic variance from the existing $\mathsf{RBC}$ interval for~RDD.
\end{corollary}

As previously, the asymptotic relative length of the confidence intervals is the square root of the ratio given in Corollary~\ref{cor RDD}. Because $\mathcal{K}_{+,\mathsf{rd}}<\mathcal{K}_{+,\mathsf{RBC}}$ and $\mathcal{K}_{-,\mathsf{rd}}<\mathcal{K}_{-,\mathsf{RBC}}$ for all the kernels considered (as seen in Table~\ref{table relative variances boundary}), it follows from Corollary~\ref{cor RDD} that our new confidence intervals $\CIrd$ and $\CIrdrbc$ are asymptotically shorter compared to the existing \textsf{RBC} intervals for RDD. Indeed, when the same kernel is applied on both sides of the cutoff, as is common in applied work, the result in Corollary~\ref{cor RDD} simplifies to that obtained in Corollary~\ref{corollary v^2_mP comparisons} and displayed in Table~\ref{table relative variances boundary}. Specifically, the new intervals are shorter than the existing \textsf{RBC} intervals by the same amount as in Table~\ref{table relative variances boundary}.

It is likely that these results can be extended to other types of RDDs, e.g., fuzzy or kinked~RDD. Although relatively straightforward conceptually, such extensions are nontrivial. For example, the fuzzy RDD estimator is a ratio of two sharp RDD estimators. Thus, limit theory would require joint convergence of those two estimators, which in turn requires a generalization of the prepivoting theory of Cavaliere et al.\ (2024) to vector-valued statistics. With such theory in hand, the fuzzy RDD estimator can be analyzed using the arguments above combined with the delta method.

\section{monte carlo} 
\label{Section Monte Carlo}

We now discuss the finite sample performance of the proposed CIs and compare them with the CIs based on existing \textsf{RBC} using Monte Carlo simulation. We also include the invalid (i.e., not prepivoted) CIs for comparison. We consider two distinct inference problems: a nonparametric regression curve evaluated at both an interior and a boundary point and a sharp RDD.

We report results for $5,000$ Monte Carlo replications and nominal level~$0.95$. Estimators are based on the Epanechnikov kernel for the nonparametric regression setup and on the triangular kernel for the RDD setup, as those represent popular kernel choices. Two relevant bandwidth choices are considered: the infeasible MSE-optimal bandwidth~($h$), as a theoretical benchmark, and the feasible (plug-in based) coverage-error-optimal bandwidth~($\hat h$) of Calonico et al.\ (2018, 2020, 2022), representing the reference bandwidth if the aim is to minimize the coverage error of \textsf{RBC} intervals. For each bandwidth choice, we report the average bandwidth~($\bar h$) across Monte Carlo replications. We report empirical coverage probabilities and average lengths for four methods: $\CIgp$, $\CIlp$, $\CIrbc$, and~$\CImplp$. The first two methods are based on the interval $\CIboot$ in~\eqref{eq general invalid BS}, using the GP and LP bootstrap schemes, respectively. $\CIrbc$ is the interval \eqref{eq RBC int} and $\CImplp$ is the interval defined in Section~\ref{sec adaptive}. For all methods, we use HC3 residuals. The bootstrap intervals are all based on a Gaussian wild bootstrap scheme, and hence implemented analytically without resampling; see Section~\ref{section equivalence}. For the same reason, results for the prepivoted GP interval $\CIpgp$ defined in Section~\ref{section GP} would be identical to $\CIrbc$, and are thus not reported.

\begin{table}[t]
\small 
\caption{coverage and length of 95\% confidence intervals - nonparametric regression}
\vskip -6pt
\label{Table nonpar}
\begin{tabular*}{\textwidth}{@{\extracolsep{\fill}}lrlrrrrrrrrr}
\toprule
& & & & \multicolumn{4}{c}{Coverage} & \multicolumn{4}{c}{Length} \\
\cline{5-8} \cline{9-12}
\multicolumn{1}{c}{eval.} & \multicolumn{1}{c}{$n$} & \multicolumn{1}{c}{$h$} & \multicolumn{1}{c}{$\bar{h}$} & 
\multicolumn{1}{c}{\textsf{GP}} & 
\multicolumn{1}{c}{\textsf{LP}} & 
\multicolumn{1}{c}{\textsf{RBC}} & 
\multicolumn{1}{c}{\textsf{mPLP}} & 
\multicolumn{1}{c}{\textsf{GP}} & 
\multicolumn{1}{c}{\textsf{LP}} & 
\multicolumn{1}{c}{\textsf{RBC}} & 
\multicolumn{1}{c}{\textsf{mPLP}}   \\ 
\midrule
int & 250  & $h$ & 0.189   & 81.5 &	88.9 &	93.6 &	94.2 &	0.635 &	0.635 &	0.915 &	0.765 \\
& ~ & $\hat h$ & 0.359  &  80.3 &	79.3	& 93.1	& 86.5	& 0.461	& 0.461 &	0.664 & 	0.554\\
& 500  & $h$ & 0.165  &  82.1 & 89.2 & 94.2 & 94.4 & 0.479 & 0.479 & 0.690 & 0.573 \\
& ~ & $\hat h$ & 0.303  &  81.6 & 85.4 & 94.1 & 91.1 & 0.353 & 0.353 & 0.510 & 0.422 \\
& 1000  & $h$ &   0.143  &  81.9 & 90.0 & 94.4 & 94.9 & 0.361 & 0.361 & 0.521 & 0.431 \\
& ~ & $\hat h$  & 	0.256    &  81.9 & 87.7 & 94.6 & 93.8 & 0.270 & 0.270 & 0.390 & 0.322 \\
& 2000  & $h$ &    0.125 &83.3&90.7&95.1&95.4&0.273&0.273&0.394&0.326\\
& & $\hat h$  &    0.217 &82.5&89.4&94.9&94.5&0.207&0.207&0.299&0.247\\
\midrule
bnd & 250  & $h$ & 0.353   &  77.7 & 85.6 & 90.9 & 92.2 & 1.277 & 1.277 & 1.901 & 1.596 \\
& ~ & $\hat h$ & 0.373    & 77.8 & 83.2 & 91.0 & 91.2 & 1.271 & 1.271 & 1.895& 1.602 \\
& 500  & $h$ &  0.307   &  79.9 & 87.6 & 93.3 & 93.8 & 0.961 & 0.961 & 1.426 & 1.191 \\
& ~ & $\hat h$ & 0.327   &  80.6 & 85.6 & 93.2 & 92.8 & 0.946  & 0.946 & 1.403 & 1.174 \\
& 1000  & $h$ & 0.267    &  80.6 & 87.8 & 93.8 & 94.0 & 0.725 & 0.725 & 1.071 & 0.894 \\
& ~ & $\hat h$ & 0.285  & 80.7 & 86.7 & 93.9 & 93.3 & 0.711 & 0.711 & 1.051 & 0.878 \\
& 2000  & $h $ &    0.233 & 81.4&89.5&94.7&94.8&0.549&0.549&0.812&0.676\\
& & $\hat h$  & 0.257 & 81.5&88.1&94.5&94.3&0.526&0.526&0.779&0.649\\
\bottomrule
\end{tabular*}
\vskip 4pt
\end{table}

\medskip

\noindent \textsc{nonparametric regression.} We consider i.i.d.\ data generated as $y_i =g(x_i)+\varepsilon_i$ with $x_i\sim U_{[-1,1]}$ and $\varepsilon_i \sim N(0,\sigma^2)$, where $\sigma=1$ and 
\begin{equation*}
g(x)= \frac{ \sin(3\pi x/2)}{1+18x^2 (\mathrm{sign}(x)+1)}.
\end{equation*}
This DGP was previously considered in Berry, Carroll, and Ruppert (2001), Hall and Horowitz (2013), and Calonico et al.\ (2018,  2022), among others. We consider inference both at an interior point, $\mathsf{x}=-1/3$, and a boundary point,~$\mathsf{x}=-1$, denoted int and bnd, respectively, in Table~\ref{Table nonpar}.

\begin{table}[t]
\small 
\caption{coverage and length of 95\% confidence intervals - rdd}
\vskip -6pt
\label{Table rd}
\begin{tabular*}{\textwidth}{@{\extracolsep{\fill}}crlrrrrrrrrr}
\toprule
& & & & \multicolumn{4}{c}{Coverage} & \multicolumn{4}{c}{Length} \\
\cline{5-8} \cline{9-12}
\multicolumn{1}{c}{DGP} & \multicolumn{1}{c}{$n$} & \multicolumn{1}{c}{$h$} & \multicolumn{1}{c}{$\bar{h}$} & 
\multicolumn{1}{c}{\textsf{GP}} & 
\multicolumn{1}{c}{\textsf{LP}} & 
\multicolumn{1}{c}{\textsf{RBC}} & 
\multicolumn{1}{c}{\textsf{mPLP}} & 
\multicolumn{1}{c}{\textsf{GP}} & 
\multicolumn{1}{c}{\textsf{LP}} & 
\multicolumn{1}{c}{\textsf{RBC}} & 
\multicolumn{1}{c}{\textsf{mPLP}}   \\ 
\midrule
1 & 500  & $h $ &    0.082 & 80.6 & 86.7 & 93.3 & 93.2 & 0.345 & 0.345 & 0.589 & 0.470 \\
& & $\hat h$  & 0.073 & 79.6 & 86.0 & 93.5 & 94.0 & 0.373 & 0.373 & 0.674 & 0.539 \\
& 1000  & $h $ &    0.072 & 81.1 & 87.8 & 94.3 & 94.2 & 0.249 & 0.249 & 0.388 & 0.318 \\
& & $\hat h$  & 0.060 & 81.0 & 86.8 & 93.8 & 93.9 & 0.276 & 0.276 & 0.441 & 0.358 \\
& 2000  & $h $ &    0.063 & 81.2 & 88.0 & 94.5 & 94.7 & 0.185 & 0.185 & 0.279 & 0.231 \\
& & $\hat h$  & 0.049 & 80.4 & 87.8 & 94.4 & 94.4 & 0.210 & 0.210 & 0.321 & 0.265 \\
& 4000  & $h $ &     0.054 & 81.3 & 88.3 & 94.8 & 95.1 & 0.138 & 0.138 & 0.205 & 0.170 \\
& & $\hat h$  &  0.041 & 81.2 & 88.2 & 94.8 & 94.8 & 0.160 & 0.160 & 0.240 & 0.199 \\
\midrule
2 & 500  & $h $ &    0.260 & 80.7 & 88.0 & 94.4 & 94.2 & 0.183 & 0.183 & 0.274 & 0.228 \\
& & $\hat h$  & 0.126 & 80.2 & 86.9 & 94.1 & 94.0 & 0.271 & 0.271 & 0.434 & 0.354 \\
& 1000  & $h $ &    0.226 & 81.6 & 89.0 & 94.7 & 94.8 & 0.136 & 0.136 & 0.202 & 0.168 \\
& & $\hat h$  & 0.115 & 81.9 & 88.7 & 94.9 & 95.1 & 0.194 & 0.194 & 0.297 & 0.244 \\
& 2000  & $h $ &    0.197 & 82.9 & 88.5 & 94.8 & 94.8 & 0.102 & 0.102 & 0.150 & 0.125 \\
& & $\hat h$  & 0.101 & 82.0 & 88.8 & 94.7 & 94.7 & 0.144 & 0.144 & 0.214 & 0.178 \\
& 4000  & $h $ &     0.172 & 81.0 & 89.1 & 94.7 & 94.7 & 0.077 & 0.077 & 0.113 & 0.094 \\
& & $\hat h$  &  0.087 & 82.4 & 89.2 & 94.8 & 95.1 & 0.108 & 0.108 & 0.160 & 0.134 \\
\bottomrule
\end{tabular*}
\vskip 4pt
\end{table}

Results for samples of size $n \in \{250, 500, 1000, 2000\}$ are presented in Table~\ref{Table nonpar}. The numerical evidence supports the anticipated decrease in interval lengths of~17\%, even for small sample sizes, suggesting that the asymptotic efficiency of \textsf{mPLP} is rapidly achieved as \( n \) increases. We observe that \textsf{RBC} and \textsf{mPLP} perform similarly in terms of coverage under both bandwidth choices, with empirical coverage probabilities remaining close to the nominal level. A deviation from the nominal level is detected for \textsf{mPLP} with $\hat{h}$, when $\mathsf{x}$ is an interior point and $n$ is small, but this deviation rapidly vanishes as the sample size increases. Finally, because they are not robust to the `large' bandwidth choices considered, non-prepivoted methods exhibit much smaller average lengths, resulting in severe undercoverage even for large~$n$.

\medskip

\noindent \textsc{rdd.} In the context of RDD, we generate i.i.d.\ data from $y_i = g (x_i)+\varepsilon_i$ with $x_i\sim 2 \mathcal{B}(2,4) -1$ and $\varepsilon_i \sim N(0,\sigma^2)$, where $\mathcal B$ is the Beta distribution and~$\sigma=0.1295$. We consider two choices for the regression function~$g$ based on the datasets in Ludwig and Miller (2007) and Lee (2008), as in Calonico et al. (2014). Specifically, the two DGPs are
\begin{align*}
\text{DGP1:} \quad g (x) &= \begin{cases}
    3.71 + 2.30x  + 3.28 x^2 + 1.45 x^3 + 0.23 x^4 + 0.03 x^5 &  \hspace{2.69cm} \text{if } x < 0 , \\
0.26 + 18.49x - 54.81 x^2 +74.30 x^3 - 45.02x^4 + 9.83x^5 & \hspace{2.69cm} \text{if } x \geq  0 ,
\end{cases}\\
\text{DGP2:} \quad g(x) &= \begin{cases}
    0.48 + 1.27x  - 0.5 \cdot  7.18 x^2 + 0.7 \cdot  20.21 x^3 + 1.1 \cdot  21.54 x^4 + 1.5 \cdot 7.33 x^5 &\text{if } x < 0 , \\
0.52 + 0.84x - 0.1 \cdot 3.00 x^2 - 0.3 \cdot  7.99 x^3 - 0.1 \cdot  9.01x^4 + 3.56x^5 &\text{if } x \geq  0 ,
\end{cases}
\end{align*}
where, as in Calonico et al.\ (2014), the fifth-order polynomial from the Lee (2008) data in DGP2 is modified to increase the curvature of $g$ and hence increase bias.

Results for samples of size $n\in\{500,1000, 2000,4000\}$ (total sample sizes are twice those in the nonparametric regression example) are presented in Table~\ref{Table rd}. The results confirm the conclusions drawn for the nonparametric regression example: the \textsf{mPLP} method produces shorter intervals than \textsf{RBC} for all the considered sample sizes and bandwidth choices, with coverage levels of both prepivoted methods close to the nominal value. As expected, non-prepivoted methods fail to deliver valid inference.

\begin{table}[t]
\caption{practical implementation guide for \textsf{mPLP}}
\label{tab:guidance}
\vskip -6pt
\begin{tabularx}{\textwidth}{@{}lX@{}}
\toprule
Choice & Recommendation \\
\midrule
Bandwidth & Any that can be used with \textsf{RBC}; e.g.\ coverage-error-optimal \\
\addlinespace
Kernel & Any standard kernel; efficiency gains range from 14\%--17\% across common choices (Epanechnikov, triangular, uniform, biweight, triweight) \\
\addlinespace
Polynomial order & $p=1$ (local linear) recommended; higher orders supported \\
\addlinespace
Interior vs.\ boundary/RDD \quad & Use \textsf{mPLP} (adapts to boundary automatically) \\
\addlinespace
Computation & Fully analytic; no resampling required \\
\addlinespace
Software & \texttt{R} packages at \url{https://pppackages.github.io} \\
\bottomrule
\end{tabularx}
\vskip 4pt
\end{table}

\section{guidance for applied researchers}
\label{Section guidance}

The \textsf{mPLP} method proposed in this paper can be used as an alternative or complement to the standard \textsf{RBC} interval in any nonparametric regression or RDD application. Its implementation requires no additional tuning parameters beyond those already needed for \textsf{RBC}: the same bandwidth~$h$, kernel~$K$, and polynomial order~$p$ are used throughout. \textsf{R} packages implementing our procedures are available at \url{https://pppackages.github.io}.

Table~\ref{tab:guidance} summarizes the key choices and considerations for practitioners. We offer the following specific recommendations.

\medskip
\noindent\textsc{bandwidth.} Our method is compatible with any bandwidth selection rule used with \textsf{RBC}, including the coverage-error-optimal bandwidth of Calonico et al.\ (2018) implemented in the \texttt{rdrobust} package, the MSE-optimal bandwidth, or cross-validation. No other bandwidth is required.

\medskip
\noindent\textsc{kernel.} Our efficiency results hold for all standard kernels. For practitioners already using \textsf{RBC} with the Epanechnikov or triangular kernel, the two most common choices,\textsf{mPLP} delivers 17\% and 16\% shorter intervals, respectively (Tables~\ref{table relative variances} and~\ref{table relative variances boundary}). There is no reason to switch kernels when moving from \textsf{RBC} to~\textsf{mPLP}.

\medskip
\noindent\textsc{polynomial order.} As with \textsf{RBC}, local linear estimation ($p=1$) is the standard choice and the one that requires least smoothness of the conditional mean function. Higher-order polynomials are supported by the theory and can be implemented if one can assume additional smoothness.

\medskip
\noindent\textsc{interior vs.\ boundary/rdd.} We recommend using the modified \textsf{mPLP} interval for both interior evaluation points, boundary points, and RDD (where the cutoff is a boundary point) because it adapts automatically to the boundary and requires no additional input from the user.

\medskip
\noindent\textsc{computation.} Because the bootstrap moments (mean and variance) entering the \textsf{mPLP} interval are available in closed form as functions of the kernel weights and residuals, implementation is fully analytic and requires no resampling.

\section{conclusions}
\label{Section conclusions}

This paper proposes novel procedures for inference in nonparametric regression and regression-discontinuity designs based on non-standard implementations of the bootstrap. New confidence intervals based on the concept of prepivoting (Beran 1987, 1988; Cavaliere et al., 2024) are shown to deliver asymptotically correct coverage under general conditions that allow for the presence of bias and thus do not require undersmoothing. We show that prepivoting different choices of bootstrap DGPs yield different robust bias correction (RBC)-type confidence intervals. This connection between the prepivoting and robust bias correction approaches allows us to identify the specific bootstrap DGP underlying the RBC approach of Calonico et al.\ (2014, 2018). More importantly, it enables us to propose a new alternative approach based on a local polynomial bootstrap algorithm that delivers improved inference. Specifically, the prepivoted local polynomial method yields 14--17\% shorter intervals than existing RBC intervals, depending only on the choice of kernel function.

It is worth noting that, because the conditional moments (expectation and variance) of our reference bootstrap statistics can be computed analytically, the practical implementation of our improved bias correction procedure does not require simulating any bootstrap samples. Although we could use resampling instead of the analytical formulas to make the implementation more automatic, this would be much more computationally costly.

The application of prepivoting to inference in the presence of nonnegligible bias has potential applications beyond those explored here. A natural extension is to sieve regression estimators, which have a long tradition in the nonparametrics literature (see, e.g., Andrews, 1991; Huang, 2003; Chen, 2007; Belloni et al., 2015; Chen and Christensen, 2015). Although bias is often assumed away through undersmoothing conditions, Cattaneo et al.\ (2020) have recently proposed RBC inference methods for the particular case of partitioning-based nonparametric estimators. It would be interesting to explore whether prepivoting can improve inference in this setting. 

Other potential applications include two-step semiparametric estimators where the first step involves nonparametric regression, potentially affecting the asymptotic distribution of the second step estimator (e.g., Andrews, 1994; Newey, 1994; Chen, Linton, and van Keilegom, 2003). While this literature has primarily focused on how to adjust the variance of the second step estimator, Cattaneo and Jansson (2018) show that asymptotic bias emerges under `small bandwidth' asymptotics when the first step uses kernel regression. Although they show that a particular bootstrap automatically corrects for this bias, their results assume away smoothing bias. Exploring whether prepivoting can improve inference in this setting would be an interesting contribution. Finally, extensions to time series and high-dimensional frameworks (Gupta and Seo, 2023) or spatial data (Hallin et al., 2004) are also of interest.

\appendix
\makeatletter
\renewcommand{\thetable}{\thesection.\@arabic\c@table}
\@addtoreset{table}{section}
\makeatother
\makeatletter
\renewcommand{\thefigure}{\thesection.\@arabic\c@figure}
\@addtoreset{figure}{section}
\makeatother
\makeatletter
\renewcommand{\thetable}{\thesection.\@arabic\c@table}
\@addtoreset{table}{section}
\makeatother

\section*{APPENDIX}

\section{notation}
\label{app:notation}

The following notation is used throughout the paper. We let $\mathbb{I}_{\{\cdot\}}$ denote the indicator function and $\iota_i$ a column vector of conforming, context-dependent dimension, with one in the $(i+1)$-th entry and zeros in the remaining entries. The $p$-th order derivative of a function $f$ is denoted~$f^{(p)}$. The Euclidean norm is denoted~$|A|$. For two intervals, $I_1,I_2 \subseteq \RR $ and a scalar $r \in \RR $, we use the notation $I_1=I_2 + r$ to denote that the Hausdorff distance between $I_1$ and $I_2$ is~$r$. Unless specified otherwise, all limits are for $n\to\infty$. The standard Gaussian cdf is denoted by~$\Phi$ and its $u$-quantile by~$z_{u}$. The uniform distribution on $[0,1]$ is denoted~$U_{[0,1]}$. If $F$ is a cdf, $F^{-1}$ denotes the right-continuous generalized inverse, i.e., $F^{-1}(u):=\sup \{v\in\RR :F(v)\leq u\}$ for $u\in\RR $.
 
For a bootstrap sequence $Y_n^{\ast}$ we use $Y_n^{\ast}=\opstar(1)$ to mean that, for any $\epsilon>0$, $\PP^{\ast}(|Y_n^{\ast}|>\epsilon)\plowto 0$, where $\PP^{\ast}$ denotes the probability measure conditional on the original data~$\mathscr{D}_n$. We use $Y_n^{\ast}\dstarto \xi$ to mean that, for all continuity points $u\in\RR $ of $G(u):=\PP(\xi\leq u)$, it holds that $\PP^{\ast}(Y_n^{\ast}\leq u)-G(u)\plowto 0$. Finally, $\E^*$ and $\var^*$ denote expectation and variance computed under~$\PP^*$.

\section{preliminaries}
\label{app:definition}


Consider the $p$-th order local polynomial estimator $\hat{g}_n(x)= \iota_{0}^{\prime}\hat{\beta}_{p,n}(x)$ of~\eqref{eq LP estimator}. Defining $y:=(y_1 , \ldots , y_n)^{\prime}$, $W(x):=h^{-1}\diag(K((x_{1}-x)/h),\ldots, K((x_n-x)/h))$, $H_p := \diag (1,h^{-1},\ldots,h^{-p})$, and $Z_p(x):=(r_p((x_1-x)/h),\ldots,r_p((x_n-x)/h))^\prime$, we can write $\hat\beta_{p,n}(x)$ in closed form as
\begin{equation*}
\hat\beta_{p,n}(x) = H_p (Z_p^{\prime}(x)W(x)Z_p(x))^{-1} Z_p^{\prime}(x)W(x)y.
\end{equation*}
As noted in Section~\ref{section preliminaries}, $\hat{g}_n(x)$ can also be defined by means of the weights
\begin{equation*}
w_i (x) := \iota_0' (Z_p ' (x) W(x) Z_p(x)/n)^{-1} r_p ((x_i - x)/h) K((x_i - x)/h),
\end{equation*}
so that $\hat{g}_n (x)=(nh)^{-1}\sum_{i=1}^{n}w_i(x)y_i$.  

Let $Z_p$ and $W$, respectively, denote $Z_p(x)$ and $W(x)$ evaluated at $x=\mathsf{x}$. Then, $C_n$ of \eqref{eq:Bn} can be written as $C_n := \iota _0^{\prime }(Z_p^{\prime }WZ_p)^{-1}Z_p^{\prime}W Z_{p+1} \iota_{p+1}$. Moreover, with $\mathcal{X}=\mathbb{R}$ if $\mathsf{x}$ is interior and $\mathcal{X}=[0,\infty)$ if $\mathsf{x}$ is on the boundary, $C:= \plim C_n= \iota_0' ( \int_{\mathcal{X}} K(s) r_p (s) r_p(s)' ds )^{-1} \int_{\mathcal{X}} K(u) r_p (u) u^{p+1}du$. Alternatively, using the notation in Section~\ref{Section main-improved}, we can write $C:=\int_{\mathcal{X}} \mathsf{w}(u)u^{p+1}du$.

\subsection{moments, weights, and equivalent kernels}

\label{app:kernels}

In this appendix, we define some relevant sample and population functions related to the moments of $K$ and~$K^2$, used to derive our main results, as well as the weight functions and equivalent kernels of the (bias-corrected) estimators. 

We first define quantities related to moments of~$K$. Specifically, $\gamma_{j,n} (x) := (nh)^{-1} \sum_{i=1}^n K ((x_i-x)/h) ((x_i-x)/h)^j$, 
\begin{equation*}
\dot\gamma_j := f(\mathsf{x}) \int_{\mathcal{X}} K(u) u^j du, \quad  \gamma_{\mathsf{E},j,n} (x) := \int_{-1}^1 K (u ) u^j f(x+uh) du.
\end{equation*}
For the moments of~$K^2$, we let $\tilde\psi_{j,n} (x) := (nh)^{-1} \sum_{i=1}^n K ((x_i-x)/h)^2 ((x_i-x)/h)^j \tilde\varepsilon_i^2$ and 
\begin{equation*}
\dot\psi_j := f(\mathsf{x}) \sigma^2(\mathsf{x}) \int_{\mathcal{X}} K(u)^2 u^j du.
\end{equation*}
Finally, we define the $(p+1) \times (p+1)$ matrices $\Gamma_p (x)$, $\Gamma_{\mathsf{E},p} (x)$, $\dot\Gamma_p$, $\tilde\Psi_p (x)$, $\dot\Psi_p$ with $(i,j)$-th entries given by $\gamma_{i+j-2,n} (x)$, $\gamma_{\mathsf{E},i+j-2,n} (x)$, $\dot\gamma_{i+j-2}$, $\tilde\psi_{i+j-2,n} (x)$, $\dot\psi_{i+j-2}$, respectively.

We now introduce the main weighting functions and equivalent kernels. First, the weights for a $p$-th order local polynomial estimator are
\begin{equation*}
w_i (x) :=  \iota_0^{\prime} \Gamma_p^{-1}(x) r_p \left(\frac{x_i - x}{h} \right)K\left(\frac{x_i - x}{h} \right) ,    
\end{equation*}
where the associated equivalent kernel is given by $\mathsf{w}(u) := \iota_0^{\prime} (\int_{\mathcal{X}} r_p(v) r_p^{\prime}(v)K(v)dv )^{-1} r_p(u) K(u)$. For boundary points, we also define $\mathsf{w}_{\mathsf{bnd}}(u,s) := \iota_0^{\prime} (\int_{-s}^1 r_p(v) r_p^{\prime}(v)K(v)dv )^{-1} r_p(u) K(u)$, such that $\mathsf{w}_{\mathsf{bnd}}(u,0)=\mathsf{w} (u)$.  Moreover, we let $\mathsf{w}_{\mathsf{E}}(u,s) :=\iota_0^{\prime} ( \int_{-1}^1 r_p(v)r_p^{\prime}(v)K(v)f(s+vh)dv )^{-1}r_p(u)K(u)$.

For the \textsf{PGP} method, the weights of the bias estimator are 
\begin{equation*}
w_{\mathsf{GP}\text{-}\mathsf{bc},i} (x) := C_n \iota_{p+1}^{\prime} \Gamma_{p+1}^{-1}(x) r_{p+1} \left(\frac{x_i - x}{h} \right) K\left(\frac{x_i - x}{h} \right) 
\end{equation*}
with associated equivalent kernel $\mathsf{w}_{\mathsf{GP}\text{-}\mathsf{bc}} (u) := C \iota_{p+1}^{\prime}(\int_{\mathcal{X}} r_{p+1}(s) r_{p+1}^{\prime}(s)K(s) ds)^{-1} r_{p+1}(u) K(u)$. Thus, the bias-corrected weights are $w_{\mathsf{PGP},i} (x) := w_i(x) - w_{\mathsf{GP}\text{-}\mathsf{bc},i} (x)$ with equivalent kernel $\mathsf{w}_{\mathsf{PGP}} (u) := \mathsf{w} (u)- \mathsf{w}_{\mathsf{GP}\text{-}\mathsf{bc}} (u)$. Moreover, by the equivalence between the \textsf{PGP} and \textsf{RBC} methods, we let $\mathsf{w}_{\mathsf{RBC}} (u) := \mathsf{w}_{\mathsf{PGP}} (u) $.

For the (\textsf{m})\textsf{PLP} method, the weights of the bias estimator are
\begin{equation*}
w_{\mathsf{LP}\text{-}\mathsf{bc},i} (x) := \frac{1}{nh} \sum_{j=1}^n w_j(x) w_i(x_j) - w_i(x)
\end{equation*}
with associated equivalent kernels $\mathsf{w}_{\mathsf{LP}\text{-}\mathsf{bc}} (u) := \mathsf{w}_{\mathsf{conv}} (u) - \mathsf{w} (u)$ and $\mathsf{w}_{\mathsf{LP}\text{-}\mathsf{bc},\mathsf{bnd}} (u) := \mathsf{w}_{\mathsf{conv},\mathsf{bnd}} (u) - \mathsf{w} (u)$, where $\mathsf{w}_{\mathsf{conv}} (u) := \int_{\mathcal{X}} \mathsf{w} (r) \mathsf{w} (u-r) dr$ and $\mathsf{w}_{\mathsf{conv},\mathsf{bnd}} (u) := \int_{\mathcal{X}} \mathsf{w}_{\mathsf{bnd}} (r,0) \mathsf{w}_{\mathsf{bnd}} (u-r,r) dr$.  Therefore, the bias-corrected weights are $w_{\mathsf{PLP},i}(x):=w_i(x)-w_{\mathsf{LP}\text{-}\mathsf{bc},i} (x)$ and $w_{\mathsf{mPLP},i}(x):=w_i(x)-Q_n w_{\mathsf{LP}\text{-}\mathsf{bc},i} (x)$ with equivalent kernels $\mathsf{w}_{\mathsf{PLP}} (u) := \mathsf{w} (u) -  \mathsf{w}_{\mathsf{LP}\text{-}\mathsf{bc}} (u)$ and $\mathsf{w}_{\mathsf{mPLP}} (u) := \mathsf{w} (u) - Q\mathsf{w}_{\mathsf{LP}\text{-}\mathsf{bc},\mathsf{bnd}} (u)$.

\subsection{bias expansions}
\label{app:lemmas}

In this appendix we collect some results regarding the bias term $B_n$ and its expansion for both interior and boundary points. Recall that throughout we consider only left boundary points. Proofs of the results in this appendix are given in Section~\ref{Supp:Sec-extra} of the Supplementary Material.

We first formalize the asymptotic behavior of $C_n$ and~$C_{\mathsf{LP},n}$.

\begin{lemma}
\label{lemmaKernelConstants}
Let Assumptions~\ref{Ass_g} and~\ref{Ass_K,h} be satisfied.\\
(i) If $\mathsf{x}$ is an interior point, then $C_n\plowto C$ and $C_{\mathsf{LP},n}\plowto C$, where $C:= \int_{\mathcal{X}}\mathsf{w} (u) u^{p+1} du$.\\
(ii) If $\mathsf{x}$ is a boundary point, then $C_n\plowto C$ and $C_{\mathsf{LP},n}\plowto C_{\mathsf{LP}}$, where
\begin{equation*}
C_{\mathsf{LP}}:=\int_{0}^1\mathsf{w}_{\mathsf{bnd}}(s,0) \int_{-s}^1\mathsf{w}_{\mathsf{bnd}}(u,s)u^{p+1}du ds \neq C.
\end{equation*}
\end{lemma}

The next two lemmas derive expansions of $\hat{B}_{\mathsf{GP},n}$ and $\hat{B}_{\mathsf{LP},n}$ for interior and boundary points. As we will see in Lemma~\ref{lemmaLPbias}, the key consequence of Lemma~\ref{lemmaKernelConstants} is that the probability limit of $A_n$ is zero if $\mathsf{x}$ is an interior point.

\begin{lemma}
\label{lemmaGPbias}
Under Assumptions~\ref{Ass_g} and~\ref{Ass_K,h}, for any evaluation point~$\mathsf{x}$,
\begin{equation*}
    \hat{B}_{\mathsf{GP},n}-B_n =\xi_{2n,\mathsf{GP}}+\op(1), \quad \xi_{2n,\mathsf{GP}}:=(nh)^{-1/2} \sum_{i=1}^n w_{\mathsf{GP}\text{-}\mathsf{bc},i} (\mathsf{x})\varepsilon_i .
\end{equation*}
\end{lemma}

\begin{lemma}
\label{lemmaLPbias}
Under Assumptions~\ref{Ass_g} and~\ref{Ass_K,h}, for any evaluation point~$\mathsf{x}$,
\begin{equation*}
    \hat{B}_{\mathsf{LP},n}-B_n - A_n =\xi_{2n,\mathsf{LP}}+\op(1), \quad \xi_{2n,\mathsf{LP}}:=(nh)^{-1/2} \sum_{i=1}^n w_{\mathsf{LP}\text{-}\mathsf{bc},i} (\mathsf{x})\varepsilon_i,
\end{equation*}
where $A_n:=\sqrt{nh^{2p+3}}{g^{(p+1)}(\mathsf{x})}({C}_{\mathsf{LP},n}-C_n)/(p+1)!$ and:\\
(i)~If $\mathsf{x}$ is an interior point, $A_n \plowto 0$;\\
(ii)~If $\mathsf{x}$ is a boundary point, $A_n \plowto \sqrt{\kappa}g^{(p+1)}(\mathsf{x})(C_{\mathsf{LP}}-C) /(p+1)!$, where $C_{\mathsf{LP}} \neq C$.
\end{lemma}

Let $T_n - B_n = \xi_{1n} := (nh)^{-1/2} \sum_{i=1}^n w_i (\mathsf{x}) \varepsilon_i$. The next two lemmas show how the equivalent kernels defined in Appendix~\ref{app:kernels} can be used in asymptotic approximations of the bias-corrected statistics, $T_n - \hat B_{\mathsf{GP},n} = \xi_{1n} - \xi_{2n,\mathsf{GP}}$ and $T_n - \hat B_{\mathsf{LP},n} = \xi_{1n} - \xi_{2n,\mathsf{LP}}$. 

\begin{lemma}
\label{lemmaequivkerGP}
Let Assumptions~\ref{Ass_g} and~\ref{Ass_K,h} be satisfied. Then,
\begin{align*}
    \left(\begin{matrix}
        \xi_{1n} \\
        \xi_{2n,\mathsf{GP}}
    \end{matrix}\right) 
    &= \frac{1}{f(\mathsf{x})\sqrt{nh}} \sum_{i=1}^n     
    \left(\begin{matrix}
        \mathsf{w} \left( \frac{x_i - \mathsf{x}}{h} \right) \\
        \mathsf{w}_{\mathsf{GP}\text{-}\mathsf{bc}}\left( \frac{x_i - \mathsf{x}}{h} \right)
    \end{matrix}\right) \varepsilon_i + \op(1) .
\end{align*}
\end{lemma}

\begin{lemma}
\label{lemmaequivker2}
Let Assumptions~\ref{Ass_g} and~\ref{Ass_K,h} be satisfied.\\
(i) If $\mathsf{x}$ is an interior point, then
\begin{align*}
    \left(\begin{matrix}
        \xi_{1n} \\
        \xi_{2n,\mathsf{LP}}
    \end{matrix}\right) 
    &= \frac{1}{f(\mathsf{x})\sqrt{nh}} \sum_{i=1}^n     
    \left(\begin{matrix}
        \mathsf{w} \left( \frac{x_i - \mathsf{x}}{h} \right) \\
        \mathsf{w}_{\mathsf{LP}\text{-}\mathsf{bc}}\left( \frac{x_i - \mathsf{x}}{h} \right)
    \end{matrix}\right) \varepsilon_i + \op(1) .
\end{align*}
(ii) If $\mathsf{x}$ is a boundary point, then
\begin{align*}
    \left(\begin{matrix}
        \xi_{1n} \\
        \xi_{2n,\mathsf{LP}}
    \end{matrix}\right) 
    &= \frac{1}{f(\mathsf{x})\sqrt{nh}} \sum_{i=1}^n    
     \left(\begin{matrix}
        \mathsf{w} \left( \frac{x_i - \mathsf{x}}{h}\right) \\
        \mathsf{w}_{\mathsf{LP}\text{-}\mathsf{bc},\mathsf{bnd}} \left( \frac{x_i - \mathsf{x}}{h} \right)
    \end{matrix}\right) \varepsilon_i + \op(1) .
\end{align*}
\end{lemma}

\section{proofs of main results}
\label{app:main proofs}

In this appendix we prove our results. Throughout, for simplicity, we give all proofs for the case of the bias-corrected HC0 residuals,
\begin{equation}
\label{eq resid}
\tilde\varepsilon_i := y_i - r_{p+1} (x_i - \mathsf{x})^\prime \hat\beta_{p+1,n}.
\end{equation}
The proofs with $k\in\{1,2,3\}$ are notationally more cumbersome, but very similar in spirit; see also Calonico et al.\ (2018, 2022).

\subsection{proofs from section~\ref{section RBC and prepivoting}}

\subsubsection*{proof of theorem~\ref{th 1}}

\noindent \textsc{part}~(i). We write the quantile function of $T_n^*$ as $\hat{L}_n^{-1}(u) = \hat{B}_n + \hat{L}^{-1}_{1,n}(u)$, where $\hat{L}_{1,n}(u) := \PP^* (\xi^*_{1n} \leq u)$ and $\xi_{1n}^* := T_n^* - \hat{B}_n$. Under Assumption~\ref{assn 1}(i) and using the continuous mapping theorem, we have $\hat{L}_{1,n}^{-1}(u) \plowto v \Phi^{-1}(u) =: L_1^{-1}(u)$ pointwise for all~$u$. Because $\hat{L}_{1,n}^{-1}(u)$ is monotonic in~$u$ and $L_1^{-1}(u)$ is continuous in~$u$, the convergence is uniform in~$u$. By the definition of $\hat{H}_n$ in \eqref{Hhat_def} we have, for any $\alpha \in (0,1)$, $\hat{H}_n^{-1}(\alpha) = \Phi (\hat{m}_n \Phi^{-1}(\alpha)) = \Phi (\hat{m}_n z_\alpha )$. Combining these results we find that 
\begin{equation*}
(nh)^{-1/2} \hat{L}_n^{-1}(\hat{H}_n^{-1}(\alpha)) = (nh)^{-1/2} \hat{B}_n + (nh)^{-1/2} \hat{v}_n \hat{m}_n z_\alpha +\op((nh)^{-1/2} ),
\end{equation*}
which proves part~(i) because $\hat{v}_n \hat{m}_n = \hat{v}_{\mathsf{P},n}$ by definition of~$\hat{m}_n$.

\medskip \noindent \textsc{part}~(ii). If $T_n^*$ is Gaussian conditionally on~$\mathscr{D}_n$, then $\hat{L}_n^{-1}(u) = \hat{B}_n + \hat{v}_n \Phi^{-1}(u)$, and the remainder of the proof follows as in part~(i).

\medskip \noindent \textsc{part}~(iii). This result follows from \eqref{CIp coverage} if we show that $\hat{H}_n$ is a uniformly consistent estimator of~$H$. Under Assumption~\ref{assn 1}(i) we have $\hat{p}_n := \PP^* (T_n^* \leq T_n ) = \Phi ((T_n - \hat{B}_n)/v) + \op(1)$, so that, applying also Assumption~\ref{assn 1}(ii),
\begin{align*}
H_n (u) := \PP (\hat{p}_n \leq u) &= \PP ( \Phi ((T_n - \hat{B}_n)/v ) \leq u) + o(1) 
= \PP ( (T_n - \hat{B}_n)/v_{\mathsf{P}}) \leq (v/v_{\mathsf{P}}) \Phi^{-1}(u)) +o(1) \\
&\to \Phi ( (v/v_{\mathsf{P}}) \Phi^{-1}(u)) =: H(u).
\end{align*}
Thus, $\sup_{u\in\RR } |\hat{H}_n(u) - H(u)| \plowto 0$ if $\hat{m}_n \plowto v_{\mathsf{P}}/v$, which holds by assumption. $\hfill \square$

\subsubsection*{proof of theorem~\ref{Th RBC PGP equivalence}}

\noindent By Theorem~\ref{th 1}, the result follows by showing that Assumption~\ref{assn 1} holds with $(T_n^{\ast},\hat B_n, \hat{v}_n,\hat{v}_{\mathsf{P},n})$ replaced by $(T_{\mathsf{GP},n}^{\ast},\hat B_{\mathsf{GP},n}, \hat{v}_{\mathsf{GP},n},\hat{v}_{\mathsf{PGP},n})$.

Let $\mathsf{x}$ be an interior point. We first show that Assumption~\ref{assn 1}(i) is satisfied under Assumptions~\ref{Ass_g}--\ref{Ass_K,h}. Note that
\begin{equation*}
T^*_{\mathsf{GP},n} -\hat B_{\mathsf{GP},n} = \xi_{1n}^* = \frac{1}{\sqrt{nh}} \sum_{i=1}^n w_i(\mathsf{x}) \varepsilon_i^* ,
\end{equation*}
where $\E^*[\xi_{1n}^*]=0$. By Lemma~\ref{lemmaequivker} we can write $\xi_{1n}^{\ast}= f(\mathsf{x})^{-1} s_{1n}^*(1+ \op (1))$, where 
\begin{equation*}
s_{1n}^* = \frac{1}{\sqrt{nh}} \sum_{i=1}^n \mathsf{w} \left( \frac{x_i-\mathsf{x}}{h}\right) \tilde\varepsilon_i e_i^*.
\end{equation*}
The conditional mean and variance are $\E^* [ s_{1n}^* ] = 0$ and $\var^* [ s_{1n}^* ] = (nh)^{-1} \sum_{i=1}^n \mathsf{w}((x_i-\mathsf{x})/h)^2 \tilde\varepsilon_i^2 = f(\mathsf{x})^2 \iota_0^{\prime} \dot\Gamma_p^{-1}\tilde\Psi_p \dot\Gamma_p^{-1}\iota_0$.
From Lemma~\ref{auxlemma2}, $\tilde\Psi_p \plowto \dot\Psi_p$, so that $\var^*[s_{1n}^*]\plowto f(\mathsf{x})^2 v_1^2 $ with $v_1^2 :=\iota_0^{\prime} \dot\Gamma_p^{-1} \dot\Psi_p \dot\Gamma_p^{-1} \iota_0$. We prove asymptotic normality using Lyapunov's CLT. To this end, we first show
\begin{equation} \label{P1}
s_{1n}^* = \frac{1}{\sqrt{nh}} \sum_{i=1}^n \mathsf{w} \left( \frac{x_i-\mathsf{x}}{h}\right) \varepsilon_i e_i^* + \opstar (1) .
\end{equation}
To see this, it is sufficient to note that $(nh)^{-1/2} \sum_{i=1}^n \mathsf{w} ((x_i-\mathsf{x})/h) (\tilde\varepsilon_i - \varepsilon_i) e_i^* = f(\mathsf{x}) \iota_0^{\prime}\dot\Gamma_p^{-1}\eta_n^*$, where $\eta_n^* := (nh)^{-1/2} \sum_{i=1}^n K(( x_i-\mathsf{x})/h) r_p ((x_i-\mathsf{x})/h) (\tilde\varepsilon_i - \varepsilon_i) e_i^*$ has conditional mean $\E^* [ \eta_n^* ] = 0$ and conditional variance $\var^* [ \eta_n^*] = (nh)^{-1} \sum_{i=1}^n K^2 (( x_i-\mathsf{x})/h) r_p ((x_i-\mathsf{x})/h) r_p^{\prime} ((x_i-\mathsf{x})/h) (\tilde\varepsilon_i - \varepsilon_i)^2$, which has $(i,j)$-th element equal to~$\tilde\psi_{i+j-2,n,2}$ defined in the proof of Lemma~\ref{auxlemma2}. Hence, \eqref{P1} follows by Chebychev's inequality because $\tilde\psi_{i+j-2,n,2}\plowto 0$ as shown in the proof of Lemma~\ref{auxlemma2}.

Clearly, $\E^* [(nh)^{-1/2} \sum_{i=1}^n \mathsf{w} ((x_i-\mathsf{x})/h) \varepsilon_i e_i^*]=0$ and $\var^*[(nh)^{-1/2} \sum_{i=1}^n \mathsf{w} ((x_i-\mathsf{x})/h) \varepsilon_i e_i^*] \plowto v_1^2 f(\mathsf{x})^2$; see also the proof of Lemma~\ref{auxlemma2}. To verify Lyapunov's condition we find, for $\delta>1$,
\begin{equation*}
\frac{1}{(nh)^\delta} \sum_{i=1}^n \E^* [ |\mathsf{w} ((x_i-\mathsf{x})/h) \varepsilon_ie_i^* |^{2\delta} ]
\leq \frac{c}{(nh)^\delta} \sum_{i=1}^n |\mathsf{w}((x_i-\mathsf{x})/h)|^{2\delta} |\varepsilon_i|^{2\delta} .
\end{equation*}
By Markov's inequality, the law of iterated expectations, and $\sup_i \E[|\varepsilon_i|^{2\delta}]<\infty$ (taking $\delta \in (1,2]$),
\begin{align*}
\PP \left( \sum_{i=1}^n |\mathsf{w}((x_i-\mathsf{x})/h)|^{2\delta}  |\varepsilon_i|^{2\delta} \geq \epsilon \right)
\leq \frac{1}{\epsilon} \E \left[ \sum_{i=1}^n |\mathsf{w}((x_i-\mathsf{x})/h)|^{2\delta} |\varepsilon_i|^{2\delta} \right]
\leq \frac{c}{\epsilon} \sum_{i=1}^n \E [ |\mathsf{w}((x_i-\mathsf{x})/h)|^{2\delta} ] .
\end{align*}
By the i.i.d.\ assumption, the right-hand side is bounded by a constant times
\begin{align}
\nonumber
n \E [ | \mathsf{w} ((x_1-\mathsf{x})/h) |^{2\delta} ] 
&= n f(\mathsf{x})^{2\delta} \int_{\mathsf{x}-h}^{\mathsf{x}+h} |\iota_0^{\prime} \dot\Gamma_p^{-1} r_p((x_1-\mathsf{x})/h) K((x_1-\mathsf{x})/h) |^{2\delta} f(x_1) dx_1 \\
&= nh f(\mathsf{x})^{2\delta} \int_{\mathcal{X}} |\iota_0^{\prime} \dot\Gamma_p^{-1} r_p(u) K(u) |^{2\delta} f(\mathsf{x}+uh) du = O(nh),
\label{Lyapunov}
\end{align}
which verifies Lyapunov's condition because $\delta>1$.

Next, we show that Assumption~\ref{assn 1}(ii) is satisfied under Assumptions~\ref{Ass_g}--\ref{Ass_K,h}. By definition of $\hat{B}_{\mathsf{GP},n}$ and Lemmas~\ref{lemmaGPbias} and~\ref{lemmaequivkerGP},
\begin{equation*}
    \left(\begin{matrix}
        T_n - B_n \\
        \hat{B}_{\mathsf{GP},n}- B_n 
    \end{matrix}\right) 
=         \frac{1}{f(\mathsf{x})\sqrt{nh}} \sum_{i=1}^n\left(\begin{matrix}
        \mathsf{w}    \left( \frac{x_i - \mathsf{x}}{h} \right)\\
        \mathsf{w}_{\mathsf{GP}\text{-}\mathsf{bc}}    \left( \frac{x_i - \mathsf{x}}{h} \right)
    \end{matrix}\right)  \varepsilon_i  + \op(1) 
= \frac{1}{f(\mathsf{x})} \mathsf{s}_{\mathsf{GP},n} + \op(1),
\end{equation*} 
where $\mathsf{s}_{\mathsf{GP},n} := (nh)^{-1/2} \sum_{i=1}^n ( \mathsf{w} ( ({x_i - \mathsf{x}})/{h} ), \mathsf{w}_{\mathsf{GP}\text{-}\mathsf{bc}}( ({x_i - \mathsf{x}})/{h} ) )^{\prime} \varepsilon_i$. Assumption~\ref{assn 1}(ii) is then verified by showing asymptotic normality of~$\mathsf{s}_{\mathsf{GP},n}$. Clearly, $\E [\mathsf{s}_{\mathsf{GP},n}]=0$ by the law of iterated expectations. The asymptotic variance of $\mathsf{s}_{\mathsf{GP},n}$ is $V_{\mathsf{GP}}$, where
\begin{align*}
V_{\mathsf{GP},11} &= \lim_{n\to \infty} \E \left[ \frac{1}{nh} \sum_{i=1}^n \mathsf{w}(({x_i - \mathsf{x}})/h)^2 \varepsilon_i^2 \right] 
= \lim_{n\to \infty} \frac{1}{nh} \sum_{i=1}^n \E [ \mathsf{w}(({x_i - \mathsf{x}})/h)^2 \sigma^2 (x_i) ] \\
&= \lim_{n\to \infty}\frac{1}{h} \int_{\mathsf{x}-h}^{\mathsf{x}+h} \mathsf{w}(({x_1 - \mathsf{x}})/h)^2 \sigma^2 (x_1) f(x_1) dx_1
=\sigma^2 (\mathsf{x})f(\mathsf{x}) \int_{\mathcal{X}} \mathsf{w}(u)^2 du .
\end{align*}
Similarly, $V_{\mathsf{GP},12} = V_{\mathsf{GP},21} = \sigma^2 (\mathsf{x}) f(\mathsf{x}) \int_{\mathcal{X}} \mathsf{w}(u) \mathsf{w}_{\mathsf{GP}\text{-}\mathsf{bc}}(u)du$ and $V_{\mathsf{GP},22}= \sigma^2 (\mathsf{x}) f(\mathsf{x}) \int_{\mathcal{X}} \mathsf{w}_{\mathsf{GP}\text{-}\mathsf{bc}}(u)^2 du$. Because $\mathsf{s}_{\mathsf{GP},n} $ is a sum of independent bivariate random variables, we use the Cram\'{e}r-Wold device and verify Lyapunov's condition. Thus, we prove that, for any $(\zeta_1 , \zeta_2 ) \neq (0,0)$ and some $\delta >1$,
\begin{equation*}
\frac{1}{(nh)^{\delta}}\sum_{i=1}^n\E [ | (\zeta_1 \mathsf{w} ( ({x_i - \mathsf{x}})/{h} )+ \zeta_2 \mathsf{w}_{\mathsf{GP}\text{-}\mathsf{bc}} ( ({x_i - \mathsf{x}})/{h} ))\varepsilon_i|^{2\delta}] \to 0.
\end{equation*}
By Minkowski's inequality we can prove this separately for each summand. For $\mathsf{w}$, the result follows directly from \eqref{Lyapunov}, and the corresponding result for $\mathsf{w}_{\mathsf{GP}\text{-}\mathsf{bc}}$ is proven in the same way and therefore omitted. This concludes the proof of asymptotic normality of~$\mathsf{s}_{\mathsf{GP},n}$ and hence of Theorem~\ref{Th RBC PGP equivalence}. 

If $\mathsf{x}$ is a boundary point, the proof follows by nearly identical steps, so we omit the details. The main difference between the interior and boundary point cases is the asymptotic variance of $(T_n - B_n, \hat B_{\mathsf{GP},n} - B_n)^{\prime}$. Specifically, if $\mathsf{x}$ is a boundary point, this asymptotic variance changes because its entries have integrals evaluated on $[0,\infty)$ instead of $\mathbb{R}$.

\subsection{proofs from section~\ref{Section main}}
\label{sec proofs Main}

\subsubsection*{proof of lemma \ref{Lemma v2plp}}

We use \eqref{eq resid} and $r_{p+1}(x) = r_{p+1}(x/h)H_{p+1}^{-1}$ to decompose $\hat{v}_{\mathsf{PLP},n}^2 = \hat{v}_{\mathsf{PLP},1n}^2 +  \hat{v}_{\mathsf{PLP},2n}^2 + \hat{v}_{\mathsf{PLP},3n}^2 + 2\hat{v}_{\mathsf{PLP},4n}^2$, where we show that
\begin{align*}
\hat{v}_{\mathsf{PLP},1n}^2 &:= \frac{1}{f(\mathsf{x})^2} \frac{1}{nh} \sum_{i=1}^n  \mathsf{w}_{\mathsf{PLP}} \left(\frac{x_i-\mathsf{x}}{h}\right)^2 \varepsilon_i^2 
=\frac{\sigma^2(\mathsf{x}) }{f(\mathsf{x})}\int_{\mathcal{X}} \mathsf{w}_{\mathsf{PLP}} (u)^2 du + \op(1) 
=: v_{\mathsf{PLP}}^2 +\op(1) ,\\
\hat{v}_{\mathsf{PLP},2n}^2 &:= \frac{1}{nh} \sum_{i=1}^n  \left( w_{\mathsf{PLP},i} (\mathsf{x})^2 - \frac{1}{f(\mathsf{x})^2}\mathsf{w}_{\mathsf{PLP}} \left(\frac{x_i-\mathsf{x}}{h}\right)^2 \right) \varepsilon_i^2 = \op(1) , \\
\hat{v}_{\mathsf{PLP},3n}^2 &:= \frac{1}{nh}\sum_{i=1}^n  w_{\mathsf{PLP},i}  (\mathsf{x})^2 \left( g (x_i) - r_{p+1}^{\prime} \left(\frac{x_i-\mathsf{x}}{h}\right) H_{p+1}^{-1}\hat\beta_{p+1,n} \right)^2 = \op(1) , \\
\hat{v}_{\mathsf{PLP},4n}^2 &:= \frac{1}{nh}\sum_{i=1}^n  w_{\mathsf{PLP},i} (\mathsf{x})^2 \left( g (x_i) - r_{p+1}^{\prime} \left(\frac{x_i-\mathsf{x}}{h}\right) H_{p+1}^{-1}\hat\beta_{p+1,n}\right) \varepsilon_i = \op(1) ,
\end{align*}
which suffices to prove the required result.

Proof for~$\hat{v}_{\mathsf{PLP},1n}^2$. By Assumption~\ref{Ass_g}, the mean is
\begin{align*}
\E[\hat{v}_{\mathsf{PLP},1n}^2] &= \frac{1}{hf(\mathsf{x})^2}\int_{\mathsf{x}-2h}^{\mathsf{x}+2h}  \mathsf{w}_{\mathsf{PLP}} \left( \frac{x_1 - \mathsf{x}}{h} \right)^2 \sigma^2(x_1) f(x_1) dx_1 \\
&=\frac{\sigma^2(\mathsf{x}) }{f(\mathsf{x})}\int_{\mathcal{X}} \mathsf{w}_{\mathsf{PLP}} (u)^2 du + o(1) 
=: v_{\mathsf{PLP}}^2 +o(1)
\end{align*}
and the variance is
\begin{align*}
\var [\hat{v}_{\mathsf{PLP},1n}^2] &= 
\frac{1}{(nh)^2 f (\mathsf{x})^4} \sum_{i=1}^n \left( \E \left[ \mathsf{w}_{\mathsf{PLP}} \left(\frac{x_i-\mathsf{x}}{h}\right)^4 \varepsilon_i^4 \right] - \left( \E \left[ \mathsf{w}_{\mathsf{PLP}} \left(\frac{x_i-\mathsf{x}}{h}\right)^2 \varepsilon_i^2 \right] \right) \right) \\
&\leq \frac{1}{(nh)^2 f (\mathsf{x})^4} \sum_{i=1}^n ch \int_{\mathsf{x}-h}^{\mathsf{x}+h} \mathsf{w}_{\mathsf{PLP}}(u)^4 f(\mathsf{x}+uh)du \\
&\quad + \frac{1}{(nh)^2 f (\mathsf{x})^4} \sum_{i=1}^n \left( h \int_{\mathsf{x}-h}^{\mathsf{x}+h} \mathsf{w}_{\mathsf{PLP}}(u)^2 \sigma^2 (\mathsf{x}+uh) f(\mathsf{x}+uh)du \right)^2
= O \left(\frac{1}{nh}\right).
\end{align*}
This shows $L_2$-convergence of~$\hat{v}_{\mathsf{PLP},1n}^2$.

Proof for~$\hat{v}_{\mathsf{PLP},2n}^2$. Follows by direct application of Lemma~\ref{new supp lemma 3} with $\zeta_i = \varepsilon_i$.

Proof for~$\hat{v}_{\mathsf{PLP},3n}^2$. By the mean-value theorem, and using $\iota _{j}^{\prime }\beta_{p+1} = g^{(j)}(\mathsf{x})/j!$, we have
\begin{align}
g (x_i) &= 
\sum_{j=1}^p (x_i-\mathsf{x})^j g^{(j)}(\mathsf{x})/j!
+ (x_i-\mathsf{x})^{p+1} g^{(p+1)}(\tilde{x}_i)/(p+1)! \nonumber \\
&= r_{p+1}^{\prime} \left(\frac{x_i-\mathsf{x}}{h}\right) H_{p+1}^{-1} \beta_{p+1}
+ (x_i-\mathsf{x})^{p+1} (g^{(p+1)}(\tilde{x}_i) - g^{(p+1)}(\mathsf{x}))/(p+1)!
\label{MVT on g}
\end{align}
for an intermediate value $\tilde{x}_i$ such that $|\tilde{x}_i - \mathsf{x}| \leq |x_i - \mathsf{x}|$. Then we can write $\hat{v}_{\mathsf{PLP},3n}^2$ as
\begin{align*}
&\frac{1}{nh} \sum_{i=1}^n  w_{\mathsf{PLP},i} (\mathsf{x})^2 \left( r_{p+1}^{\prime} \left(\frac{x_i-\mathsf{x}}{h}\right)H_{p+1}^{-1} (\beta_{p+1} - \hat\beta_{p+1,n}) \right)^2 \\
& + \frac{1}{nh} \sum_{i=1}^n  w_{\mathsf{PLP},i}(\mathsf{x})^2 \left(\frac{x_i-\mathsf{x}}{h}\right)^{2p+2} h^{2p+2} \left(\frac{g^{(p+1)}(\tilde{x}_i) - g^{(p+1)}(\mathsf{x})}{(p+1)!} \right)^2  \\
& + \frac{2}{nh} \sum_{i=1}^n  w_{\mathsf{PLP},i} (\mathsf{x})^2 \left(\frac{x_i-\mathsf{x}}{h}\right)^{p+1}r_{p+1}^{\prime} \left(\frac{x_i-\mathsf{x}}{h}\right) H_{p+1}^{-1} (\beta_{p+1} - \hat\beta_{p+1,n}) h^{p+1} \left(\frac{g^{(p+1)}(\tilde{x}_i) - g^{(p+1)}(\mathsf{x})}{(p+1)!} \right) .
\end{align*}
By the fact that $|\hat\beta_{s,p+1,n} - \beta_{s,p+1}|= \Op ((nh^{2s-1})^{-1/2})$, where $\hat\beta_{s,p+1,n}$ and $\beta_{s,p+1}$ are the $s$-th elements of $\hat\beta_{p+1,n}$ and $\beta_{p+1}$, respectively, we obtain $|H_{p+1}^{-1}(\hat\beta_{p+1,n} - \beta_{p+1})| = \Op ((nh)^{-1/2}) = \op(1)$. Moreover, noting that $w_{\mathsf{PLP},i} (\mathsf{x}) \neq 0$ only when $|\tilde{x}_i-\mathsf{x}| \leq | x_i-\mathsf{x}| \leq 2h$, the Hölder condition on $g^{(p+1)}$ in Assumption~\ref{Ass_g}(iii) implies that $\sup_{i:w_{\mathsf{PLP},i} (\mathsf{x}) \neq 0} |g^{(p+1)}(\tilde{x}_i) - g^{(p+1)}(\mathsf{x})| \leq \sup_{x: |x-\mathsf{x}|\leq 2h}|g^{(p+1)}(x) - g^{(p+1)}(\mathsf{x})| \leq \sup_{x: |x-\mathsf{x}|\leq 2h} c |x-\mathsf{x}|^\eta = o(1)$. Thus, for some power~$k\geq 0$, we can bound all the terms in $\hat{v}_{\mathsf{PLP},3n}^2$ by
\begin{align*}
\op(1) \frac{1}{nh} \sum_{i=1}^n  w_{\mathsf{PLP},i} (\mathsf{x})^2 \left|\frac{x_i-\mathsf{x}}{h}\right|^k
= \op(1) \frac{1}{nhf(\mathsf{x})^2} \sum_{i=1}^n \mathsf{w}_{\mathsf{PLP}} \left( \frac{x_i - \mathsf{x}}{h} \right)^2 \left|\frac{x_i-\mathsf{x}}{h}\right|^k +\op(1),
\end{align*}
where the equality is by Lemma~\ref{new supp lemma 3} with $\zeta_i = |(x_i-\mathsf{x})/h|^k$. The main term satisfies 
\begin{equation*}
\frac{1}{nhf(\mathsf{x})^2} \sum_{i=1}^n \mathsf{w}_{\mathsf{PLP}} \left( \frac{x_i - \mathsf{x}}{h} \right)^2 \left|\frac{x_i-\mathsf{x}}{h}\right|^k = \frac{1}{f(\mathsf{x})} \int_{\mathcal{X}} \mathsf{w}_{\mathsf{PLP}} (u)^2 |u|^k du +\op (1),
\end{equation*}
which proves the result for~$\hat{v}_{\mathsf{PLP},3n}^2$. 

Proof for~$\hat{v}_{\mathsf{PLP},4n}^2$. By the same arguments as in the proof for $\hat{v}_{\mathsf{PLP},3n}^2$ we find that all the terms in $\hat{v}_{\mathsf{PLP},4n}^2$ are bounded, for some~$k$, by
\begin{align*}
\op(1) \frac{1}{nh} \sum_{i=1}^n  w_{\mathsf{PLP},i} (\mathsf{x})^2 \left|\frac{x_i-\mathsf{x}}{h}\right|^k |\varepsilon_i|
= \op(1) \frac{1}{nhf(\mathsf{x})^2} \sum_{i=1}^n \mathsf{w}_{\mathsf{PLP}} \left( \frac{x_i - \mathsf{x}}{h} \right)^2 \left|\frac{x_i-\mathsf{x}}{h}\right|^k |\varepsilon_i| +\op(1),
\end{align*}
where the equality is by Lemma~\ref{new supp lemma 3} with $\zeta_i =  |(x_i-\mathsf{x})/h|^k |\varepsilon_i|$, and the result for~$\hat{v}_{\mathsf{PLP},4n}^2$ follows as for~$\hat{v}_{\mathsf{PLP},3n}^2$.  $\hfill \square$

\subsubsection*{proof of theorem~\ref{Th RBC PLP equivalence}}

\noindent By Theorem~\ref{th 1}, the result follows by showing that Assumption~\ref{assn 1} holds with $(T_n^{\ast},\hat B_n, \hat{v}_n,\hat{v}_{\mathsf{P},n})$ replaced by $(T_{\mathsf{LP},n}^{\ast},\hat B_{\mathsf{LP},n}, \hat{v}_{\mathsf{LP},n},\hat{v}_{\mathsf{PLP},n})$. First, Assumption~\ref{assn 1}(i) is satisfied because $T^*_{\mathsf{LP},n} -\hat B_{\mathsf{LP},n}  = \xi_{1n}^* = (nh)^{-1/2} \sum_{i=1}^n w_i(\mathsf{x}) \varepsilon_i^*$ is identical to the corresponding term in the proof of Theorem~\ref{Th RBC PGP equivalence}.

We now prove that Assumption~\ref{assn 1}(ii) holds. By definition of $\hat{B}_{\mathsf{LP},n}$ and Lemma~\ref{lemmaLPbias},
\begin{equation}
    \left(\begin{matrix}
        T_n - B_n \\
        \hat{B}_{\mathsf{LP},n}- B_n 
    \end{matrix}\right) 
=   \left(\begin{matrix}
        0 \\
        A_n
    \end{matrix} \right) 
+  \frac{1}{\sqrt{nh}} \sum_{i=1}^n\left(\begin{matrix}
        w_i(\mathsf{x})\\
        w_{\mathsf{LP}\text{-}\mathsf{bc},i} (\mathsf{x})
    \end{matrix}\right)  \varepsilon_i  + \op(1) .
\end{equation}
By Lemma~\ref{lemmaLPbias}(i), $A_n = \op(1)$ because $\mathsf{x}$ is an interior point, and by Lemma~\ref{lemmaequivker2}(i) we have that
\begin{align*}
        \frac{1}{\sqrt{nh}} \sum_{i=1}^n\left(\begin{matrix}
        w_i(\mathsf{x})\\
        w_{\mathsf{LP}\text{-}\mathsf{bc},i} (\mathsf{x})
    \end{matrix}\right)  \varepsilon_i 
&=         \frac{1}{f(\mathsf{x})\sqrt{nh}} \sum_{i=1}^n\left(\begin{matrix}
        \mathsf{w}    \left( \frac{x_i - \mathsf{x}}{h} \right)\\
        \mathsf{w}_{\mathsf{LP}\text{-}\mathsf{bc}}    \left( \frac{x_i - \mathsf{x}}{h} \right)
    \end{matrix}\right)  \varepsilon_i  + \op(1) 
= \frac{1}{f(\mathsf{x})} \left[\begin{matrix}
        1 & 0 \\
        -1 & 1
    \end{matrix}\right]    \mathsf{s}_{\mathsf{LP},n} + \op(1),
\end{align*} 
where $\mathsf{s}_{\mathsf{LP},n} := (nh)^{-1/2} \sum_{i=1}^n ( \mathsf{w} ( ({x_i - \mathsf{x}})/{h} ), \mathsf{w}_{\mathsf{conv}}( ({x_i - \mathsf{x}})/{h} ) )^{\prime} \varepsilon_i$. Because $\mathsf{s}_{\mathsf{LP},n}$ is a sum of independent bivariate random variables, we can verify Lyapunov's condition in the same way as in the proof of Theorem~\ref{Th RBC PGP equivalence}. The condition for the term involving $\mathsf{w}$ was verified in the proof of Theorem~\ref{Th RBC PGP equivalence}, so we are left with proving that, for some $\delta >1$,
\begin{equation*}
\frac{1}{(nh)^{\delta}}\sum_{i=1}^n\E [ | \mathsf{w}_{\mathsf{conv}} ( ({x_i - \mathsf{x}})/{h} )\varepsilon_i|^{2\delta}] \to 0.
\end{equation*}
By the i.i.d.\ assumption and taking $\delta \in (1,2]$, the left-hand side is
\begin{align*}
\frac{n}{(nh)^{\delta}} \E [ | \mathsf{w}_{\mathsf{conv}} ( ({x_1 - \mathsf{x}})/{h} )\varepsilon_1|^{2\delta}]
&\leq c \frac{n}{(nh)^{\delta}} \E [ | \mathsf{w}_{\mathsf{conv}} ( ({x_1 - \mathsf{x}})/{h} ) |^{2\delta}] \\
&= c\frac{n}{(nh)^{\delta}} \int_{\mathbb{S}_x} | \mathsf{w}_{\mathsf{conv}} ( ({x_1 - \mathsf{x}})/{h} )|^{2\delta} f(x_1)dx_1 \\
&= c\frac{1}{(nh)^{\delta-1}} \int_{\mathcal{X}} | \mathsf{w}_{\mathsf{conv}} (u)|^{2\delta} f(\mathsf{x}+uh) du \to 0.
\end{align*}

We are left with defining the asymptotic variance of $\mathsf{s}_{\mathsf{LP},n}$, which is needed to define the asymptotic variance of $(T_n - B_n ,\hat{B}_{\mathsf{LP},n}- B_n)^{\prime}$. To do so, we let $\Omega_{\mathsf{LP},n} := \lim_{n \to \infty} \var [\mathsf{s}_{\mathsf{LP},n}]$ and note that $\omega_{\mathsf{LP},11} = V_{\mathsf{GP},11} = \sigma^2 (\mathsf{x}) f(\mathsf{x})\int_{\mathcal{X}} \mathsf{w}(u)^2 du$, and, similarly, $\omega_{\mathsf{LP},12} =\sigma^2 (\mathsf{x}) f(\mathsf{x})\int_{\mathcal{X}} \mathsf{w} (u) \mathsf{w}_{\mathsf{conv}} (u) du$ and $\omega_{\mathsf{LP},22}= \sigma^2 (\mathsf{x}) f(\mathsf{x})\int_{\mathcal{X}} \mathsf{w}_{\mathsf{conv}} (u)^2 du$. Hence, the convergence $\mathsf{s}_{\mathsf{LP},n} \dlowto N(0,\Omega_{\mathsf{LP}})$ implies that $ (T_n - B_n,\hat{B}_{\mathsf{LP},n}- B_n)^{\prime} \dlowto N(0,V_{\mathsf{LP}})$ with 
\begin{equation} \label{VLPformula}
            V_{\mathsf{LP}} := \frac{1}{f (\mathsf{x})^2} \left[\begin{matrix}
        1 & 0 \\
        -1 & 1
    \end{matrix}\right] \Omega_{\mathsf{LP}} \left[\begin{matrix}
        1 & -1 \\
        0 & 1
    \end{matrix}\right] 
\end{equation}
concluding the proof of Theorem~\ref{Th RBC PLP equivalence}. $\hfill \square$

\subsubsection*{proof of corollary \ref{corollary v^2_P comparisons}}

From \eqref{VLPformula} we obtain
\begin{align*}
V_{\mathsf{LP},11} &= \frac{\sigma^2(\mathsf{x})}{f(\mathsf{x})} \int_{\mathcal{X}} \mathsf{w} (u)^2du, \\
V_{\mathsf{LP},12} &= \frac{\sigma^2(\mathsf{x})}{f(\mathsf{x})}\int_{\mathcal{X}} \mathsf{w}(u) (\mathsf{w}_{\mathsf{conv}}(u) -\mathsf{w} (u) )du, \\    
V_{\mathsf{LP},22} &= \frac{\sigma^2(\mathsf{x})}{f(\mathsf{x})}  \int_{\mathcal{X}} (\mathsf{w} (u)- \mathsf{w}_{\mathsf{conv}} (u))^2du,
\end{align*}
which implies that $v_{\mathsf{PLP}}^2 := V_{\mathsf{LP},11} + V_{\mathsf{LP},22} - 2V_{\mathsf{LP},12} = (\sigma^2(\mathsf{x})/f(\mathsf{x}))\mathcal{K}_{\mathsf{PLP}}$, where 
\begin{equation*}
\mathcal{K}_{\mathsf{PLP}} := \int_{\mathcal{X}} (2\mathsf{w} (u)- \mathsf{w}_{\mathsf{conv}} (u))^2 du = \int_{\mathcal{X}} \mathsf{w}_{\mathsf{PLP}} (u)^2 du .
\end{equation*}
Similarly, by the proof of Theorem \ref{Th RBC PGP equivalence}, we have that $v_{\mathsf{RBC}}^2 = (\sigma^2(\mathsf{x})/f(\mathsf{x}))\mathcal{K}_{\mathsf{RBC}}$, where 
\begin{equation*}
\mathcal{K}_{\mathsf{RBC}} := \int_{\mathcal{X}} (\mathsf{w} (u) - \mathsf{w}_{\mathsf{GP}\text{-}\mathsf{bc}} (u))^2 du = \int_{\mathcal{X}} \mathsf{w}_{\mathsf{RBC}} (u)^2 du ,
\end{equation*}
which concludes the proof. $\hfill \square$

\subsection{proofs from section~\ref{Section boundary}}

\subsubsection*{proof of lemma~\ref{Lemma v2mplp}}

The proof is nearly identical to that of Lemma~\ref{Lemma v2plp}, except for two differences. First, we apply Lemma~\ref{new supp lemma 4} instead of Lemma~\ref{new supp lemma 3} to show that $\hat{v}_{\mathsf{mPLP}}^2 = \hat{v}_{\mathsf{mPLP},1n}^2 + \op (1)$. Second, to define the asymptotic variance we need to distinguish between interior and boundary points. Specifically, in the interior case we have $\hat{v}_{\mathsf{mPLP},1n}^2 := \hat{v}_{\mathsf{PLP},1n}^2 \plowto v^2_{\mathsf{PLP}} =: v^2_{\mathsf{mPLP}}$, and in the boundary case we have
\begin{equation*}
\hspace{1.45cm} \hat{v}_{\mathsf{mPLP},1n}^2 := \frac{1}{nhf(\mathsf{x})^2} \sum_{i=1}^n \mathsf{w}_{\mathsf{mPLP}} \left(\frac{x_i-\mathsf{x}}{h}\right)^2 \varepsilon_i^2 \plowto \frac{\sigma^2 (\mathsf{x})}{f(\mathsf{x})} \int_{\mathcal{X}} \mathsf{w}_{\mathsf{mPLP}}(u)^2du =: v^2_{\mathsf{mPLP}}. \qquad\quad  \square
\end{equation*}

\subsubsection*{proof of theorem~\ref{Th RBC mPLP equivalence}}

By Theorem~\ref{th 1}, the proof follows by showing that Assumption~\ref{assn 1} holds with $(T_n^{\ast},\hat B_n, \hat{v}_n,\hat{v}_{\mathsf{P},n})$ replaced by $(T_{\mathsf{mLP},n}^{\ast},\hat B_{\mathsf{mLP},n}, \hat{v}_{\mathsf{mLP},n},\hat{v}_{\mathsf{mPLP},n})$. 

If $\mathsf{x}$ is an interior point, $Q_n = 1 + \op(1)$ by Lemma~\ref{lemmaKernelConstants} so that $(T_{\mathsf{mLP},n}^{\ast},\hat B_{\mathsf{mLP},n}, \hat{v}_{\mathsf{mLP},n},\hat{v}_{\mathsf{mPLP},n}) = (T_{\mathsf{LP},n}^{\ast},\hat B_{\mathsf{LP},n}, \hat{v}_{\mathsf{LP},n},\hat{v}_{\mathsf{PLP},n}) + \opstar (1)$. Thus, Assumption~\ref{assn 1} holds by Theorem~\ref{Th RBC PLP equivalence}.

If $\mathsf{x}$ is a boundary point, $Q_n = C /C_{\mathsf{LP,bnd}} + \op(1)$ by Lemma~\ref{lemmaKernelConstants}. In this case, Assumption~\ref{assn 1}(i) is verified by noting that $T_{\mathsf{mLP},n}^{\ast} - \hat B_{\mathsf{mLP},n} = Q_n \xi_{1n}^* \dstarto N(0,v^2_{\mathsf{mLP}})$ with $v^2_{\mathsf{mLP}}:= Q^2 v_1^2$, which follows directly from the proof of Theorem~\ref{Th RBC PLP equivalence}. To show that Assumption~\ref{assn 1}(ii) is satisfied, we first apply \eqref{eq:BmLP-B} and Lemma~\ref{lemmaequivker2}(ii) to obtain
\begin{align*}
    \left(\begin{matrix}
        T_n - B_n\\
        \hat{B}_{\mathsf{mLP},n} - B_n
    \end{matrix}\right) = \frac{1}{f(\mathsf{x})\sqrt{nh}} \sum_{i=1}^n     \left(\begin{matrix}
        \mathsf{w}  \left( \frac{x_i - \mathsf{x}}{h}\right) \\
        Q\mathsf{w}_{\mathsf{LP}\text{-}\mathsf{bc},\mathsf{bnd}} \left( \frac{x_i - \mathsf{x}}{h} \right)
    \end{matrix}\right) \varepsilon_i + \op(1) 
     = \frac{1}{f(\mathsf{x})} \left[\begin{matrix}
        1 & 0 \\
        -Q & Q
    \end{matrix}\right]    \mathsf{s}_{\mathsf{LP},\mathsf{bnd},n}+ \op(1),
\end{align*}
where $\mathsf{s}_{\mathsf{LP},\mathsf{bnd},n}  := (nh)^{-1/2} \sum_{i=1}^n (
        \mathsf{w}  ( ({x_i - \mathsf{x}})/{h} ), 
        \mathsf{w}_{\mathsf{conv,bnd}}( ({x_i - \mathsf{x}})/{h}))^{\prime}  \varepsilon_i$. Asymptotic normality of $\mathsf{s}_{\mathsf{LP},\mathsf{bnd},n}$ follows from the Cram\'{e}r-Wold device and Lyapunov's condition, similarly to $\mathsf{s}_{\mathsf{GP},n}$ in the proof of Theorem~\ref{Th RBC PGP equivalence} and $\mathsf{s}_{\mathsf{LP},n}$ in the proof of Theorem~\ref{Th RBC PLP equivalence} by noting that, for any $(\zeta_1,\zeta_2)$ and $\delta \in (1,2]$, 
$\E [ | (\zeta_1 \mathsf{w}  ( ({x_i - \mathsf{x}})/{h} )+ \zeta_2 \mathsf{w}_{\mathsf{conv},\mathsf{bnd}} ( ({x_i - \mathsf{x}})/{h} )\varepsilon_i|^{2\delta}]=O(h)$.
The asymptotic variance of $\mathsf{s}_{\mathsf{LP},\mathsf{bnd},n}$, denoted by~$\Omega_{\mathsf{LP},\mathsf{bnd}}$, can therefore be derived in the same way as in the proof of Theorem~\ref{Th RBC PLP equivalence}, noting that integrals are taken over $[0,\infty)$ because $\mathsf{x}$ is a (left) boundary point, and has elements $\omega_{\mathsf{LP},\mathsf{bnd},11} =\sigma^2 (\mathsf{x}) f(\mathsf{x})\int_{\mathcal{X}} \mathsf{w}(u)^2 du$, $\omega_{\mathsf{LP},\mathsf{bnd},12} =\sigma^2 (\mathsf{x}) f(\mathsf{x})\int_{\mathcal{X}} \mathsf{w} (u) \mathsf{w}_{\mathsf{conv},\mathsf{bnd}} (u)   du$, and $
        \omega_{\mathsf{LP},\mathsf{bnd},22} =\sigma^2 (\mathsf{x}) f(\mathsf{x})\int_{\mathcal{X}}  \mathsf{w}_{\mathsf{conv},\mathsf{bnd}} (u)^2 du$. Finally, the convergence of $\mathsf{s}_{\mathsf{LP},\mathsf{bnd},n}$ implies that $(T_n - B_n, \hat{B}_{\mathsf{mLP},n} - B_n)^{\prime} \dlowto N(0,V_{\mathsf{mLP}})$, where 
\begin{equation*}
        V_{\mathsf{mLP}}:= \frac{1}{f^2 (\mathsf{x})} \left[\begin{matrix}
        1 & 0 \\
        -Q & Q
    \end{matrix}\right] \Omega_{\mathsf{LP},\mathsf{bnd}}\left[\begin{matrix}
        1 & -Q \\
        0 & Q
    \end{matrix}\right] ,
\end{equation*}
which concludes the proof of Theorem~\ref{Th RBC mPLP equivalence}. $\hfill \square$

\subsubsection*{proof of corollary \ref{corollary v^2_mP comparisons}}

If $\mathsf{x}$ is an interior point, the proof is identical to the proof of Corollary~\ref{corollary v^2_P comparisons} because $Q=1$ implies that the asymptotic variances of $(T_n - B_n, \hat{B}_{\mathsf{LP},n} - B_n)$ and $(T_n - B_n, \hat{B}_{\mathsf{mLP},n} - B_n)$ coincide.

If $\mathsf{x}$ is on the boundary, then $Q \neq 1$ and analogous results as in the proof of Corollary~\ref{corollary v^2_P comparisons} yields
\begin{align*}
    &V_{\mathsf{mLP},1}^2 = \frac{\sigma^2(\mathsf{x})}{f(\mathsf{x})} \int_{\mathcal{X}} \mathsf{w}  (u)^2 du, \\
    &V_{\mathsf{mLP},12} = \frac{\sigma^2(\mathsf{x})}{f(\mathsf{x})} Q  \int_{\mathcal{X}} \mathsf{w} (u) (\mathsf{w}_{\mathsf{conv},\mathsf{bnd}} (u)  -  \mathsf{w}  (u)) du, \\    
    &V_{\mathsf{mLP},2}^2 = \frac{\sigma^2(\mathsf{x})}{f(\mathsf{x})} Q^2  \int_{\mathcal{X}} (\mathsf{w}  (u) - \mathsf{w}_{\mathsf{conv},\mathsf{bnd}} (u) )^2 du,
\end{align*}
which implies that $v_{\mathsf{mPLP}}^2=(\sigma^2(\mathsf{x})/f(\mathsf{x}))\mathcal{K}_{\mathsf{mPLP}}$ where 
\begin{equation*}
\mathcal{K}_{\mathsf{mPLP}} :=  \int_{\mathcal{X}} ((Q+1)\mathsf{w}(u) - Q\mathsf{w}_{\mathsf{conv},\mathsf{bnd}} (u))^2 du = \int_{\mathcal{X}} \mathsf{w}_{\mathsf{mPLP}} (u)^2 du.
\end{equation*}
Similarly, by the proof of Theorem \ref{Th RBC PGP equivalence}, we have that $v_{\mathsf{RBC}}^2 = (\sigma^2(\mathsf{x})/f(\mathsf{x}))\mathcal{K}_{\mathsf{RBC}}$, where $\mathcal{K}_{\mathsf{RBC}}$ is defined in the proof of Corollary \ref{corollary v^2_P comparisons}, thus concluding the proof. $\hfill \square$

\subsubsection*{proof of theorem \ref{Th RBC mPLP equivalence rdd}}

\noindent \textsc{part}~(i). Let $\hat{v}_{+,\mathsf{mPLP},n}^2 := (nh)^{-1} \sum_{i=1}^n w_{+,\mathsf{mPLP},i} (\mathsf{x})^2 \tilde \varepsilon_i^2 $, $\hat{v}_{-,\mathsf{mPLP},n}^2 := (nh)^{-1} \sum_{i=1}^n w_{-,\mathsf{mPLP},i} (\mathsf{x})^2 \tilde \varepsilon_i^2 $, where $w_{+,\mathsf{mPLP},i} (\mathsf{x})$ and $w_{-,\mathsf{mPLP},i} (\mathsf{x})$ are defined like $w_{\mathsf{mPLP},i} (\mathsf{x})$ replacing the kernel function with $K_{+}$ and $K_{-}$, respectively. Then, $\hat{v}_{\mathsf{rd},n}^2 := \hat{v}_{+,\mathsf{mPLP},n}^2 +\hat{v}_{-,\mathsf{mPLP},n}^2$. Because the cutoff truncation induced by $K_{+}$ and $K_{-}$ implies that $\mathsf{x}$ is a (strict) boundary point for both $\hat{v}_{+,\mathsf{mPLP},n}^2$ and $\hat{v}_{-,\mathsf{mPLP},n}^2$, and because observations on different sides of the cutoff are independent, we can apply Lemma~\ref{Lemma v2mplp} to each term in~$\hat{v}_{\mathsf{rd},n}^2$. Specifically, by Lemma~\ref{Lemma v2mplp}, $\hat{v}_{\mathsf{rd},n}^2 \plowto v_{\mathsf{rd}}^2 := f(\mathsf{x})^{-1} (\sigma_{+}^2(\mathsf{x}) + \sigma_{-}^2(\mathsf{x})) \mathcal{K}_{\mathsf{mPLP}}$.

\medskip \noindent \textsc{parts}~(ii)--(iv). In view of the decompositions $T_n=T_{+,n}-T_{-,n}$ and $T_n^{\ast}=T_{+,n}^{\ast}-T_{-,n}^{\ast}$, and the independence of observations on different sides of the cutoff, the results follow by application of Theorem~\ref{Th RBC mPLP equivalence} on both sides of the cutoff. $\hfill \square$

\subsubsection*{proof of corollary \ref{cor RDD}}

Suppose a different kernel is applied to the left and to the right of the cutoff, and redefine $K_{+}$, $K_{-}$, and subsequent quantities accordingly. In this case, by slight modification of the proof of Theorem~\ref{Th RBC mPLP equivalence rdd}(i), we have that $v_{\mathsf{rd}}^2= f(\mathsf{x})^{-1} (\sigma_{+}^2(\mathsf{x}) \mathcal{K}_{+,\mathsf{rd}} + \sigma_{-}^2(\mathsf{x}) \mathcal{K}_{-,\mathsf{rd}})$.

For the \textsf{RBC} approach, note that validity of the method for the RDD extends in the same manner as for the \textsf{mPLP} bootstrap, with the asymptotic properties of the test statistic being derived from the boundary point case; see also Calonico et al.~(2014). Hence, from the proof of Corollary~\ref{corollary v^2_mP comparisons}, we have that $v_{\mathsf{RBC}\text{-}\mathsf{\mathsf{rd}}}^2 = f(\mathsf{x})^{-1}(\sigma_{+}^2(\mathsf{x})\mathcal{K}_{+,\mathsf{RBC}} + \sigma_{-}^2(\mathsf{x})\mathcal{K}_{-,\mathsf{RBC}})$. $\hfill \square$

\phantomsection
\section*{References}
\addcontentsline{toc}{section}{References}

\begin{description}

\item \textsc{Andrews, D.W.K.}\ (1991): Asymptotic normality of series estimators for nonparametric and semiparametric regression models. \emph{Econometrica} 59, 307--345.

\item \textsc{Andrews, D.W.K.}\ (1994): Asymptotics for semiparametric econometric models via stochastic equicontinuity. \emph{Econometrica} 62, 43--72.

\item \textsc{Bartalotti, O., Calhoun, G., and He, Y.}\ (2017): Bootstrap confidence intervals for sharp regression discontinuity designs. In \emph{Regression Discontinuity Designs: Theory and Applications}, 421--453. Emerald Publishing Limited.

\item \textsc{Belloni, A., Chernozhukov, V., Chetverikov, D., and Kato, K.}\ (2015): Some new asymptotic theory for least squares series: Pointwise and uniform results. \emph{Journal of Econometrics} 186, 345--366.

\item \textsc{Beran, R.}\ (1987): Prepivoting to reduce level error of confidence sets. \emph{Biometrika} 74, 457--468.

\item \textsc{Beran, R.}\ (1988): Prepivoting test statistics: A bootstrap view of asymptotic refinements. \emph{Journal of the American Statistical Association} 83, 687--697.

\item \textsc{Berry, S.M., Carroll, R.J., and Ruppert, D.}\ (2002): Bayesian smoothing and regression splines for measurement error problems. \emph{Journal of the American Statistical Association} 97, 160--169.

\item \textsc{Calonico, S., Cattaneo, M.D., and Farrell, M.H.}\ (2018): On the effect of bias estimation on coverage accuracy in nonparametric inference, \emph{Journal of the American Statistical Association} 113, 767--779.

\item \textsc{Calonico, S., Cattaneo, M.D., and Farrell, M.H.}\ (2020): Optimal bandwidth choice for robust bias-corrected inference in regression discontinuity designs. \emph{Econometrics Journal} 23, 192--210.

\item \textsc{Calonico, S., Cattaneo, M.D., and Farrell, M.H.}\ (2022): Coverage error optimal confidence intervals for local polynomial regression. \emph{Bernoulli} 28, 2998--3022.

\item \textsc{Calonico, S., Cattaneo, M.D., and Titiunik, R.}\ (2014): Robust nonparametric confidence intervals for regression-discontinuity designs, \emph{Econometrica} 82, 2295--2326.

\item \textsc{Cattaneo, M.D., Farrell, M.H., and Feng, Y.}\ (2020): Large sample properties of partitioning-based series estimators. \emph{Annals of Statistics} 48, 1718--1741,

\item \textsc{Cattaneo, M.D., and Jansson, M.}\ (2018): Kernel-based semiparametric estimators: Small bandwidth asymptotics and bootstrap consistency. \emph{Econometrica} 86, 955--995.

\item \textsc{Cavaliere, G., and Georgiev, I.}\ (2020): Inference under random limit bootstrap measures. \emph{Econometrica} 88, 2547--2574.

\item \textsc{Cavaliere, G., Gon\c{c}alves, S., Nielsen, M.\O ., and Zanelli, E.}\ (2024): Bootstrap inference in the presence of bias. \emph{Journal of the American Statistical Association} 119, 2908--2918.

\item \textsc{Chen, X.}\ (2007): Large sample sieve estimation of semi-nonparametric models. In: Heckman, J.J., Leamer, E.E.\ (Eds.), Handbook of Econometrics, vol.\ 6B, 5549-–5632.

\item \textsc{Chen, X., and Christensen, T.M.}\ (2015): Optimal uniform convergence rates and asymptotic normality for series estimators under weak dependence and weak conditions. \emph{Journal of Econometrics} 188, 447--465.

\item \textsc{Chen, X., Linton, O. and Van Keilegom, K.}\ (2003): Estimation of semiparametric models when the criterion function is not smooth. \emph{Econometrica} 71, 1591--1608.

\item \textsc{Fan, J., and Gijbels, I.}\ (1996): Local Polynomial Modelling and Its Applications. \emph{Monographs on Statistics and Applied Probability}, vol.\ 66. Chapman \& Hall.

\item \textsc{Gupta, A., and Seo, M.H.}\ (2023): Robust inference on infinite and growing dimensional time-series regression. \emph{Econometrica} 91, 1333--1361.

\item \textsc{Hahn, J., Todd, P., and van der Klaauw, W.}\ (2001): Identification and estimation of treatment effects with a regression-discontinuity design. \emph{Econometrica} 69, 201--209.

\item \textsc{Hall, P.}\ (1992): Effect of bias estimation on coverage accuracy of bootstrap confidence intervals for a probability density. \emph{Annals of Statistics} 20, 675--694.

\item \textsc{Hall, P., and Horowitz, J.}\ (2013): A simple bootstrap method for constructing nonparametric confidence bands for functions. \emph{Annals of Statistics} 41, 1892--1921.

\item \textsc{Hall, P., and Marron, J.S.}\ (1990): On variance estimation in nonparametric regression. \emph{Biometrika} 77, 415--419.

\item \textsc{Hallin, M., Lu, Z., and Tran, L.T.}\ (2004): Local linear spatial regression. \emph{Annals of Statistics} 32, 2469--2500.

\item \textsc{H\"{a}rdle, W., and Bowman, A.W.}\ (1988): Bootstrapping in nonparametric regression: Local adaptive smoothing and confidence bands. \emph{Journal of the American Statistical Association} 83, 102--110.

\item \textsc{Härdle, W., and Marron, J.S.}\ (1991): Bootstrap simultaneous error bars for nonparametric regression. \emph{Annals of Statistics} 19, 778--796.

\item \textsc{He, Y., and Bartalotti, O.}\ (2020): Wild bootstrap for fuzzy regression discontinuity designs: Obtaining robust bias-corrected confidence intervals. \emph{Econometrics Journal} 23, 211--231.

\item \textsc{Huang, J.Z.}\ (2003): Local asymptotics for polynomial spline regression. \emph{Annals of Statistics} 31, 1600--1635.

\item \textsc{Imbens, G.W., and Kalyanaraman, K.}\ (2012): Optimal bandwidth choice for the regression discontinuity estimator. \emph{Review of Economic Studies} 79, 933--959.

\item \textsc{Imbens, G.W., and Lemieux, T.}\ (2008): Regression discontinuity designs: A guide to practice. \emph{Journal of Econometrics} 142, 615--635.

\item \textsc{Lee, D.S.}\ (2008): Randomized experiments from non-random selection in US House elections. \emph{Journal of Econometrics} 142, 675--697.

\item \textsc{Ludwig, J., and Miller, D.L.}\ (2007): Does Head Start improve children's life changes? Evidence from a regression discontinuity design. \emph{Quarterly Journal of Economics} 122, 159--208.

\item \textsc{Newey, W.K.}\ (1994): The asymptotic variance of semiparametric estimators. \emph{Econometrica} 62, 1349--1382.    

\item\textsc{Ruppert, D., and Wand, M.P.}\ (1994): Multivariate locally weighted least squares regression. \emph{Annals of Statistics} 22, 1346--1370.

\end{description}

\newpage
\newcounter{suppsection}
\renewcommand{\thesuppsection}{S.\arabic{suppsection}}

\newcommand{\suppsection}[1]{%
  \refstepcounter{suppsection}%
  \section*{S.\arabic{suppsection} \quad #1}%
  \addcontentsline{toc}{section}{S.\arabic{suppsection} \quad #1}%
  \label{Supp:Sec-\arabic{suppsection}}%
}

\counterwithout{lemma}{suppsection}
\renewcommand{\thelemma}{S.\arabic{lemma}}
\setcounter{lemma}{0}

\counterwithout{table}{suppsection}
\renewcommand{\thetable}{S.\arabic{table}}
\setcounter{table}{0}

\counterwithout{equation}{suppsection}
\renewcommand{\theequation}{S.\arabic{equation}}
\setcounter{equation}{0}

\phantomsection
\addcontentsline{toc}{section}{Supplementary Material}

\begin{center}
{\LARGE \vspace*{1in}}

{\huge Supplementary Material:}

{\LARGE \vspace*{0.5in}}

{\huge{{Improved Inference for Nonparametric Regression}}}

\bigskip
\bigskip

{\Large by}

\bigskip
\bigskip

{\Large G.\ Cavaliere, S.\ Gon\c{c}alves, M.\O .\ Nielsen, and E.\ Zanelli}

{\Large \bigskip}

{\Large \bigskip}

{\Large
\date{November 25, 2025}
}
\end{center}

\bigskip

{\LARGE \vspace*{1in}}

This supplementary material for Cavaliere, Gon\c{c}alves, Nielsen, and Zanelli (2025) contains two sections. Section~\ref{Supp:Sec-new-additional-lemmas} collects auxiliary approximation lemmas that are used throughout. Section~\ref{Supp:Sec-extra} contains the proofs of the bias expansion lemmas in Appendix~\ref{app:lemmas}. References are included at the end of the supplement.

\setcounter{page}{1}%
\newpage

\suppsection{approximation lemmas} 
\label{Supp:Sec-new-additional-lemmas}

\begin{lemma}
\label{auxlemma1}
Let Assumptions~\ref{Ass_g} and~\ref{Ass_K,h} be satisfied. For $j=0,1,2,\ldots$:\\
(i)~$\sup_{x \in \mathbb{S}_x} |\gamma_{j,n}(x) - \gamma_{\mathsf{E},j,n} (x)| = \op (1)$ and (ii)~$\gamma_{\mathsf{E},j,n} (\mathsf{x}) \to \dot\gamma_j$.
\end{lemma}

\noindent\textsc{proof of lemma \ref{auxlemma1}}. Part~(i). Suppose first that $K$ is globally Lipschitz continuous. Then $K(u) u^j$ is globally Lipschitz continuous for all $j=0,1,\ldots$, the conditions in Masry (1996, Theorem~2) apply; see also Hansen (2008). Therefore, 
\begin{equation*}
\sup_{x\in\mathbb{S}_x} | \gamma_{j,n}(x) - \E [\gamma_{j,n}(x)] | = \Op\left( \sqrt{\frac{\log n}{nh}} \right) =\op(1).
\end{equation*}

Next, suppose $K$ is not globally Lipschitz continuous such that Theorem~2 of Masry (1996) does not apply. In this case, $K(u)=0$ for $|u| \geq 1$ and it is Lipschitz continuous on $(-1,1)$.

Since $\mathbb{S}_x$ is compact, it can be divided into a finite number $L(n)$ of intervals $I_{k,n}$ with centers $x_{k,n}$ and length $\ell_n = c_0 /L(n)$ for some $c_0 > 0$ and $k =1, \ldots, L(n)$. Therefore,
\begin{align*}
\sup_{x\in\mathbb{S}_x} | \gamma_{j,n}(x) - \E [\gamma_{j,n}(x)] |  &= \max_{1\leq k \leq L(n)} \sup_{x \in \mathbb{S}_x \cap I_{k,n}} | \gamma_{j,n}(x) - \E [\gamma_{j,n}(x)] | \\
&\leq \max_{1\leq k \leq L(n)} \sup_{x \in \mathbb{S}_x \cap I_{k,n}} | \gamma_{j,n}(x) - \gamma_{j,n}(x_{k,n}) |\\
& \quad + \max_{1\leq k \leq L(n)} | \gamma_{j,n}(x_{k,n}) - \E [\gamma_{j,n}(x_{k,n})] | \\
& \quad +  \max_{1\leq k \leq L(n)} \sup_{x \in \mathbb{S}_x \cap I_{k,n}} | \E [\gamma_{j,n}(x_{k,n})] - \E [\gamma_{j,n}(x)] |\\
&=: \mathsf{Q}_{1n} + \mathsf{Q}_{2n} + \mathsf{Q}_{3n}.
\end{align*}
Let $\ell_n = ((\log n)h^3/n)^{1/2}$. Note that the Lipschitz condition is not needed for $\mathsf{Q}_{2n}$ because the function is only evaluated at $x_{k,n}$, so we can conclude that $\mathsf{Q}_{2n} = \Op(\ell_n h^{-2}) =\Op((\log n)^{1/2}/(nh)^{1/2})$ by the proof of Theorem~2 of Masry (1996).

We now prove that $\mathsf{Q}_{1n}=\Op((\log n)^{1/2}/(nh)^{1/2})$. For $j=0,1,\ldots$, let $\mathsf{k}_j(u) := u^j \mathbb{I}_{\{ |u| < 1 \}}$. Similarly to Cattaneo, Jansson and Ma (2024), we will use the fact that
\begin{align}
\label{k property}
|\mathsf{k}_j(u)  - \mathsf{k}_j(v) | &\leq c_1 |u-v| + c_2 \mathbb{I}_{ \{ ||u|-1| \leq |u-v| \} }, \\
\label{K property}
|K(u)  - K(v) | &\leq c_1 |u-v| + c_2 \mathbb{I}_{ \{ ||u|-1| \leq |u-v| \} },
\end{align}
where $c_1$ and $c_2$ are bounded positive constants. For all $j=0,1,\ldots$, 
\begin{align*}
\mathsf{Q}_{1n} &\leq \max_{1\leq k \leq L(n)} \sup_{x \in \mathbb{S}_x \cap I_{k,n}} \frac{1}{nh} \sum_{i=1}^n  \left| K\left( \frac{x_i - x}{h} \right) \left( \frac{x_i - x}{h} \right)^j - K\left( \frac{x_i - x_{k,n}}{h} \right) \left( \frac{x_i - x_{k,n}}{h} \right)^j \right| \\
& \leq \max_{1\leq k \leq L(n)} \sup_{x \in \mathbb{S}_x \cap I_{k,n}}\frac{1}{nh} \sum_{i=1}^n   K\left( \frac{x_i - x}{h} \right) \left| \mathsf{k}_j \left( \frac{x_i - x}{h} \right) - \mathsf{k}_j\left( \frac{x_i - x_{k,n}}{h} \right)   \right| \\
& \quad + \max_{1\leq k \leq L(n)} \sup_{x \in \mathbb{S}_x \cap I_{k,n}}\frac{1}{nh} \sum_{i=1}^n   \left| \mathsf{k}_j \left( \frac{x_i - x_{k,n}}{h} \right) \right|  \left|K\left( \frac{x_i - x}{h} \right) - K\left( \frac{x_i - x_{k,n}}{h} \right)  \right| \\
&=: \mathsf{Q}_{11n} + \mathsf{Q}_{12n}.
\end{align*}
By \eqref{k property} and \eqref{K property} we have 
\begin{align*}
\mathsf{Q}_{11n} &\leq \Op(\ell_n h^{-2}) + c_2 \max_{1\leq k \leq L(n)} \sup_{x \in \mathbb{S}_x \cap I_{k,n}} \frac{1}{nh} \sum_{i=1}^n K \left(\frac{x_i - x}{h}\right) \mathbb{I}_{ \{ ||x_i - x_{k,n}|-h| \leq |x - x_{k,n}| \} } ,\\
\mathsf{Q}_{12n} &\leq \Op(\ell_n h^{-2}) + c_2 \max_{1\leq k \leq L(n)} \sup_{x \in \mathbb{S}_x \cap I_{k,n}} \frac{1}{nh} \sum_{i=1}^n   \left| \mathsf{k}_j \left( \frac{x_i - x_{k,n}}{h} \right) \right|  \mathbb{I}_{ \{ ||x_i - x_{k,n}|-h| \leq |x - x_{k,n}| \} } .
\end{align*}
By uniform boundedness of $\mathsf{k}$ and $K$ on~$\mathbb{R}$, and since $\mathbb{I}_{ \{ ||x_i - x_{k,n}|-h| \leq |x - x_{k,n}| \} } \leq \mathbb{I}_{ \{ ||x_i - x_{k,n}|-h| \leq \ell_n \} }$, to show that $\mathsf{Q}_{1n}=\Op((\log n)^{1/2}/(nh)^{1/2})$, it suffices to show that $\max_{1\leq k \leq L(n)} (nh)^{-1}\mathsf{S}_{k,n} = \Op(\ell_n h^{-2})$, where $\mathsf{S}_{k,n} := \sum_{i=1}^n  \mathbb{I}_{ \{ ||x_i - x_{k,n}|-h| \leq \ell_n \} } \sim \text{Bin} (n,\mathsf{p}(x_{k,n}))$ with $\mathsf{p}(x) := \mathbb{P} (||x_1 - x|-h| \leq \ell_n)$. By the union bound, $\mathbb{P} ( \max_{1\leq k \leq L(n)} h(\ell_n n)^{-1}\mathsf{S}_{k,n} > \epsilon ) \leq L(n) \max_{1\leq k \leq L(n)} \mathbb{P} ( h(\ell_n n)^{-1} \mathsf{S}_{k,n} > \epsilon )$. Note that $\mathbb{E}[\mathsf{S}_{k,n}] =n\mathsf{p}(x_{k,n})$, where $\sup_{x\in\mathbb{S}_x} \mathsf{p}(x) \leq c_3 \ell_n = o(h)$ for some $c_3>0$. For any $\epsilon >0$ we then have, for $\delta \in (0, \epsilon )$ and $n$ sufficiently large, that $\epsilon > \delta +  \mathsf{p}(x_{k,n}) h / \ell_n $ such that 
\begin{align*}
\mathbb{P} ( h(\ell_n n)^{-1}\mathsf{S}_{k,n} >\epsilon) &\leq \mathbb{P} ( \mathsf{S}_{k,n} - n\mathsf{p}(x_{k,n}) > \delta \ell_n n h^{-1} ) \\
&\leq \exp \left( - \frac{\frac{1}{2} (\delta \ell_n n h^{-1})^2}{n \mathsf{p}(x_{k,n})(1-\mathsf{p}(x_{k,n}))  + \frac{1}{3} \delta \ell_n n h^{-1}} \right) 
= O \left( \exp\left(- \frac{3}{2} \delta \ell_n n h^{-1} \right) \right) ,
\end{align*}
again for $n$ sufficiently large, where the second inequality is Bernstein's and the final equality is because $\sup_{x\in\mathbb{S}_x} \mathsf{p}(x) \leq c_3 \ell_n$. Since $L_n = c_0 / \ell_n$, it then follows that
\begin{equation*}
\mathbb{P} \left( \max_{1\leq k \leq L(n)} \frac{h}{\ell_n n}\mathsf{S}_{k,n} > \epsilon\right) \leq L(n) \max_{1\leq k \leq L(n)} \mathbb{P} \left( \frac{h}{\ell_n n} \mathsf{S}_{k,n} > \epsilon \right) 
\leq \frac{c_0}{\ell_n} O \left( \exp\left(- \frac{3}{2} \delta \ell_n n h^{-1} \right) \right) ,
\end{equation*} 
which is $o(1)$ because
\begin{align*}
 - \log \ell_n - \frac{3}{2} \delta \ell_n n h^{-1} 
&= \sqrt{(\log n) nh } \left(  -\frac{1}{2} \frac{\log\log n}{\sqrt{(\log n) nh}} - \frac{3}{2} \frac{\log h}{\sqrt{(\log n) nh}}+ \frac{1}{2} \Big( \frac{\log n }{nh}\Big)^{1/2} - \frac{3}{2} \delta\right)
\to - \infty
\end{align*}
by Assumption~\ref{Ass_K,h}(ii), noting that the term in the large parenthesis converges to~$-3\delta /2$. This shows the result for~$\mathsf{Q}_{1n}$. 

The result for $\mathsf{Q}_{3n}$ follows analogously, since, by Jensen's inequality, for all $j=0,1,\ldots$, 
\begin{align*}
\mathsf{Q}_{3n} & \leq \max_{1\leq k \leq L(n)} \sup_{x \in \mathbb{S}_x \cap I_{k,n}} \mathbb{E} \left[ \frac{1}{nh} \sum_{i=1}^n   K\left( \frac{x_i - x}{h} \right) \left| \mathsf{k}_j \left( \frac{x_i - x}{h} \right) - \mathsf{k}_j\left( \frac{x_i - x_{k,n}}{h} \right)   \right| \right] \\
& \quad + \max_{1\leq k \leq L(n)} \sup_{x \in \mathbb{S}_x \cap I_{k,n}} \mathbb{E} \left[ \frac{1}{nh} \sum_{i=1}^n \left| \mathsf{k}_j \left( \frac{x_i - x_{k,n}}{h} \right) \right| \left| K \left( \frac{x_i - x}{h} \right) - K\left( \frac{x_i - x_{k,n}}{h} \right) \right| \right] \\
&=: \mathsf{Q}_{31n} + \mathsf{Q}_{32n} .
\end{align*}
By \eqref{k property} and \eqref{K property}, we have that $\mathsf{Q}_{31n} \leq c_4 \max_{1\leq k \leq L(n)} \sup_{x \in \mathbb{S}_x \cap I_{k,n}} h^{-1}  \mathsf{p}(x_{k,n}) +O( \ell_n h^{-2})$ for some $c_4 >0$ and $\mathsf{Q}_{32n} \leq  c_5 \max_{1\leq k \leq L(n)} \sup_{x \in \mathbb{S}_x \cap I_{k,n}} h^{-1} \mathsf{p}(x_{k,n})+O( \ell_n h^{-2})$ for some $c_5>0$. Because $\mathsf{p}(x_{k,n}) = O(\ell_n)$ it follows that $\mathsf{Q}_{3n} = O( \ell_n h^{-2}) = O((\log n)^{1/2}/(nh)^{1/2})$.

Finally we derive the mean,
\begin{align*}
\E [\gamma_{j,n}(x)] 
&=\frac{1}{nh} \sum_{i=1}^n \E \left[ K\left(\frac{x_i-x}{h}\right) \left(\frac{x_i-x}{h}\right)^j \right]
= \frac{1}{h} \E \left[ K\left(\frac{x_1-x}{h}\right) \left(\frac{x_1-x}{h}\right)^j \right] \\
&=  \frac{1}{h} \int_{x-h}^{x+h} K\left(\frac{x_1-x}{h}\right) \left(\frac{x_1-x}{h}\right)^j f(x_1) dx_1 
=  \int_{-1}^1 K (u ) u^j f(x+uh) du =: \gamma_{\mathsf{E},j,n}(x) ,
\end{align*}
where the second equality is by the i.i.d.\ assumption and the third uses the bounded support of~$K$.

\medskip \noindent \textsc{part}~(ii). If $\mathsf{x}$ is an interior point, then 
$\int_{-1}^1 K (u ) u^j f(\mathsf{x}+uh) du \to f(\mathsf{x})\int_{-1}^1 K (u) u^j du$. If $\mathsf{x}$ is a (left) boundary point, then
$\int_{-1}^1 K(u) u^j f(\mathsf{x}+uh) du 
= \int_0^1 K(u) u^j f(\mathsf{x}+uh) du
\to f(\mathsf{x})\int_0^1 K(u) u^j du 
$. Therefore, for every $\mathsf{x}$, we have that $\E [ \gamma_{j,n} ] = \dot\gamma_{j}+o(1)$. $\hfill \square$

\begin{lemma}
\label{auxlemma1Gamma}
Let Assumptions~\ref{Ass_g} and~\ref{Ass_K,h} be satisfied. Then:\\
(i)~$\sup_{x \in \mathbb{S}_x} |\Gamma_p(x) - \Gamma_{\mathsf{E},p} (x)]| = \op (1)$ and (ii)~$\Gamma_{\mathsf{E},p} (\mathsf{x}) \to \dot\Gamma_p$.
\end{lemma}

\noindent\textsc{proof of lemma \ref{auxlemma1Gamma}}. These results follow by noting that the $(i,j)$-th elements of $\Gamma_p (x)$, $\Gamma_{\mathsf{E},p} (x)$, $\dot\Gamma_p$ are $\gamma_{i+j-2,n}(x)$, $\gamma_{\mathsf{E},i+j-2,n}(x)$, $\dot\gamma_{i+j-2,n}$, respectively, and applying Lemma~\ref{auxlemma1}. $\hfill \square$

\begin{lemma} 
\label{auxlemma2}
Let Assumptions~\ref{Ass_g} and~\ref{Ass_K,h} be satisfied. For $j=0,1,2,\ldots$, $\tilde\psi_{j,n} = \dot\psi_j + \op (1)$.
\end{lemma}

\noindent\textsc{proof of lemma \ref{auxlemma2}}. We first prove that $\psi_{j,n} = \dot\psi_j + \op (1)$ with $\psi_{j,n} := (nh)^{-1}\sum_{i=1}^n K ((x_i-\mathsf{x})/h)^2 ((x_i-\mathsf{x})/h)^j \sigma^2 (x_i)$.

If $\mathsf{x}$ if an interior point, note that 
\begin{align*}
\E [ \psi_{j,n} ] 
&= \frac{1}{h}  \E \left[ K^2 \left( \frac{x_1-\mathsf{x}}{h}\right)\left( \frac{x_1-\mathsf{x}}{h}\right)^j \sigma^2 (x_1) \right]  \\
&=  \frac{1}{h} \int_{\mathsf{x}-h}^{\mathsf{x}+h} K^2 \left( \frac{x_1-\mathsf{x}}{h}\right)\left( \frac{x_1-\mathsf{x}}{h}\right)^j f(x_1) \sigma^2 (x_1) dx_1  \\
&= \int_{-1}^1 K^2 ( u )u^j f(\mathsf{x}+uh) \sigma^2 (\mathsf{x}+uh) du  
\to \sigma^2 (\mathsf{x})f(\mathsf{x})\int_{-1}^1 K^2 ( u ) u^j  du,
\end{align*}
where the first equality is by the i.i.d.\ assumption and the second uses bounded support of~$K$. 
Similarly, if $\mathsf{x}$ is a boundary point, $\int_{-1}^1 K^2 (u) u^j f(\mathsf{x}+uh) \sigma^2 (\mathsf{x}+uh) du \xrightarrow{}\sigma^2 (\mathsf{x}) f(\mathsf{x}) \int_0^1 K^2 (u) u^j du $. Therefore, for every $\mathsf{x}$, we have that $\E [ \psi_{j,n} ] = \dot\psi_{j}+o(1)$. Moreover,
\begin{align*}
\E [ \psi_{j,n}^2 ] &= \E \left[\frac{1}{nh}  \sum_{i=1}^n K^2 \left( \frac{x_i-\mathsf{x}}{h}\right)\left( \frac{x_i-\mathsf{x}}{h}\right)^j \sigma^2 (x_i) \right]^2\\
&= \frac{1}{(nh)^2}  \sum_{i=1}^n \E \left[ K^4 \left( \frac{x_i-\mathsf{x}}{h}\right)\left( \frac{x_i-\mathsf{x}}{h}\right)^{2j} \sigma^4(x_i) \right]+ \\
&\quad + \frac{1}{(nh)^2}  \sum_{i\not=i'}  \E \left[ K^2 \left( \frac{x_i-\mathsf{x}}{h}\right)\left( \frac{x_i-\mathsf{x}}{h}\right)^{j} \sigma^2 (x_i) \right] \E \left[ K^2 \left( \frac{x_{i'}-\mathsf{x}}{h}\right)\left( \frac{x_{i'}-\mathsf{x}}{h}\right)^{j} \sigma^2 (x_{i'}) \right]   \\
&= \Op \left( \frac{1}{nh}\right) + \dot\psi_j^2 ,
\end{align*}
which shows $L_2$-convergence.  

We now show that $\tilde\psi_{j,n} = \psi_{j,n} +\op (1)$ for $j=0,1,\ldots$. Insert $\tilde\varepsilon_i$ from \eqref{eq resid} and decompose
\begin{align*}
\tilde\psi_{j,n} ={}& \frac{1}{nh} \sum_{i=1}^n K^2\left(\frac{x_i-\mathsf{x}}{h}\right) \left(\frac{x_i-\mathsf{x}}{h}\right)^j \varepsilon_i^2 \\
&+\frac{1}{nh} \sum_{i=1}^n K^2\left(\frac{x_i-\mathsf{x}}{h}\right) \left(\frac{x_i-\mathsf{x}}{h}\right)^j \left( g (x_i) - r_{p+1}^{\prime} \left(\frac{x_i-\mathsf{x}}{h}\right) H_{p+1}^{-1} \hat\beta_{p+1,n}\right)^2 \\
&+ \frac{2}{nh} \sum_{i=1}^n K^2\left(\frac{x_i-\mathsf{x}}{h}\right) \left(\frac{x_i-\mathsf{x}}{h}\right)^j \left( g (x_i) - r_{p+1}^{\prime} \left(\frac{x_i-\mathsf{x}}{h}\right)H_{p+1}^{-1} \hat\beta_{p+1,n} \right) \varepsilon_i \\
&=: \tilde\psi_{j,n,1} + \tilde\psi_{j,n,2} + \tilde\psi_{j,n,3}.
\end{align*}
We show that $\tilde\psi_{j,n,1} = \psi_{j,n} + \op (1)$ and $\tilde\psi_{j,n,2} = \op (1)$. It then follows from the Cauchy-Schwarz inequality that $\tilde\psi_{j,n,3} = \op (1)$ because $\psi_{j,n} = \Op (1)$. 

Consider first~$\tilde\psi_{j,n,2}$. By mean-value argument in \eqref{MVT on g} we can write $\tilde\psi_{j,n,2}$ as
\begin{align*}
&\frac{1}{nh} \sum_{i=1}^n K^2\left(\frac{x_i-\mathsf{x}}{h}\right) \left(\frac{x_i-\mathsf{x}}{h}\right)^j \left( r_{p+1}^{\prime} \left(\frac{x_i-\mathsf{x}}{h}\right)H_{p+1}^{-1} (\beta_{p+1} - \hat\beta_{p+1,n}) \right)^2 \\
& + \frac{1}{nh} \sum_{i=1}^n K^2\left(\frac{x_i-\mathsf{x}}{h}\right) \left(\frac{x_i-\mathsf{x}}{h}\right)^{2p+2+j} h^{2p+2} \left( \frac{g^{(p+1)}(\tilde{x}_i) - g^{(p+1)}(\mathsf{x})}{(p+1)!} \right)^2 \\
& + \frac{2}{nh} \sum_{i=1}^n K^2\left(\frac{x_i-\mathsf{x}}{h}\right) \left(\frac{x_i-\mathsf{x}}{h}\right)^{p+1+j} r_{p+1}^{\prime} \left(\frac{x_i-\mathsf{x}}{h}\right) H_{p+1}^{-1}(\beta_{p+1}- \hat\beta_{p+1,n}) h^{p+1} \left( \frac{g^{(p+1)}(\tilde{x}_i) - g^{(p+1)}(\mathsf{x})}{(p+1)!} \right) \\
&=: \tilde\psi_{j,n,21} + \tilde\psi_{j,n,22} + \tilde\psi_{j,n,23}. 
\end{align*}
By the fact that $|\hat\beta_{s,p+1,n} - \beta_{s,p+1}|= \Op ((nh^{2s-1})^{-1/2})$, where $\hat\beta_{s,p+1,n}$ and $\beta_{s,p+1}$ are the $s$-th elements of $\hat\beta_{p+1,n}$ and $\beta_{p+1}$, respectively, we obtain $|H_{p+1}^{-1}(\hat\beta_{p+1,n} - \beta_{p+1})| = \Op ((nh)^{-1/2}) = \op(1)$. Moreover, noting that $K ((x_i-\mathsf{x})/h) \neq 0$ only when $|\tilde{x}_i-\mathsf{x}| \leq | x_i-\mathsf{x}| \leq h$, the Hölder condition on $g^{(p+1)}$ in Assumption~\ref{Ass_g}(iii) implies that $\sup_{i:K ((x_i-\mathsf{x})/h) \neq 0} |g^{(p+1)}(\tilde{x}_i) - g^{(p+1)}(\mathsf{x})| \leq \sup_{x: |x-\mathsf{x}|\leq h}|g^{(p+1)}(x) - g^{(p+1)}(\mathsf{x})| \leq \sup_{x: |x-\mathsf{x}|\leq 2h} c |x-\mathsf{x}|^\eta = o(1)$. Thus, for some power~$k\geq 0$, we can bound all the terms in $\tilde\psi_{j,n,2}$ by $\op(1)(nh)^{-1} \sum_{i=1}^n K^2 ((x_i-\mathsf{x})/h) ((x_i-\mathsf{x})/h)^k = \op(1)$ by Assumption~\ref{Ass_K,h}(i).

The proof is concluded by showing that $\tilde\psi_{j,n,1} = \psi_{j,n} + \op (1)$. We first notice that, by the law of iterated expectations, $\E [ \tilde\psi_{j,n,1}] = \E [ \psi_{j,n}]$. Also by the law of iterated expectations,
\begin{align*}
\E &[ ( \tilde\psi_{j,n,1} -\psi_{j,n})^2 ] \\
&= \E \left[ \frac{1}{(nh)^2} \sum_{i_1=1}^n \sum_{i_2=1}^n K^2\left(\frac{x_{i_1}-\mathsf{x}}{h}\right) K^2\left(\frac{x_{i_2}-\mathsf{x}}{h}\right) \left(\frac{x_{i_1}-\mathsf{x}}{h}\right)^j \left(\frac{x_{i_2}-\mathsf{x}}{h}\right)^j (\varepsilon_{i_1}^2 - \sigma^2(x_{i_1}) (\varepsilon_{i_2}^2  - \sigma^2(x_{i_2}) \right] \\
&= \E \left[ \frac{1}{(nh)^2} \sum_{i=1}^n  K^4\left(\frac{x_i-\mathsf{x}}{h}\right) \left(\frac{x_i-\mathsf{x}}{h}\right)^{2j} (\E [\varepsilon_i^4|x_i] -\sigma^4 (x_i)) \right]. 
\end{align*}
By Assumption~\ref{Ass_g}, $\sup_{i=1,\ldots , n}|\E [\varepsilon_i^4|x_i] -\sigma^4 (x_i)|<\infty$, and the proof is then completed in the same way as in the proof for $\psi_{j,n}$ above. $\hfill \square$

\begin{lemma} \label{bound lemma}
Let Assumptions~\ref{Ass_g} and~\ref{Ass_K,h} be satisfied. Then: \\
(i) $\sup_{i=1,\ldots,n} |w_i(\mathsf{x})| = \Op(1)$, \\
(ii) $\sup_{i=1,\ldots,n} |(nh)^{-1} \sum_{j=1}^n w_i (x_j)w_j(\mathsf{x})| = \Op(1)$, \\
(iii) $\sup_{i=1,\ldots,n} | w_{\mathsf{PLP},i} (\mathsf{x})| = \Op(1)$, \\
(iv) $\sup_{i=1,\ldots,n} | w_{\mathsf{mPLP},i} (\mathsf{x})| = \Op(1)$.
\end{lemma}

\noindent\textsc{proof of lemma~\ref{bound lemma}}. Part~(i): By Lemma~\ref{auxlemma1Gamma}, $\Gamma_p^{-1} (\mathsf{x})=\Op(1)$. The result then follows by uniform boundedness of $K$ and $r_p$ on~$\mathbb{S}_x$.

\medskip \noindent \textsc{part}~(ii). Using part~(i) we find that $| \sum_{j=1}^n w_i (x_j)w_j(\mathsf{x})| \leq \sum_{j=1}^n \mathbb{I}_{\{|x_j-\mathsf{x}| \leq h\}} |w_j(\mathsf{x})| |w_i (x_j)| = \Op(1)\sum_{j=1}^n \mathbb{I}_{\{|x_j-\mathsf{x}| \leq h\}} |w_i (x_j)|$. By Lemma~\ref{auxlemma1Gamma} we can replace $\Gamma_p^{-1}(x_j)$ by $(\E [\Gamma_p(x_j)])^{-1}$, which is bounded uniformly in $x_j \in [\mathsf{x}-h,\mathsf{x}+h]$ for sufficiently large~$n$ because $K$ and $r_p$ are bounded on $\mathbb{S}_x$ and $f$ is bounded away from zero in a neighborhood of~$\mathsf{x}$. We are left with the term $\sup_{i=1,\ldots,n}(nh)^{-1}\sum_{j=1}^n \mathbb{I}_{\{|x_j-\mathsf{x}| \leq h\}} r_p((x_i-x_j)/h) K((x_i-x_j)/h) \leq c (nh)^{-1}\sum_{j=1}^n \mathbb{I}_{\{|x_j-\mathsf{x}| \leq h\}}$, where the inequality is again due to boundedness of $K$ and~$r_p$. The right-hand side is $\Op(1)$ because the second moment is
\begin{align*}
\E \bigg[ \Big( \frac{1}{nh} \sum_{j=1}^n \mathbb{I}_{\{|x_j-\mathsf{x}| \leq h\}}\Big)^2 \bigg] 
&= \frac{1}{(nh)^2}\sum_{j=1}^n \PP (|x_j-\mathsf{x}| \leq h) + \frac{2}{(nh)^2}\sum_{j=2}^n \sum_{m=1}^{j-1}\PP (|x_j-\mathsf{x}| \leq h)\PP (|x_m-\mathsf{x}| \leq h) \\
&= \frac{1}{nh^2} \int_{\mathsf{x}-h}^{\mathsf{x}+h} f(x_1)dx_1 + \frac{n-1}{nh^2}\Big(\int_{\mathsf{x}-h}^{\mathsf{x}+h} f(x_1)dx_1\Big)^2 =O(1).
\end{align*}

\medskip \noindent \textsc{parts}~(iii)--(iv). Follow directly by application of parts~(i) and~(ii) and the fact that $Q_n = \Op(1)$. $\hfill \square$

\begin{lemma} 
\label{lemmaequivker}
Let Assumptions~\ref{Ass_g} and~\ref{Ass_K,h} be satisfied. Then
\begin{align*}
\sup_{i=1,\ldots ,n} &\left| w_i (\mathsf{x}) - \frac{1}{f(\mathsf{x})} \mathsf{w} \left( \frac{x_i - \mathsf{x}}{h} \right) \right| = \op (1), \\
\sup_{x\in\mathbb{S}_x}\sup_{i=1,\ldots ,n} &\left| w_i (x) - \mathsf{w}_{\mathsf{E}} \left( \frac{x_i - x}{h},x \right) \right| = \op (1).
\end{align*}
\end{lemma}

\noindent\textsc{proof of Lemma \ref{lemmaequivker}}. The results follow immediately from Lemma~\ref{auxlemma1Gamma} and the definition of~$w_i(x)$. $\hfill \square$


\begin{lemma} \label{new supp lemma 1}
Let Assumptions~\ref{Ass_g} and~\ref{Ass_K,h} be satisfied, and let $\zeta_i$ be such that $\phi^2(x_i) := \E[\zeta_i^2|x_i]$ for a continuous and bounded function $\phi: \mathbb{S}_x \to \mathbb{R}$ and $\E[\zeta_i^4|x_i]$ is uniformly bounded. For an interior point $\mathsf{x}$ it holds that
\begin{align}
\label{new supp lemma 1 eq1}
\frac{1}{nh}\sum_{i=1}^n  \left(w_{\mathsf{PLP},i}  (\mathsf{x}) - \frac{1}{f(\mathsf{x})}\mathsf{w}_{\mathsf{PLP}} \left( \frac{x_i - \mathsf{x}}{h} \right) \right)^2\zeta_i^2  &= \op(1) , \\
\label{new supp lemma 1 eq2}
\frac{1}{nh}\sum_{i=1}^n  \left(w_{\mathsf{PLP},i}  (\mathsf{x}) + \frac{1}{f(\mathsf{x})}\mathsf{w}_{\mathsf{PLP}} \left( \frac{x_i - \mathsf{x}}{h} \right) \right)^2 \zeta_i^2  &= \Op(1) .
\end{align}
\end{lemma}

\noindent \textsc{proof of lemma \ref{new supp lemma 1}}. We will prove \eqref{new supp lemma 1 eq1} by showing that 
\begin{align*}
\mathsf{W}_{1n} :=&\frac{1}{nh}\sum_{i=1}^n  \left(w_i  (\mathsf{x}) - \frac{1}{f(\mathsf{x})}\mathsf{w} \left( \frac{x_i - \mathsf{x}}{h} \right) \right)^2\zeta_i^2 = \op(1) , \\
\mathsf{W}_{2n} :=&\frac{1}{nh}\sum_{i=1}^n  \left(\frac{1}{nh} \sum_{j=1}^n w_{j} (\mathsf{x})w_{i} (x_j) - \frac{1}{f(\mathsf{x})}\mathsf{w}_{\mathsf{conv}} \left( \frac{x_i - \mathsf{x}}{h} \right) \right)^2\zeta_i^2   = \op(1) , \\
\mathsf{W}_{12n} :=&\frac{1}{nh}\sum_{i=1}^n  \left(w_i  (\mathsf{x}) - \frac{1}{f(\mathsf{x})}\mathsf{w} \left( \frac{x_i - \mathsf{x}}{h} \right) \right) \left( \frac{1}{nh} \sum_{j=1}^n w_{j} (\mathsf{x})w_{i} (x_j) - \frac{1}{f(\mathsf{x})}\mathsf{w}_{\mathsf{conv}} \left( \frac{x_i - \mathsf{x}}{h} \right)  \right)\zeta_i^2   = \op(1) .
\end{align*}
By Lemma~\ref{lemmaequivker}, $\mathsf{W}_{1n} = \op(1) (nh)^{-1} \sum_{i: |x_i - \mathsf{x}|<h} \zeta_i^2$, and by Chebychev's inequality and the law of iterated expectations, $(nh)^{-1} \sum_{i: |x_i - \mathsf{x}|<h} \zeta_i^2 = \Op(1)$. 

Next, by the uniform convergence in Lemma~\ref{lemmaequivker}, we find that
\begin{align*}
\mathsf{W}_{2n}&=\frac{1}{nhf(\mathsf{x})^2}\sum_{i=1}^n \left( \frac{1}{nh }\sum_{j=1}^n  \mathsf{w} \left( \frac{x_j - \mathsf{x}}{h} \right) \mathsf{w}_{\mathsf{E}}\left( \frac{x_i - x_j}{h} ,x_j \right) - \mathsf{w}_{\mathsf{conv}}\left( \frac{x_i - \mathsf{x}}{h} \right) \right)^2\zeta_i^2 + \op(1) \\
&=\frac{1}{nhf(\mathsf{x})^2}\sum_{i=1}^n  \left( \frac{1}{nh }\sum_{j=1, j \not=i}^n  \mathsf{w} \left( \frac{x_j - \mathsf{x}}{h} \right) \mathsf{w}_{\mathsf{E}}\left( \frac{x_i - x_j}{h} ,x_j \right) - \mathsf{w}_{\mathsf{conv}}\left( \frac{x_i - \mathsf{x}}{h} \right) \right)^2\zeta_i^2 + \op(1) \\
&=: \frac{1}{f(\mathsf{x})^2}R_n +  \op(1),
\end{align*}
where the second equality follows because the sum
\begin{align*}
&\frac{1}{nh}\sum_{i=1}^n \left( \frac{1}{nh } \mathsf{w} \left( \frac{x_i - \mathsf{x}}{h} \right) \mathsf{w}_{\mathsf{E}}\left( 0 ,x_i \right) \right)^2\zeta_i^2 \\
&+ \frac{1}{nh}\sum_{i=1}^n  \left( \frac{1}{nh }\sum_{j=1, j \not=i}^n  \mathsf{w} \left( \frac{x_j - \mathsf{x}}{h} \right) \mathsf{w}_{\mathsf{E}}\left( \frac{x_i - x_j}{h} ,x_j \right) - \mathsf{w}_{\mathsf{conv}}\left( \frac{x_i - \mathsf{x}}{h} \right) \right)
\left( \frac{1}{nh } \mathsf{w} \left( \frac{x_i - \mathsf{x}}{h} \right) \mathsf{w}_{\mathsf{E}}\left( 0 ,x_i \right) \right) \zeta_i^2 
\end{align*}
is $\Op((nh)^{-1})$ by $L_1$-convergence. We will also show that $R_n \plowto 0$ by $L_1$-convergence. Specifically, by the law of iterated expectations, 
\begin{align*}
\E [R_n] &= \E\left[ \frac{1}{nh} \sum_{i=1}^n \left(\frac{1}{nh}\sum_{j=1, j \not=i}^n  \mathsf{w} \left( \frac{x_j - \mathsf{x}}{h} \right) \mathsf{w}_{\mathsf{E}}\left( \frac{x_i - x_j}{h} ,x_j \right) - \mathsf{w}_{\mathsf{conv}}\left( \frac{x_i - \mathsf{x}}{h} \right) \right)^2 \phi^2 (x_i)  \right] \\
&=:  \E [R_{1n}]  +  \E [R_{2n}]  - 2  \E [R_{12n}] .
\end{align*}
We start by deriving the limit of $ \E [R_{1n}]$, which is
\begin{align*}
\E [R_{1n}] &= \E \left[   \frac{1}{nh} \sum_{i=1}^n \left(\frac{1}{nh }\sum_{j=1, j \not=i}^n  \mathsf{w} \left( \frac{x_j - \mathsf{x}}{h} \right)  \mathsf{w}_{\mathsf{E}}\left( \frac{x_i - x_j}{h}, x_j\right) \right)^2 \phi^2 (x_i) \right]\\
&= \E \left[ \frac{1}{(nh)^3} \underset{\text{all distinct}}{\sum_{i,j,j^{\prime}=1}^n} \mathsf{w} \left( \frac{x_j - \mathsf{x}}{h} \right)  \mathsf{w}_{\mathsf{E}}\left( \frac{x_i - x_j}{h}, x_j\right) \mathsf{w} \left( \frac{x_{j^{\prime}} - \mathsf{x}}{h} \right)  \mathsf{w}_{\mathsf{E}}\left( \frac{x_i - x_{j^{\prime}} }{h}, x_{j^{\prime}} \right)  \phi^2 (x_i) \right]\\
&\quad + \E \left[ \frac{1}{(nh)^3} \sum_{i,j=1, i \not=j}^n   \mathsf{w} \left( \frac{x_j - \mathsf{x}}{h} \right)^2  \mathsf{w}_{\mathsf{E}}\left( \frac{x_i - x_j}{h}, x_j\right)^2  \phi^2 (x_i) \right] .
\end{align*}
The second term on the right-hand side is
\begin{align*}
&\frac{n(n-1)}{(nh)^3} \int_{\mathbb{S}_x} \int_{\mathbb{S}_x} \mathsf{w} \left( \frac{x_2 - \mathsf{x}}{h} \right)^2  \mathsf{w}_{\mathsf{E}}\left( \frac{x_1 - x_2}{h}, x_2\right)^2 \phi^2 (x_1) f(x_1) f (x_2) dx_2 dx_1 \\
&=\frac{n(n-1)}{n^3h}\int_{\mathcal{X}} \int_{\mathcal{X}} \mathsf{w} \left( u\right)^2  \mathsf{w}_{\mathsf{E}}\left( s-u, \mathsf{x}+uh\right)^2   \phi^2 (\mathsf{x}+sh) f(\mathsf{x}+sh) f (\mathsf{x}+uh) du ds  \\
&=\frac{n(n-1)}{n^3h} \left( \phi^2 (\mathsf{x}) f(\mathsf{x})^2 \int_{\mathcal{X}} \int_{\mathcal{X}}   \mathsf{w} \left( u\right)^2  \mathsf{w}\left( s-u\right)^2 du ds + o(1) \right) = O\left(\frac{1}{nh}\right),    
\end{align*}
and the main term in $\E [R_{1n}]$ is
\begin{align*}
&\E \left[ \frac{1}{(nh)^3} \underset{\text{all distinct}}{\sum_{i,j,j^{\prime}=1}^n} \mathsf{w} \left( \frac{x_j - \mathsf{x}}{h} \right)  \mathsf{w}_{\mathsf{E}}\left( \frac{x_i - x_j}{h}, x_j\right) \mathsf{w} \left( \frac{x_{j^{\prime}} - \mathsf{x}}{h} \right)  \mathsf{w}_{\mathsf{E}}\left( \frac{x_i - x_{j^{\prime}} }{h}, x_{j^{\prime}} \right)  \phi^2 (x_i) \right] \\
&= \frac{(n-1)(n-2)}{n^2h^3} \int_{\mathbb{S}_x}  \left(  \int_{\mathsf{x} - h}^{\mathsf{x} + h}   \mathsf{w} \left( \frac{x_2 - \mathsf{x}}{h} \right)  \mathsf{w}_{\mathsf{E}}\left( \frac{x_1 - x_2}{h}, x_2\right)    f(x_2)   dx_2 \right)^2 \phi^2 (x_1) f(x_1) dx_1\\
&= \phi^2  (\mathsf{x})f (\mathsf{x})\int_{\mathcal{X}} \left(  \int_{\mathcal{X}} \mathsf{w} (u)  \mathsf{w} (s-u)    du \right)^2  ds + o(1) \to \mathsf{R},
\end{align*}
where $\mathsf{R} := \phi^2 (\mathsf{x})f (\mathsf{x})\int_{\mathcal{X}} \mathsf{w}_{\mathsf{conv}}(s)^2 ds$. Similarly, we find that 
\begin{align*}
\E [R_{2n}] &= \E \left[ \frac{1}{nh} \sum_{i=1}^n \left(\int_{\mathcal{X}} \mathsf{w} (u)  \mathsf{w}\left( \frac{x_i - \mathsf{x}}{h}-u\right) du  \right)^2  \phi^2  (x_i)  \right] \\
&=  \frac{1}{h} \int_{\mathbb{S}_x} \Biggl(\int_{\mathcal{X}}    \mathsf{w} \left( u\right)  \mathsf{w}\left( \frac{x_1 - \mathsf{x}}{h}-u\right)     du  \Biggr)^2  \phi^2  (x_1)  f (x_1) dx_1 = \mathsf{R} + o(1)
\end{align*}
and 
\begin{align*}
\E [R_{12n}] &= \E \left[ \frac{1}{(nh)^2} \sum_{i=1}^n \mathsf{w}_{\mathsf{conv}} \left( \frac{x_i - \mathsf{x}}{h} \right)  \sum_{j=1, j \not=i}^n  \mathsf{w} \left( \frac{x_j - \mathsf{x}}{h} \right)  \mathsf{w}_{\mathsf{E}}\left( \frac{x_i - x_j}{h}, x_j\right)  \phi^2  (x_i) \right] \\
&= \frac{n-1}{nh^2} \int_{\mathbb{S}_x}\int_{\mathsf{x}-h}^{\mathsf{x}+h} \mathsf{w}_{\mathsf{conv}} \left( \frac{x_1 - \mathsf{x}}{h} \right) \mathsf{w} \left( \frac{x_2 - \mathsf{x}}{h} \right)  \mathsf{w}_{\mathsf{E}}\left( \frac{x_1 - x_2}{h}, x_2\right)  \phi^2  (x_1) f(x_1) f (x_2)dx_1 dx_2  \\ 
&= \phi^2 (\mathsf{x}) f (\mathsf{x})  \int_{\mathcal{X}} \mathsf{w}_{\mathsf{conv}} (r) \left( \int_{\mathcal{X}} \mathsf{w} (r-s) \mathsf{w} (s) ds \right) dr + o(1) = \mathsf{R} + o(1),
\end{align*}
which shows that $R_n \plowto 0$ and hence that $\mathsf{W}_{2n}\plowto 0$.

Next, by the Cauchy-Schwarz inequality, $\mathsf{W}_{12n}= \mathsf{W}_{1n}^{1/2} \mathsf{W}_{2n}^{1/2}=\op(1)$. 

Finally, using the result in \eqref{new supp lemma 1 eq1}, to prove \eqref{new supp lemma 1 eq2} it is sufficient to show that 
\begin{equation*}
\mathsf{W}_{3n} := \frac{4}{nhf(\mathsf{x})}\sum_{i=1}^n  w_{\mathsf{PLP},i} (\mathsf{x}) \mathsf{w}_{\mathsf{PLP}} \left( \frac{x_i - \mathsf{x}}{h} \right)  \zeta_i^2  = \Op(1) .
\end{equation*}
By Lemma~\ref{bound lemma}, $|\mathsf{W}_{3n}| = \Op((nh)^{-1})\sum_{i=1}^n  |\mathsf{w}_{\mathsf{PLP}} ( (x_i - \mathsf{x})/h )|  \zeta_i^2$. To conclude the proof, we apply the law of iterated expectations and find $\E[(nh)^{-1}\sum_{i=1}^n |\mathsf{w}_{\mathsf{PLP}} ( (x_i - \mathsf{x})/h )|  \zeta_i^2] = \E[(nh)^{-1}\sum_{i=1}^n |\mathsf{w}_{\mathsf{PLP}} ( (x_i - \mathsf{x})/h )|  \phi^2(x_i)] = f(\mathsf{x})^{-2}\phi^2(\mathsf{x}) \int_{\mathcal{X}} |\mathsf{w}_{\mathsf{PLP}} (u) |du + o(1) = O(1)$ by the same method as above. $\hfill \square$

\begin{lemma} \label{new supp lemma 2}
Let Assumptions~\ref{Ass_g} and~\ref{Ass_K,h} be satisfied, and let $\zeta_i$ be such that $\phi^2(x_i) := \E[\zeta_i^2|x_i]$ for a continuous and bounded function $\phi: \mathbb{S}_x \to \mathbb{R}$ and $\E[\zeta_i^4|x_i]$ is uniformly bounded. Then:\\
(i) If $\mathsf{x}$ is an interior point,
\begin{align}
\label{new supp lemma 2 eq3}
\frac{1}{nh}\sum_{i=1}^n  \left(w_{\mathsf{mPLP},i}  (\mathsf{x}) - \frac{1}{f(\mathsf{x})}\mathsf{w}_{\mathsf{PLP}} \left( \frac{x_i - \mathsf{x}}{h} \right) \right)^2\zeta_i^2  &= \op(1) , \\
\label{new supp lemma 2 eq4}
\frac{1}{nh}\sum_{i=1}^n  \left(w_{\mathsf{mPLP},i}  (\mathsf{x}) + \frac{1}{f(\mathsf{x})}\mathsf{w}_{\mathsf{PLP}} \left( \frac{x_i - \mathsf{x}}{h} \right) \right)^2 \zeta_i^2  &= \Op(1) .
\end{align}
(ii) If $\mathsf{x}$ is a boundary point,
\begin{align}
\label{new supp lemma 2 eq1}
\frac{1}{nh}\sum_{i=1}^n  \left(w_{\mathsf{mPLP},i}  (\mathsf{x}) - \frac{1}{f(\mathsf{x})}\mathsf{w}_{\mathsf{mPLP}}  \left( \frac{x_i - \mathsf{x}}{h} \right) \right)^2\zeta_i^2  &= \op(1), \\
\label{new supp lemma 2 eq2}
\frac{1}{nh}\sum_{i=1}^n  \left(w_{\mathsf{mPLP},i}  (\mathsf{x}) + \frac{1}{f(\mathsf{x})}\mathsf{w}_{\mathsf{mPLP}}  \left( \frac{x_i - \mathsf{x}}{h} \right) \right)^2 \zeta_i^2  &= \Op(1) .
\end{align}
\end{lemma}

\noindent \textsc{proof of lemma \ref{new supp lemma 2}}. We prove \eqref{new supp lemma 2 eq3} by showing that
 \begin{align*}
\mathsf{Z}_{1n}&:=\frac{1}{nh}\sum_{i=1}^n \left( (1+Q_n)w_i(\mathsf{x}) -  2f(\mathsf{x})^{-1}\mathsf{w}\left(\frac{x_i - \mathsf{x}}{h}\right) \right)^2  \zeta_i^2 = \op(1) , \\
\mathsf{Z}_{2n}&:=\frac{1}{nh}\sum_{i=1}^n \left( Q_n\frac{1}{nh} \sum_{j=1}^n w_j(\mathsf{x}) w_i(x_j) - f(\mathsf{x})^{-1}\mathsf{w}_{\mathsf{conv}}\left(\frac{x_i - \mathsf{x}}{h}\right) \right)^2 \zeta_i^2  = \op(1) ,\\
\mathsf{Z}_{12n}&:= \frac{1}{nh}\sum_{i=1}^n \left( (1+Q_n)w_i(\mathsf{x}) -  2f(\mathsf{x})^{-1}\mathsf{w}\left(\frac{x_i - \mathsf{x}}{h}\right) \right)  \\
& \quad \times \left( Q_n\frac{1}{nh} \sum_{j=1}^n w_j(\mathsf{x}) w_i(x_j) - f(\mathsf{x})^{-1}\mathsf{w}_{\mathsf{conv}}\left(\frac{x_i - \mathsf{x}}{h}\right) \right) \zeta_i^2 = \op(1) .
\end{align*} 
By Lemma~\ref{lemmaKernelConstants} and Lemma \ref{bound lemma}, we have that $\mathsf{Z}_{1n} = 4 \mathsf{W}_{1n} + \op(1)$, $\mathsf{Z}_{2n} = \mathsf{W}_{2n} + \op(1)$, and $\mathsf{Z}_{12n} = 2 \mathsf{W}_{12n} + \op(1)$, where $\mathsf{W}_{1n}$, $\mathsf{W}_{2n}$, and $\mathsf{W}_{12n}$ are defined in the proof of Lemma~\ref{new supp lemma 1}. Therefore, the proof of \eqref{new supp lemma 2 eq3} is concluded because we showed in the proof of Lemma~\ref{new supp lemma 1} that $\mathsf{W}_{1n}$, $\mathsf{W}_{2n}$, and $\mathsf{W}_{12n}$ are all~$\op(1)$.

Next, we prove \eqref{new supp lemma 2 eq4}. To do so, by \eqref{new supp lemma 2 eq3}, it is sufficient to show that 
\begin{align*}
\mathsf{Z}_{3n} &:= \frac{4}{nhf(\mathsf{x})}\sum_{i=1}^n  w_{\mathsf{mPLP},i} (\mathsf{x}) \mathsf{w}_{\mathsf{PLP}} \left( \frac{x_i - \mathsf{x}}{h} \right)  \zeta_i^2  = \Op(1).
\end{align*}
By Lemma \ref{bound lemma}, $|\mathsf{Z}_{3n}| = \Op((nh)^{-1})\sum_{i=1}^n  |\mathsf{w}_{\mathsf{PLP}} \left( (x_i - \mathsf{x})/h \right)|  \zeta_i^2$. The result then follows by Markov's inequality and the law of iterated expectations because
\begin{equation*}
\E \left[ \frac{1}{nh}\sum_{i=1}^n \left| \mathsf{w}_{\mathsf{PLP}} \left( \frac{x_i - \mathsf{x}}{h} \right) \right| \zeta_i^2 \right] = \phi^2(\mathsf{x}) \int_{\mathcal{X}} |\mathsf{w}_{\mathsf{PLP}}(u) |du + o(1) = O(1).
\end{equation*}

Next, we will prove \eqref{new supp lemma 2 eq1} in the same way as \eqref{new supp lemma 2 eq3}, by showing that 
\begin{align*}
\mathsf{Z}_{\mathsf{bnd},1n} &:= \frac{1}{nh}\sum_{i=1}^n  \left( (1+Q_n)w_i  (\mathsf{x}) - \frac{1+Q}{f(\mathsf{x})}\mathsf{w}  \left( \frac{x_i - \mathsf{x}}{h}\right) \right)^2\zeta_i^2 = \op(1) , \\
\mathsf{Z}_{\mathsf{bnd},2n} &:=\frac{1}{nh}\sum_{i=1}^n \left(\frac{ Q_n}{nh} \sum_{j=1}^n w_{j} (\mathsf{x})w_{i} (x_j) - \frac{Q}{f(\mathsf{x})}\mathsf{w}_{\mathsf{conv},\mathsf{bnd}} \left( \frac{x_i - \mathsf{x}}{h} \right) \right)^2\zeta_i^2   = \op(1) , \\
\mathsf{Z}_{\mathsf{bnd},12n} &:=\frac{1}{nh}\sum_{i=1}^n  \left((1+Q_n)w_i  (\mathsf{x}) - \frac{(1+Q)}{f(\mathsf{x})}\mathsf{w}  \left( \frac{x_i - \mathsf{x}}{h} \right) \right) \\
& \qquad \times \left( \frac{Q_n}{nh} \sum_{j=1}^n w_{j} (\mathsf{x})w_{i} (x_j) - \frac{Q}{f(\mathsf{x})}\mathsf{w}_{\mathsf{conv},\mathsf{bnd}} \left( \frac{x_i - \mathsf{x}}{h} \right)  \right)\zeta_i^2   = \op(1) .
\end{align*}
By Lemma~\ref{lemmaKernelConstants} and Lemma~\ref{bound lemma}, we have that $\mathsf{Z}_{\mathsf{bnd},1n} = (1+Q)^2 \mathsf{Z}^{\dagger}_{1n} + \op(1) $, $\mathsf{Z}_{\mathsf{bnd},2n} = Q^2 \mathsf{Z}^{\dagger}_{2n} + \op(1)$, and $\mathsf{Z}_{\mathsf{bnd},12n} = Q(1+Q) \mathsf{Z}^{\dagger}_{12n} + \op(1)$, where 
 \begin{align*}
\mathsf{Z}_{1n}^{\dagger}&:=\frac{1}{nh}\sum_{i=1}^n \left( w_i(\mathsf{x}) -  \frac{1}{f(\mathsf{x})}\mathsf{w}\left(\frac{x_i - \mathsf{x}}{h}\right) \right)^2  \zeta_i^2 , \\
\mathsf{Z}_{2n}^{\dagger}&:=\frac{1}{nh}\sum_{i=1}^n \left( \frac{1}{nh} \sum_{j=1}^n w_j(\mathsf{x}) w_i(x_j) - \frac{1}{f(\mathsf{x})}\mathsf{w}_{\mathsf{conv}, \mathsf{bnd}}\left(\frac{x_i - \mathsf{x}}{h}\right) \right)^2 \zeta_i^2 , \\
\mathsf{Z}_{12n}^{\dagger}&:= \frac{1}{nh}\sum_{i=1}^n    \left( w_i(\mathsf{x}) - \frac{1}{f(\mathsf{x})}\mathsf{w} \left(\frac{x_i - \mathsf{x}}{h}\right) \right) \left( \frac{1}{nh} \sum_{j=1}^n w_j(\mathsf{x}) w_i(x_j) - \frac{1}{f(\mathsf{x})}\mathsf{w}_{\mathsf{conv}, \mathsf{bnd}}\left(\frac{x_i - \mathsf{x}}{h}\right) \right) \zeta_i^2 .
\end{align*} 
Then \eqref{new supp lemma 2 eq1} follows by showing that $\mathsf{Z}_{1n}^{\dagger}$, $\mathsf{Z}_{2n}^{\dagger}$, and $\mathsf{Z}_{12n}^{\dagger}$ are all~$\op(1)$, which holds by identical arguments applied in the proof of Lemma~\ref{new supp lemma 1} for $\mathsf{W}_{1n}$, $\mathsf{W}_{2n}$, and $\mathsf{W}_{12n}$, respectively. 

Finally, the proof of \eqref{new supp lemma 2 eq2} is analogous to that of \eqref{new supp lemma 2 eq4} and is therefore omitted. $\hfill \square$


\begin{lemma} \label{new supp lemma 3}
Let Assumptions~\ref{Ass_g} and~\ref{Ass_K,h} be satisfied, and let $\zeta_i$ be such that $\phi^2(x_i) := \E[\zeta_i^2|x_i]$ for a continuous and bounded function $\phi: \mathbb{S}_x \to \mathbb{R}$ and $\E[\zeta_i^4|x_i]$ is uniformly bounded. For an interior point $\mathsf{x}$ it holds that
\begin{equation}
\label{LemS4eq}
\frac{1}{nh}\sum_{i=1}^n  \left(w_{\mathsf{PLP},i} (\mathsf{x})^2 - \frac{1}{f(\mathsf{x})^2}\mathsf{w}_{\mathsf{PLP}} \left( \frac{x_i - \mathsf{x}}{h} \right)^2 \right)\zeta_i^2 = \op(1) .
\end{equation}
\end{lemma}

\noindent \textsc{proof of lemma \ref{new supp lemma 3}}. By the Cauchy-Schwarz inequality, the absolute value of the left-hand side of \eqref{LemS4eq} is bounded by
\begin{align*}
&\Biggl| \frac{1}{nh}\sum_{i=1}^n  \left(w_{\mathsf{PLP},i}  (\mathsf{x}) - \frac{1}{f(\mathsf{x})}\mathsf{w}_{\mathsf{PLP}} \left( \frac{x_i - \mathsf{x}}{h} \right) \right)\left(w_{\mathsf{PLP},i}  (\mathsf{x}) + \frac{1}{f(\mathsf{x})}\mathsf{w}_{\mathsf{PLP}} \left( \frac{x_i - \mathsf{x}}{h} \right) \right)\zeta_i^2 \Biggr|  \\
&\leq \left( \frac{1}{nh}\sum_{i=1}^n  \left(w_{\mathsf{PLP},i}  (\mathsf{x}) - \frac{1}{f(\mathsf{x})}\mathsf{w}_{\mathsf{PLP}} \left( \frac{x_i - \mathsf{x}}{h} \right) \right)^2 \zeta_i^2  \right)^{1/2} \\
& \qquad \times \left( \frac{1}{nh}\sum_{i=1}^n\left(w_{\mathsf{PLP},i}  (\mathsf{x}) + \frac{1}{f(\mathsf{x})}\mathsf{w}_{\mathsf{PLP}} \left( \frac{x_i - \mathsf{x}}{h} \right) \right)^2\zeta_i^2  \right)^{1/2},
\end{align*}
which is $\op(1)$ by Lemma~\ref{new supp lemma 1}. $\hfill \square$

\begin{lemma} \label{new supp lemma 4}
Let Assumptions~\ref{Ass_g} and~\ref{Ass_K,h} be satisfied, and let $\zeta_i$ be such that $\phi^2(x_i) := \E[\zeta_i^2|x_i]$ for a continuous and bounded function $\phi: \mathbb{S}_x \to \mathbb{R}$ and $\E[\zeta_i^4|x_i]$ is uniformly bounded. Then:\\
(i) If $\mathsf{x}$ is an interior point,
\begin{equation}
\frac{1}{nh}\sum_{i=1}^n  \left(w_{\mathsf{mPLP},i}  (\mathsf{x})^2 - \frac{1}{f(\mathsf{x})^2}\mathsf{w}_{\mathsf{PLP}} \left( \frac{x_i - \mathsf{x}}{h} \right)^2 \right)\zeta_i^2 = \op(1).
\end{equation}
(ii) If $\mathsf{x}$ is a boundary point,
\begin{equation}
\frac{1}{nh}\sum_{i=1}^n  \left(w_{\mathsf{mPLP},i}  (\mathsf{x})^2 - \frac{1}{f(\mathsf{x})^2}\mathsf{w}_{\mathsf{mPLP}} \left( \frac{x_i - \mathsf{x}}{h} \right)^2 \right)\zeta_i^2 = \op(1).
\end{equation}   
\end{lemma}

\noindent \textsc{proof of lemma \ref{new supp lemma 4}}. Like the proof of Lemma~\ref{new supp lemma 3}, both parts follow from application of the Cauchy-Schwarz inequality, except we use Lemma~\ref{new supp lemma 2} instead of Lemma~\ref{new supp lemma 1}.  $\hfill \square$

\suppsection{proofs of bias expansion lemmas in appendix \ref{app:lemmas}}
\label{Supp:Sec-extra}

\noindent\textsc{proof of lemma \ref{lemmaKernelConstants}}. Proof for~$C_n$: By Lemma~\ref{lemmaequivker}, $C_n = f(\mathsf{x})^{-1} (nh)^{-1}\sum_{i=1}^n \mathsf{w}((x_i-\mathsf{x})/h) ((x_i-\mathsf{x})/h)^{p+1} + \op (1)$. Inserting $\mathsf{w}$ yields the vector $(nh)^{-1}\sum_{i=1}^n r_p ((x_i-\mathsf{x})/h) K((x_i-\mathsf{x})/h) ((x_i-\mathsf{x})/h)^{p+1}$. This has $j$-th element~$\gamma_{j+p,n}(\mathsf{x})$, and the results then follow from Lemma~\ref{auxlemma1}.

\medskip \noindent \textsc{part}~(i) \textsc{for}~$C_{\mathsf{LP},n}$. Let $\theta_{\mathsf{E},p}(x):= \int_{\mathcal{X}} K(u) r_p(u) u^{p+1}  f(x+uh)du$. By Lemmas~\ref{auxlemma1} and~\ref{lemmaequivker},
\begin{align*}
    C_{\mathsf{LP},n} 
&=\frac{1}{nhf(\mathsf{x})} \sum_{i=1}^n \mathsf{w} \left(\frac{x_i - \mathsf{x}}{h} \right) \frac{1}{nh} \sum_{j=1}^n \mathsf{w}_{\mathsf{E}}\left(\frac{x_j - x_i}{h} , x_i \right) \left(\frac{x_j - x_i}{h} \right)^{p+1} + \op(1) \\
&=\frac{1}{nhf(\mathsf{x})} \sum_{i=1}^n \mathsf{w} \left(\frac{x_i - \mathsf{x}}{h} \right) \iota_0^{\prime} \Gamma_{\mathsf{E},p}^{-1}(x_i)\frac{1}{nh} \sum_{j=1}^n K\left(\frac{x_j - x_i}{h}\right) r_p\left(\frac{x_j - x_i}{h}\right) \left(\frac{x_j - x_i}{h} \right)^{p+1} + \op(1) \\
&=\frac{1}{nhf(\mathsf{x})} \sum_{i=1}^n \mathsf{w} \left(\frac{x_i - \mathsf{x}}{h} \right) \iota_0^{\prime} \Gamma_{\mathsf{E},p}^{-1}(x_i)\theta_{\mathsf{E},p}(x_i)+ \op(1) =: C^{\dagger}_{\mathsf{LP},n} + \op(1) .
\end{align*}
Thus, we analyze $C^{\dagger}_{\mathsf{LP},n}$ and find that, by the i.i.d.\ assumption and Lemmas~\ref{auxlemma1} and~\ref{auxlemma1Gamma}, $\E [C^{\dagger}_{\mathsf{LP},n} ] \to \iota_0^{\prime} \Gamma_{\mathsf{E},p}^{-1}(\mathsf{x})\theta_{\mathsf{E},p}(\mathsf{x}) \int_{\mathcal{X}}  \mathsf{w} (u)  du =C$. Moreover, by the i.i.d.\ assumption,
\begin{equation*}
\E [(C^{\dagger}_{\mathsf{LP},n} )^2 ] = (\E [ C^{\dagger}_{\mathsf{LP},n} ])^2 
+ \frac{1}{n(hf(\mathsf{x}))^2} \E \left[ \mathsf{w} \left(\frac{x_1 - \mathsf{x}}{h} \right)^2 (\iota_0^{\prime} \Gamma_{\mathsf{E},p}^{-1}(x_1)\theta_{\mathsf{E},p}(x_1))^2\right] .
\end{equation*}
Since $\E [ \mathsf{w} ((x_1 - \mathsf{x})/h )^2 (\iota_0^{\prime} \Gamma_{\mathsf{E},p}^{-1}(x_1)\theta_{\mathsf{E},p}(x_1))^2]=O(h)$, it follows that $\E [(C^{\dagger}_{\mathsf{LP},n} )^2 ] = (\E [ C^{\dagger}_{\mathsf{LP},n} ])^2 +O((nh)^{-1})$, which concluding the proof of part~(i) by $L_2$-convergence.

\medskip \noindent \textsc{part}~(ii) \textsc{for}~$C_{\mathsf{LP},n}$. As in the proof of part~(i), we can write
\begin{align*}
C_{\mathsf{LP},n} 
&=\frac{1}{nhf(\mathsf{x})} \sum_{i=1}^n \mathsf{w}_{\mathsf{bnd}} \left(\frac{x_i - \mathsf{x}}{h},0 \right) \iota_0^{\prime} \Gamma_{\mathsf{E},p}^{-1}(x_i)\theta_{\mathsf{E},p}(x_i) + \op (1)=: C^{\dagger}_{\mathsf{LP},\mathsf{bnd},n} + \op (1).
\end{align*} 
In this case we find that
\begin{align*}
\E [C^{\dagger}_{\mathsf{LP},n} ] 
&= \frac{1}{f(\mathsf{x})} \int_0^1 \mathsf{w}_{\mathsf{bnd}}(s,0) \iota_0^{\prime} \Gamma_{\mathsf{E},p}^{-1}(\mathsf{x}+sh)\theta_{\mathsf{E},p}(\mathsf{x}+sh) f(\mathsf{x}+sh)ds \\
&= \frac{1}{f(\mathsf{x})} \int_0^1 \mathsf{w}_{\mathsf{bnd}}(s,0) \iota_0^{\prime}
\left(\int_{-s}^1 K (v)r_p(v)r_p^{\prime}(v) f (\mathsf{x}+sh+vh)dv \right)^{-1} \\
&\quad \times \int_{-s}^1 K(u) r_p (u) u^{p+1} f(\mathsf{x}+sh+uh)du f(\mathsf{x}+sh)ds \\
&= \int_0^1 \mathsf{w}_{\mathsf{bnd}} (s,0) \int_{-s}^1 \mathsf{w}_{\mathsf{bnd}} (u, s) u^{p+1} du ds +o(1) = C_{\mathsf{LP}} + o(1),
\end{align*}
where we used the fact that, since $\mathsf{x}$ is a (left) boundary point, $f(\mathsf{x}+a)=0$ for $a<0$. Finally, the proof is completed by noting  that $\E [(C^{\dagger}_{\mathsf{LP}, n} )^2 ] = (\E [ C^{\dagger}_{\mathsf{LP},n} ])^2 + o(1)$ by the same arguments used in part~(i). $\hfill \square$

\bigskip
\noindent\textsc{proof of lemma \ref{lemmaGPbias}.} By definition of $\hat{B}_{\mathsf{GP},n}$ and using $\tilde{g}_{\mathsf{x},n} (\mathsf{x})=\iota _0^{\prime }\hat{\beta}_{p+1,n}(\mathsf{x})$,
\begin{align}
\hat{B}_{\mathsf{GP},n} &= \sqrt{nh}\iota _0^{\prime}\big( (Z_p^{\prime }WZ_p)^{-1}Z_p^{\prime }W Z_{p+1} H_{p+1}^{-1} \hat{\beta}_{p+1,n}(\mathsf{x})-\hat{\beta}_{p+1,n}(\mathsf{x}))\big) \nonumber \\
&= \sqrt{nh}\iota _0^{\prime }(Z_p^{\prime }WZ_p)^{-1}Z_p^{\prime}W Z_{p+1} \iota_{p+1}\iota_{p+1}^{\prime} H_{p+1}^{-1} \hat{\beta}_{p+1,n}(\mathsf{x}) \nonumber \\
&= \sqrt{nh^{2p+3}} C_n \iota_{p+1}' \hat{\beta}_{p+1,n}(\mathsf{x})
=\sqrt{nh^{2p+3}}\hat{g}_n^{(p+1)}(\mathsf{x})\tfrac{C_n}{(p+1)!}.
\label{BhatGP-details}
\end{align}
From \eqref{eq:Bn} and \eqref{BhatGP-details} we have $\hat{B}_{\mathsf{GP},n} - B_n =\sqrt{nh^{2p+3}}(\hat{g}_n^{(p+1)}(\mathsf{x})-g^{(p+1)}(\mathsf{x}))C_n/(p+1)! +\op (1)$, where $\hat{g}_n^{(p+1)}(\mathsf{x}) := (p+1)! \iota_{p+1}^{\prime} \hat{\beta}_{p+1,n} = (p+1)! (nh^{p+2})^{-1} \sum_{i=1}^n w_{(p+1),i} (\mathsf{x})y_i$ with $w_{(q),i} (x) := \iota_q^{\prime} \Gamma_q^{-1}(x) r_q ((x_i-x)/h) K((x_i-x)/h)$. Let $\tilde{x}_i$ be an intermediate value between $x_i$ and~$\mathsf{x}$. Then, by the mean-value argument in \eqref{MVT on g},
\begin{align*}
\frac{(p+1)!}{nh^{p+2}} &\sum_{i=1}^n w_{(p+1),i} (\mathsf{x}) g(x_i) 
= \frac{(p+1)!}{nh^{p+2}} \sum_{i=1}^n w_{(p+1),i} (\mathsf{x}) \Biggr( r_{p+1}^{\prime} \left( \frac{x_i - \mathsf{x}}{h}\right) H^{-1}_{p+1}\beta_{p+1} \\
& \qquad +\left( \frac{x_i - \mathsf{x}}{h}\right)^{p+1} h^{p+1} \left( \frac{g^{(p+1)} (\tilde x_{i})- g^{(p+1)}(\mathsf{x})}{(p+1)!} \right) \Biggr) \\
&= g^{(p+1)} (\mathsf{x}) +  \frac{1}{nh} \sum_{i=1}^n w_{(p+1),i} (\mathsf{x})  \left( \frac{x_i - \mathsf{x}}{h}\right)^{p+1} (g^{(p+1)} (\tilde x_{i})- g^{(p+1)}(\mathsf{x}))  = g^{(p+1)} (\mathsf{x})  +\Op(h^\eta),
\end{align*}
where the last equality follows by H\"{o}lder continuity of $g^{(p+1)}$ (Assumption~\ref{Ass_g}(iii)) and the fact that $(nh)^{-1}\sum_{i=1}^n w_{(p+1),i} (\mathsf{x}) ( (x_i - \mathsf{x})/h )^{p+1} = \Op(1)$. That is, $\hat{g}_n^{(p+1)}(\mathsf{x}) = g^{(p+1)} (\mathsf{x})  +\Op(h^\eta) + (p+1)! (nh^{p+2})^{-1} \sum_{i=1}^n w_{(p+1),i} (\mathsf{x}) \varepsilon_i$ such that 
\begin{equation*}
\hat{B}_{\mathsf{GP},n} - B_n =\frac{1}{\sqrt{nh}} \sum_{i=1}^n w_{(p+1),i} (\mathsf{x}) \varepsilon_i C_n  +\op (1),
\end{equation*}
which proves the result by definition of the weights~$w_{\mathsf{GP}\text{-}\mathsf{bc},i} (\mathsf{x})$.  $\hfill \square$

\bigskip
\noindent\textsc{proof of lemma \ref{lemmaLPbias}.} By definition of $\hat{B}_{\mathsf{LP},n}$, we have that
\begin{align*}
\hat{B}_{\mathsf{LP},n} &= \sqrt{nh} \left( \frac{1}{nh} \sum_{i=1}^n w_i(\mathsf{x}) \hat{g}_n(x_i) - \hat{g}_n(\mathsf{x}) \right) \\
&= \frac{1}{\sqrt{nh}}  \sum_{i=1}^n w_i (\mathsf{x})  \left(\frac{1}{nh}  \sum_{j=1}^n w_j (x_i) g(x_j) - g(x_i) \right)    
+ \frac{1}{\sqrt{nh}} \sum_{i=1}^n w_i (\mathsf{x})  \left(\frac{1}{nh}  \sum_{j=1}^n w_j (x_i) \varepsilon_j - \varepsilon_i \right) \\
&=: B_{2n} + \xi_{2n,\mathsf{LP}}. 
\end{align*}
By the mean-value theorem,
\begin{equation*}
B_{2n} =\frac{1}{\sqrt{nh}} \sum_{i=1}^n w_i (\mathsf{x}) \left( \frac{1}{nh} \sum_{j=1}^n w_j (x_i)  \left( r_p^{\prime} \left( \frac{x_j - x_i}{h}\right) H^{-1}_{p}\beta_p(x_i)  + h^{p+1} \frac{g^{(p+1)}(\tilde{x}_{ij})}{(p+1)!}\left(\frac{x_j - x_i}{h}\right)^{p+1} \right) -  g(x_i) \right) ,
\end{equation*}
where $\tilde{x}_{ij}$ takes values in the open interval $(\min \{x_j,x_i\}, \max \{x_j,x_i\})$. By definition of the weights~$w_j$, $(nh)^{-1}\sum_{j=1}^n w_j (x_i) r_p^\prime (( x_j - x_i)/h) = \iota_0^\prime$, so that
\begin{equation*}
B_{2n} =\frac{1}{\sqrt{nh}} \sum_{i=1}^n w_i (\mathsf{x}) \frac{1}{nh} \sum_{j=1}^n w_j (x_i) h^{p+1} \frac{g^{(p+1)}(\tilde{x}_{ij}) }{(p+1)!} \left(\frac{x_j - x_i}{h}\right)^{p+1} .
\end{equation*}
Using $\max_{i,j: w_j(x_i) >0}\{|x_j - \mathsf{x}|, |x_i - \mathsf{x}|  \} \leq 2h$ and H\"{o}lder continuity of $g^{(p+1)}$ (Assumption~\ref{Ass_g}(iii)),
\begin{equation*}
B_{2n} = \sqrt{nh^{2p+3}} \frac{g^{(p+1)}(\mathsf{x})}{(p+1)!} C_{\mathsf{LP},n} (1+ O(h^{\eta}))   = \sqrt{nh^{2p+3}} \frac{g^{(p+1)}(\mathsf{x})}{(p+1)!} C_{\mathsf{LP},n} + \op(1) ,
\end{equation*}
where the last equality follows from the fact that $\sqrt{nh^{2p+3}} \frac{g^{(p+1)}(\mathsf{x})}{(p+1)!} C_{\mathsf{LP},n}=\Op(1)$, which holds by Lemma~\ref{lemmaKernelConstants} whether $\mathsf{x}$ is interior or on the boundary. Combining this with \eqref{eq:Bn} we then find
\begin{equation*}
B_{2n} - B_n =  \sqrt{nh^{2p+3}} \frac{g^{(p+1)}(\mathsf{x})}{(p+1)!} (C_{\mathsf{LP},n} - C_n ) +  \op(1) =: A_n + \op(1),
\end{equation*}
which concludes the proof of the expansion. The limits of $A_n$ follow immediately from Lemma~\ref{lemmaKernelConstants} and by noticing that $\sqrt{nh^{2p+3}} \to \sqrt{\kappa}$ by Assumption~\ref{Ass_K,h}(ii). $\hfill \square$

\bigskip
\noindent \textsc{proof of lemma \ref{lemmaequivkerGP}.} Part~(i). The result for $\xi_{1n}$ follows from $r_{1n} := (nh)^{-1/2}\sum_{i=1}^n (w_i(\mathsf{x}) - f(\mathsf{x})^{-1} \mathsf{w}((x_i-\mathsf{x})/h))\varepsilon_i \plowto 0$. This in turn follows because $\E [r_{1n}]=0$ and $\E [r_{1n}^2|\mathscr{X}_n]= \mathsf{W}_{1n} \plowto 0$; see the proof of Lemma~\ref{new supp lemma 1} with $\zeta_i^2= \sigma^2 (x_i), i=1,\ldots,n$. The result for $\xi_{2n,\mathsf{GP}}$ follows similarly because $r_{2n} := (nh)^{-1/2}\sum_{i=1}^n (w_{\mathsf{GP}\text{-}\mathsf{bc},i}(\mathsf{x}) - f(\mathsf{x})^{-1} \mathsf{w}_{\mathsf{GP}\text{-}\mathsf{bc}}((x_i-\mathsf{x})/h))\varepsilon_i \plowto 0$. This in turn follows because $\E [r_{2n}]=0$ and $\E [r_{2n}^2|\mathscr{X}_n] \plowto 0$, which can be shown in the same way as for $r_{1n}$ using the method of proof of Lemma~\ref{new supp lemma 1} and applying also Lemma~\ref{lemmaKernelConstants}.

\medskip \noindent \textsc{part}~(ii). The proof is identical to that in part~(i), but replacing interior point equivalent kernels with boundary point equivalent kernels. $\hfill \square$

\bigskip
\noindent \textsc{proof of lemma \ref{lemmaequivker2}.} We first write
\begin{equation*}
    \left(\begin{matrix}
        \xi_{1n} \\
        \xi_{2n,\mathsf{LP}}
    \end{matrix}\right) =  \frac{1}{\sqrt{nh}} \sum_{i=1}^n
    \left(\begin{matrix}
 w_{i} (\mathsf{x}) \\
 (nh)^{-1}\sum_{j=1}^n w_{j} (\mathsf{x}) w_{i} (x_j) - w_{i} (\mathsf{x})
    \end{matrix}\right) \varepsilon_i = \left[\begin{matrix}
        1 & 0 \\ -1 & 1
    \end{matrix}\right] s_n ,
\end{equation*}
where $s_n = (s_{1n}, s_{2n})^{\prime} := (nh)^{-1/2} \sum_{i=1}^n (w_i (\mathsf{x}), (nh)^{-1}\sum_{j=1}^n w_j (\mathsf{x}) w_i (x_j)  )^{\prime} \varepsilon_i$. Clearly it suffices to prove the results separately for $s_{1n}$ and~$s_{2n}$.

\medskip \noindent \textsc{part}~(i). The proof for $s_{1n}$ is given in the proof of Lemma~\ref{lemmaequivkerGP} because $s_{1n}=\xi_{1n}$. The proof for $s_{2n}$ is identical, but applies $\mathsf{W}_{2n}$ from the proof of Lemma~\ref{new supp lemma 1} instead of~$\mathsf{W}_{1n}$.

\medskip \noindent \textsc{part}~(ii). The proof is identical to that in part~(i), but applies $\mathsf{Z}_{1n}^{\dagger}$ and $\mathsf{Z}_{2n}^{\dagger}$ from the proof of Lemma~\ref{new supp lemma 2} instead of $\mathsf{W}_{1n}$ and~$\mathsf{W}_{2n}$. $\hfill \square$

\section*{additional references}

\begin{description}

\item \textsc{Cattaneo, M.D., Jansson, M., and Ma, X.}\ (2024): Local regression distribution estimators. \emph{Journal of Econometrics} 240, article 105074.

\item \textsc{Cavaliere, G., Gon\c{c}alves, S., Nielsen, M.\O ., and Zanelli, E.}\ (2025): Improved inference for nonparametric regression.

\item \textsc{Hansen, B.E.}\ (2008): Uniform convergence rates for kernel estimation with dependent data. \emph{Econometric Theory} 24, 726–-748.

\item \textsc{Masry, E.}\ (1996): Multivariate local polynomial regression for time series: Uniform strong consistency and rates. \emph{Journal of Time Series Analysis} 17, 571--599.

\end{description}

\end{document}